\documentclass[twocolumn,aps,pra]{revtex4-2}
\usepackage{graphicx,amsmath,amssymb,hyperref,csquotes}
\usepackage{float} 
\DeclareGraphicsExtensions{.pdf,.png,.jpg}

\begin{document}
	
	\title{Overlooked local interactions in the EPR Paradox}
	
	\author{N.O.~Chudak}
	\author{O.S.~Potiienko}
	\author{I.V.~Sharph}
	\author{V.P.~Smolyar}
	\affiliation{Department of Theoretical and Experimental Nuclear Physics, Odessa Polytechnic National University, Odessa, Ukraine}
	
\begin{abstract}
	Five objections to the conventional arguments underlying the EPR \enquote{paradox} are presented. It is shown that for entangled subsystems the formation of the post-measurement state necessarily involves local interactions affecting both subsystems, contradicting standard EPR assumptions. Correlations between measurements by remote apparatuses are shown to be consistent with relativistic principles. For entangled eigenstates of total momentum or total spin, eliminating redundant degrees of freedom in analogy with generalized Hamiltonian dynamics prevents the emergence of the EPR \enquote{paradox}. A different paradox is identified for a quantum charged particle, whose electric field is shown to be determined by potential configurations encoded in the quantum state rather than by actual measurement events.
\end{abstract}	
	
	\maketitle

\tableofcontents

\section{Introduction}

The considerations leading to the Einstein-Podolsky-Rosen (EPR) paradox and the concept of entangled quantum states~\cite{EPRPhysRev.47.777,schredinger_1935,schredinger_1936,BohrNPhysRev.48.696,FurryPhysRev.49.393,AharonovBohmPhysRev.108.1070} have been extensively studied~\cite{ColloquiumRevModPhys.81.1727,CANTRELL1978499,PhysRevX.13.021031,Kupczynski:2016ysv,book:1012864,PhysRevLett.113.160402,BanaszekPhysRevA.58.4345,Laloe:2001tbo,Norsen2005EinsteinsB,Home:1991kk,Selleribook:970920,Bricmont2021}.
Typically, the system under consideration consists of two interacting quantum subsystems, labeled $1$ and $2$. For convenience, we denote the composite system as \mbox{$1\otimes2$}. This notation does not imply a literal tensor product, but it reflects the fact that, in the nonrelativistic case, the Hilbert space of the composite system is the tensor product of the Hilbert spaces of the individual subsystems.

We represent the state of the system using a probability amplitude $\Psi(t, x_1, x_2)$, where $t$ is an arbitrary time, and $x_1$, $x_2$ denote the sets of dynamical variables corresponding to subsystems $1$ and $2$, respectively. For example, if the system consists of two particles, $x_1$ and $x_2$ may be interpreted as their position vectors $\vec{r}_1$ and $\vec{r}_2$, which can be measured.

Due to the interaction between subsystems $1$ and $2$, the measurement outcomes for one subsystem cannot be predicted independently of the other. Consequently, the individual subsystems do not possess independent quantum states—only the state of the composite system \mbox{$1\otimes2$} exists. Formally, this is expressed by the non-factorizability of the probability amplitude:
\begin{equation}\label{entanglement}
	\Psi(t, x_1, x_2) \ne \Psi_1(t, x_1)\Psi_2(t, x_2).
\end{equation}
Following Schr\"{o}dinger~\cite{schredinger_1935}, a state satisfying \eqref{entanglement} is said to be \emph{entangled}. Importantly, if the interaction between the subsystems ceases due to spatial separation, the entanglement typically persists. Thus, even in the absence of interaction, measurements performed on one subsystem remain correlated with those on the other.
This persistence of correlations lies at the heart of the EPR argument, which leads to the paradoxical conclusion that a measurement on subsystem~$1$ appears to instantaneously determine the state of subsystem~$2$, despite the absence of any interaction. Since the measurement apparatus interacts only with subsystem~$1$, and the subsystems are assumed to be noninteracting, it seems that the state of subsystem~$2$ is affected without any physical influence.

In this work, we further argue that, according to quantum mechanics, subsystems $1$ and $2$ are not genuinely spatially separated and interact locally, without invoking action at a distance. By the absence of spatial separation, we mean that the time evolution of the state of the system \mbox{$1\otimes2$} always proceeds in such a way that, at any given moment, there is a nonzero probability of observing the particles of the two subsystems at arbitrarily small distances from one another.
This occurs because the time evolution of the system's state involves two competing processes: (i) the drift of the regions in which the particles of subsystems $1$ and $2$ are most likely to be found, and (ii) the spreading of these regions. The former increases the probability of finding the particles at large separations, thereby decreasing the probability of finding them close together. However, the latter includes spreading toward each other, which counteracts the decay of the probability at small separations.

To analyze this interplay, we consider the case where the particles of the two subsystems are assumed to be noninteracting. The behavior of the system can then be studied using the analytic properties of the expansion coefficients of an arbitrary state in the basis of coherent states~\cite{klauder2006fundamentals}, along with the known result concerning the free time evolution of a coherent state~\cite{Flugge2012practical}. We show that under these conditions, the probability of observing the particles of different subsystems at small separations remains nonzero.
According to the Schr\"{o}dinger equation, this implies that the time derivative of the probability amplitude includes nonzero contributions from the action of the local interaction operator between particles of different subsystems. Thus, the assumption that the asymptotic time evolution of the system \mbox{$1\otimes2$} proceeds without interaction between the subsystems leads to a contradiction: the time evolution calculated under this assumption implies, in fact, that the interaction between the subsystems does influence the evolution.

Taken together, these considerations imply that in the quantum case, there is in fact no possibility for subsystems $1$ and $2$ to remain noninteracting in the asymptotic state. Indeed, in the absence of interaction, spatial separation does not occur, and the subsystems continue to interact locally. Conversely, if we assume that spatial separation emerges due to some interaction between the subsystems, this implies that such interaction plays a significant role in shaping the properties of the asymptotic state.
Accordingly, any measurement performed in such a state takes place under conditions where the subsystems continue to interact appreciably. As a result, one cannot claim that the post-measurement state is formed without any influence on one of the subsystems.

To further support this conclusion, we note that the impossibility of spatial separation in quantum mechanics can also be inferred from the analyticity properties of the probability amplitude~\cite{krantz2002primer}. Indeed, the existence of the time-evolution operator requires the existence of derivatives of all orders with respect to all variables on which the probability amplitude depends. Furthermore, the Taylor expansion in time used to define the time-evolution operator must have an infinite radius of convergence. This requirement imposes constraints on the growth of the magnitude of these derivatives with increasing order.
If we choose, among the variables on which the amplitude depends, the differences between the coordinates of particles from different subsystems, then the probability amplitude must be an analytic function of each such difference for any fixed values of the other variables. As a result, it can vanish only on a discrete set of isolated points~\cite{krantz2002primer}. Therefore, even if the amplitude vanishes at the point where the coordinate difference is zero, it must be nonzero in some neighborhood of that point. Consequently, particles from different subsystems can always be observed arbitrarily close to one another with nonzero probability.

Thus, the analyticity of the probability amplitude prohibits genuine spatial separation in the quantum case. However, the imposition of special boundary conditions may violate analyticity. For example, this situation can arise in certain thought experiments involving particles in boxes~\cite{debroglie:jpa-00236174,Hardy,Norsen2005EinsteinsB,2002AmJPh..70..307G,Broglie1964TheCI}. In such cases, it is precisely the singular nature of the system that allows the paradox to be avoided. We comment on this conclusion in more detail in Appendix~3. Let us note that it has also been anticipated in earlier discussions~\cite{Louis_de_Broglie}. We comment on it in more detail in Appendix~3.

%

We also show that the post-measurement state of the system \mbox{$1\otimes2$} is formed not only due to interaction between the subsystems themselves, but also as a result of the local interaction of each subsystem with the measuring apparatus—even if only a single apparatus is used.

As we will demonstrate in the following sections, this follows from the fact that if the pre-measurement state of the system \mbox{$1\otimes2$} includes the possibility of finding subsystem~$1$ near the apparatus and subsystem~$2$ far from it, it must also include the alternative possibility of finding subsystem~$2$ near the apparatus and subsystem~$1$ far from it. An illustration of these two possibilities is provided in Fig.~\ref{fig:exchangeterms} below.

It is important to note that the Hamiltonian describing the interaction of the system \mbox{$1\otimes2$} with the apparatus must include terms corresponding to the interaction with particles from both subsystems. The structure of the pre-measurement state described above leads to nonzero contributions from all such interaction terms in the time derivative of the probability amplitude. These contributions determine the formation of the post-measurement state.
Properly accounting for these local interactions—both between the subsystems and with the measuring apparatus—resolves the aspect of the paradox related to claims about forming the system’s state or acquiring information without any physical disturbance.

While the considerations above resolve the aspect of the paradox related to the physical mechanism of state formation, another line of reasoning frequently emphasized in the original EPR discussion~\cite{EPRPhysRev.47.777} and related works~\cite{schredinger_1935,schredinger_1936,BohrNPhysRev.48.696,FurryPhysRev.49.393,AharonovBohmPhysRev.108.1070} concerns not how the post-measurement state is formed, but rather which state is said to result from the measurement. This issue is typically addressed through the reduction postulate~\cite{Von_Neumann_1996}, on which the EPR argument crucially relies. Therefore, to properly address this second aspect of the paradox, we must refine the reduction postulate.

To clarify the refinement of the reduction postulate, it is instructive to consider a concrete example. Suppose a quantum particle $q$ is in a non-eigenstate of momentum:
\begin{equation}\label{Momentum_no_eigenstate}
	\left| q \right\rangle = \int q(t_i, \vec{p}) \left| \vec{p} \right\rangle d\vec{p}.
\end{equation}
Here, $t_i$ is some initial moment in time, $\left| \vec{p} \right\rangle$ denotes a momentum eigenstate with eigenvalue $\vec{p}$, and $q(t_i, \vec{p})$ is the corresponding probability amplitude. Let us assume that the momentum of particle $q$ is measured via a capture process in which $q$ is absorbed by a classical particle $C$, forming an excited bound state of the two particles—i.e., without any emission of photons or other particles that might carry away part of the total momentum.

We consider the case where the classical momentum $\vec{P}_{\mathrm{cl}}$ of particle $C$ before capture is zero in the reference frame associated with the observer, and that the interaction between the particles responsible for the capture begins after the initial moment $t_i$.
Although particle $C$ is classical, there is no objection to describing it quantum mechanically. Its state prior to the interaction can be treated as the classical limit of a quantum state, as discussed in Refs.~\cite{Sakurai,LandauLifshitz,feynman2010quantum,klauder2006fundamentals}. In this framework, the classical nature of particle~$C$ means that the measurement errors achievable in the experiment under consideration are much larger than the corresponding quantum uncertainties in its dynamical variables. As a result, realizations of different potential values cannot be distinguished within experimental error margins, for example, for  the momentum of particle C. Consequently, the classical momentum is treated as having definite components, typically identified with the midpoints of these uncertainty intervals.

As a result, realizations of different potential values cannot be distinguished within experimental error margins, for example, for  the momentum of particle C cannot be distinguished within experimental error margins. Consequently, the classical momentum is treated as having definite components, typically identified with the midpoints of these uncertainty intervals.

Based on these considerations, we represent the state of the classical particle corresponding to $\vec{P}_{\mathrm{cl}} = \vec{0}$ (denoted $\left| \vec{P}_{\mathrm{cl}} = \vec{0} \right\rangle$), along with some internal state $\left| I \right\rangle$ of particle $C$, as
\begin{equation}\label{Clasuchnij_stan_Pcl0}
	\begin{aligned}
		&\left| \vec{P}_{\mathrm{cl}} = \vec{0} \right\rangle \otimes \left| I \right\rangle =\\
		&= \int C(t_i, \vec{P}, \vec{P}_{\mathrm{cl}} = \vec{0}) \left| \vec{P} \right\rangle d\vec{P} \otimes \left| I \right\rangle,
	\end{aligned}
\end{equation}
where $\left| \vec{P} \right\rangle$ denotes a quantum momentum eigenstate with eigenvalue $\vec{P}$, and $C(t_i, \vec{P}, \vec{P}_{\mathrm{cl}} = \vec{0})$ is the classical limit of the quantum probability amplitude associated with $\vec{P}_{\mathrm{cl}} = \vec{0}$. The modulus $\left| C(t_i, \vec{P}, \vec{P}_{\mathrm{cl}} = \vec{0}) \right|$ is significantly nonzero only for values of $\vec{P}$ that differ from $\vec{0}$ by quantities of the order of the quantum uncertainty. Since the possibilities described by this amplitude are indistinguishable experimentally, the explicit form of the function is irrelevant.

The time evolution of the composite system consisting of particles $q$ and $C$, including the interaction responsible for capture, is described by the evolution operator $\hat{U}$:
\begin{equation}\label{U}
	\begin{aligned}
		&\hat{U} \left( \left| q \right\rangle \otimes \left| \vec{P}_{\mathrm{cl}} = \vec{0} \right\rangle \otimes \left| I \right\rangle \right)=\\
		&= \int d\vec{p} \, d\vec{P} \, C(t_i, \vec{P},\times \vec{P}_{\mathrm{cl}} = \vec{0}) \, q(t_i, \vec{p}) \\
		&\times \hat{U} \left( \left| \vec{p} \right\rangle \otimes \left| \vec{P} \right\rangle \otimes \left| I \right\rangle \right).
	\end{aligned}
\end{equation}
Since particles $q$ and $C$ interact only with each other, the total system is invariant under spatial translations. Consequently, the operator $\hat{U}$ must map each basis state in Eq.~\eqref{U} to a total momentum eigenstate with the same total momentum eigenvalue.

\begin{equation}\label{U1}
	\begin{aligned}
		&\hat{U} \left( \left| q \right\rangle \otimes \left| \vec{P}_{\mathrm{cl}} = \vec{0} \right\rangle \otimes \left| I \right\rangle \right)\\
		&= \int d\vec{p} \, d\vec{P} \, C(t_i, \vec{P}, \vec{P}_{\mathrm{cl}} = \vec{0}) \, q(t_i, \vec{p})\times \\
		&\times \left| \vec{p} + \vec{P} \right\rangle \otimes \hat{U} \left( \left| I \right\rangle \right).
	\end{aligned}
\end{equation}

Note the formal similarity of the result \eqref{U1} to those obtained in quantum measurement models considered in Refs.~\cite{Von_Neumann_1996,Hall:2016oqf}. However, in contrast to those models, the present derivation involves no approximations and does not rely on any particular interaction form between the quantum system and the apparatus.
Taking into account the properties of $C(t_i, \vec{P}, \vec{P}_{\mathrm{cl}} = \vec{0})$, we note that the integrand is significantly nonzero only in a region of width comparable to the quantum uncertainty around $\vec{p}$. Using Eq.~\eqref{Clasuchnij_stan_Pcl0}, the result can be rewritten as
\begin{equation}\label{U_result}
	\begin{aligned}
		&\hat{U} \left( \left| q \right\rangle \otimes \left| \vec{P}_{\mathrm{cl}} = \vec{0} \right\rangle \otimes \left| I \right\rangle \right)=\\
		&= \int d\vec{p} \, q(t_i, \vec{p}) \, \left| \vec{P}_{\mathrm{cl}} = \vec{p} \right\rangle \otimes \hat{U} \left( \left| I \right\rangle \right).
	\end{aligned}
\end{equation}

It can be seen from the above considerations that during the entire time evolution of the system consisting of the two particles $C$ and $q$, there was no moment in time when each component of the momentum of the quantum particle $q$ had a single definite value. At every moment, these quantum degrees of freedom can only be associated with a set of potentially possible values, and none of the members of this set can be said to be "realized" at any particular time. They remain merely potential throughout the experiment.

In contrast, the components of the momentum of the classical particle that emerges after capturing the quantum particle $q$ by the classical particle $C$ have definite values in each system of the ensemble at every moment in time. Theoretically, these definite values correspond to states of the type given by Eq.~\eqref{Clasuchnij_stan_Pcl0}. In such states, all potentially possible values of the classical dynamical variable are so close to each other that they are indistinguishable within the experimentally achievable measurement accuracy. Thus, the entire interval of potentially possible values of a classical degree of freedom appears experimentally as a single point corresponding to one definite value. This also means that for any physical phenomenon whose realization depends on the value of a classical degree of freedom, the realizations corresponding to different potentially possible values cannot be distinguished experimentally. Some of these phenomena can be used to discover what the definite value of the classical variable is. In this sense, the values of classical degrees of freedom are never merely potential — they are always realized, in contrast to the quantum case.

As can be seen from Eq.~\eqref{U_result}, the definite value of the momentum of the classical particle formed after capturing the quantum particle is equal to one of the potentially possible values of the momentum of the quantum particle in its pre-measurement state. Since the momentum of a classical particle must always have a definite value, and since this value coincides with one of the potentially possible values of the quantum particle’s momentum in its initial state, one may say that the definiteness of the classical particle’s momentum "enforces" the manifestation of one of these possible values. In this sense, any quantum measurement necessarily involves a classical apparatus, among other reasons, because the definiteness of its degrees of freedom "enforces" the manifestation of the potential possibilities contained in the state being measured.

This manifestation occurs through the classical motion of the apparatus itself, as in the case under consideration, or through the classical motion of the particles composing the apparatus, as, for example, in measurements performed using a Wilson chamber. In the latter case, the classical motion of gas molecules leads to the formation of liquid droplets, and the occurrence of this motion reflects the presence, in the set of potentially possible position vectors of the quantum particle, of those values lying within the region where the droplets form. An analogous situation arises in the present case: as shown by Eqs.~\eqref{U}--\eqref{U_result}, each term in the expansion of the initial state~\eqref{Momentum_no_eigenstate} gives rise to a distinct classical motion of the apparatus. The observation of a particular motion can be interpreted as the manifestation of the corresponding eigenvalue of the measured observable—a feature also emphasized in quantum measurement models considered in Refs.~\cite{Von_Neumann_1996,Hall:2016oqf,omnes2018interpretation,Widom_Srivastava_2007}. However, unlike those models and in contrast to the reduction postulate \cite{Von_Neumann_1996} itself, we have seen that the emergence of a particular eigenvalue does not imply that the quantum system has been projected into the corresponding eigenstate of the measured variable. Indeed, as demonstrated by the example of momentum measurement for the quantum particle~$q$, there is no moment (either during or after the measurement) at which the particle can be said to occupy a momentum eigenstate.

Since, at any point in time, only sets of potentially possible values can be associated with the quantum dynamical variables (e.g. the components of a quantum particle’s momentum) and none of these sets ever reduces to a single value, the term manifestation is used here rather than realization. That is, as a result of a measurement, we uncover that the state being measured contained a certain potential possibility; however, in the general case, we cannot say that this possibility itself has been realized, because, as argued above, what becomes realized is not the possibility itself but a particular classical motion arising from its existence, which thereby serves as evidence of it.

As follows from the above considerations, the effect observed in the apparatus during measurement generally provides no information about the post-measurement state of the quantum system.
Moreover, the very question of what the state of an individual system from a quantum ensemble is after its interaction with a classical apparatus proves to be meaningless. This conclusion stems from the interpretation of the quantum state. As is well known, there are two opposing viewpoints regarding this interpretation. According to one of them, the state is attributed to each individual member of the quantum ensemble~\cite{omnes2018interpretation}. According to the other~\cite{EINSTEIN1936349,Kupczynski10.1063/1.2399618,Kupczynski:2016ysv,RevModPhys.42.358}, the state characterizes the ensemble as a whole, but not its individual constituents. 
In our view, \textit{the appropriate interpretation of the quantum state depends on whether classical degrees of freedom are present in the system under consideration}.

If we consider a purely quantum system that possesses no classical degrees of freedom, then the quantum state should be associated with each individual system of the ensemble, rather than with the ensemble as a whole. Indeed, when dealing with a pure quantum state described by a probability amplitude, we must imagine an ensemble in which each system is prepared in the same state corresponding to that amplitude. By contrast, a mixed state described by a density matrix represents an ensemble in which different systems may occupy different quantum states, due to the way the mixture is prepared (e.g., through uncontrolled external influences). Thus, the very concept of a pure quantum state requires that the state be attributed to each individual system in the ensemble.

Another argument supporting the assignment of the state to individual systems is the existence of interference between different quantum states. Interfering alternatives~\cite{feynman2010quantum} must be present within the same system; otherwise, it would imply that the probability of a measurement outcome for one system could be influenced by alternatives associated with other systems in the ensemble. A well-known example illustrating this point is the fact that interference can occur between different states of a single photon, but not between states of different photons~\cite{Dirac1930-DIRTPO}.
A third argument, based on the gauge principle~\cite{YangMillsPhysRev.96.191}, will be discussed later in this work.

The interpretation discussed above is sufficient when addressing the standard quantum-mechanical problem of predicting the set of possible measurement outcomes and the corresponding probabilities (or probability densities) across different realizations within the ensemble. In such cases, one can avoid a detailed analysis of the measurement process and instead rely on the reduction postulate, but only to the extent that it yields statistical predictions for both the values of the dynamical variable being measured and their associated probabilities. The question of the system's state after measurement may be disregarded, as it has no bearing on these predictions.

However, if we wish to examine in detail what happens to the system during the measurement process, we must consider a different type of ensemble --- a \textit{hybrid ensemble}~\cite{Hall:2016oqf,PhysRevA.86.042120}. Such an ensemble includes not only the quantum degrees of freedom associated with the system but also the classical degrees of freedom associated with the measuring apparatus, possessing the appropriate definiteness, as discussed above.

In close analogy with the example considered earlier, a hybrid system may be formally described as an entirely quantum system, after which the classical limit is taken for those degrees of freedom for which it can be consistently applied. In this approach, the state of the system can, in principle, be represented by a probability amplitude $\Psi(t, X_c, x_q)$, where $X_c$ denotes the set of dynamical variables corresponding to the classical degrees of freedom, and $x_q$ denotes the set of quantum dynamical variables. The Schr\"{o}dinger equation for the probability amplitude $\Psi(t, X_c, x_q)$ can, in principle, be written by taking into account all interactions among all degrees of freedom. To fully specify the dynamical problem, an appropriate initial condition must also be provided.

Let us assume that this dynamical problem can indeed be solved, and that the solution $\Psi(t, X_c, x_q)$ is obtained. The question then arises: how should this probability amplitude be interpreted, given what is actually observed in experiment?
To answer this question, let us once again consider the example with particles $C$ and $q$. In the expansion~\eqref{U_result}, the probability amplitude of each basis state $\left| \vec{P}_{\mathrm{cl}} = \vec{p} \right\rangle$ in the momentum representation is significantly different from zero only within a small region around the vector~$\vec{p}$. Meanwhile, as follows from~\eqref{Momentum_no_eigenstate}, the range of possible values of $\vec{p}$ can be arbitrarily large. This means that, in the momentum representation, the probability amplitude of the state represented by the entire linear combination~\eqref{U_result} can be nonzero over a sufficiently wide range of~$\vec{p}$ values. 
Within this range, one can choose two values, $\vec{P}_{\mathrm{cl}} = \vec{p}_1$ and $\vec{P}_{\mathrm{cl}} = \vec{p}_2$, whose components differ by more than the quantum uncertainty of the classical particle's momentum. Such two values can no longer be regarded as merely potentially possible values that could coexist within the same ensemble system, as is the case for quantum degrees of freedom. Rather, these two values (or values close to them within an insignificant quantum uncertainty) must be regarded as realized in different systems of the hybrid ensemble. 

Analogously, in the general case of a hybrid system, let $X_c^{(1)}$ and $X_c^{(2)}$ be two sets of values of the classical degrees of freedom, such that at least one corresponding classical dynamical variable differs between them by more than its quantum uncertainty. In this case, $X_c^{(1)}$ and $X_c^{(2)}$ should be interpreted not as potentially possible values within a single ensemble system, but as realizations in different systems of the hybrid ensemble. Based on this reasoning, the probability amplitude $\Psi(t, X_c, x_q)$ can no longer be interpreted as a characteristic of an individual system in the ensemble, but only as a characteristic of the ensemble as a whole.
This means that the quantity $|\Psi(t, X_{c}, x_{q})|^{2}$ can be used to calculate the fraction of systems in the hybrid ensemble in which a particular set of values of the dynamical variables $X_{c}$ and $x_{q}$ is realized. However, it cannot be used to make predictions concerning an individual system of the ensemble. For example, it is meaningless to predict the probabilities of different values of $X_{c}$ being realized in a given system of the hybrid ensemble, since in that system a definite set of these values already exists, and we can only determine what these values are.

Thus, the refinement of the reduction postulate \cite{Von_Neumann_1996}, introduced earlier and required for the subsequent discussion, consists in the recognition that the manifestation of a particular eigenvalue of the measured observable in a single measurement run performed on an individual system of the ensemble generally provides no information about the state of that system after the measurement. As seen from the above considerations, for hybrid systems that include both quantum and classical degrees of freedom, the question of time evolution of the state of an individual system in the ensemble is, in principle, ill-defined. In view of the EPR considerations, the question of the post-measurement state of one subsystem following a measurement performed on the other subsystem likewise has no physical meaning. Accordingly, we shall not consider this question further in this work.

Let us emphasize the distinction between the present approach and the existing ones. In Ref.~\cite{Von_Neumann_1996}, it was asserted that a quantum system switches from a pure state to a mixture as a result of measurement. However, the analysis there referred solely to the state of the quantum system under measurement, without taking into account the classical degrees of freedom of the measuring apparatus. In contrast, our analysis concerns a hybrid system composed of an interacting quantum system and a classical apparatus. Therefore, the measurement process does not lead to a mixture, but rather to an analogue of a pure quantum state, albeit with a revised interpretation.

As already mentioned in Refs.~\cite{EINSTEIN1936349,Kupczynski10.1063/1.2399618,Kupczynski:2016ysv,RevModPhys.42.358}, the quantum state has been interpreted as a characteristic of the ensemble as a whole, rather than of its individual systems. However, those interpretations were applied to arbitrary quantum systems. In contrast, in our approach such an ensemble interpretation emerges specifically upon transitioning from a purely quantum system to a hybrid one, that is, with the inclusion of classical degrees of freedom.

Apart from these considerations, it is also essential to turn to another class of questions concerning the theoretical and experimental study of correlations between measurements performed at large spatial separations~\cite{article,Piccioni1989,1982FoPh...12.1171H,Ghirardi:1983ff,DIEKS1982271,Stapp1988,Selleribook:970920,Kupczynski:2016ysv,ClauserHorneShimonyHoltPhysRevLett.23.880,AspectGrangierRogerPhysRevLett.49.91,A_R_Wilson_1976,V_Paramananda_1987,PhysRevX.13.021031}.
Such long-distance correlations arise when an entangled state is an eigenstate of a certain dynamical variable of the composite system \mbox{$1 \otimes 2$}, e.g. of the total momentum~\cite{EPRPhysRev.47.777} or total spin~\cite{AharonovBohmPhysRev.108.1070}.  
These states may be prepared so that they are not eigenstates of the momentum or spin components of the individual subsystems $1$ and $2$. In such cases, the values of these components in a given run of a measurement are determined by the interaction between each subsystem and its measuring device, which may be placed arbitrarily far apart.

Because interactions are local and relativity forbids faster-than-light influence, spatial separation is typically taken to imply that the measurement processes for each subsystem occur independently. This raises the question of how the required correlations nevertheless arise and how they are to be reconciled with the principles of relativity.
One relevant  observation concerning this issue is that, in the cases mentioned above, the time evolution of the state of the system \mbox{$1 \otimes 2$}, whose state space is $\mathcal{H}_{1\otimes2}$, takes place within a linear subspace $\mathcal{H}_1 \subset \mathcal{H}_{1\otimes2}$ consisting of eigenstates of the total momentum or total spin. This subspace can be specified by imposing a set of linear constraints on the coefficients in the expansion of an arbitrary state in some chosen basis of $\mathcal{H}_{1\otimes2}$. Such basis expansions were considered in Refs.~\cite{EPRPhysRev.47.777,schredinger_1935}, but the existence of these constraints was not taken into account. Since a measurement interaction need not preserve them, their violation will remove the state from $\mathcal{H}_1$ and thereby destroy the corresponding correlations.

As will be discussed later, in the case of a total-spin eigenstate with all components equal to zero, measuring one spin component of one subsystem will generally drive the state out of the subspace of eigenstates of the other two components of the total spin, thereby eliminating the associated correlations.  
Therefore, a proper analysis of such correlations requires a more detailed treatment of the measurement process than is afforded by a straightforward application of the projection postulate.

Another observation, which will be used later in this work, is the existence of an analogy between the description of entangled states subject to the above-mentioned linear constraints and the description of quantized gauge fields~\cite{Gupta:1949rh,Bleuler1950ANM,Christ:1980ku,faddeev1991gauge}. 
In the case of gauge fields, the gauge-fixing conditions are imposed not on the field operators but on the elements of the state space, thereby selecting a certain linear subspace. 
In both these situations of gauge fields and the present case the time evolution of the state takes place entirely within a linear subspace. Therefore, a natural next step is to introduce variables on which the state vector depends in such a way that the constraints are automatically satisfied~\cite{Christ:1980ku}. 
That is, any function of these variables automatically belongs to the required subspace. 
In our case, this means choosing some basis in $\mathcal{H}_1$ and considering only linear combinations of the basis states. 
The coefficients of such an expansion form the probability amplitude of a state that is guaranteed to belong to $\mathcal{H}_1$, and these coefficients depend on the quantities parametrizing the chosen basis elements of $\mathcal{H}_1$.

As an example, consider the case in which the time evolution of the state of the system \mbox{$1 \otimes 2$} takes place within the subspace \mbox{$\mathcal{H}_1 = \mathcal{H}_{\vec{P}=\vec{0}}$}, spanned by the eigenstates of the total momentum $\vec{P}$ of the system \mbox{$1 \otimes 2$} corresponding to the eigenvalue $\vec{P} = \vec{0}$. 
Since we intend to measure the momenta of the subsystems using two widely separated detectors, for the application of the reduction postulate we can take as a basis in the full space $\mathcal{H}_{1\otimes 2}$ the joint momentum eigenstates of the subsystems, 
\begin{equation}\label{PsiPP2R1R2}
	\begin{aligned}
		& \Psi \left( \vec{P}_1, \vec{P}_2; \vec{R}_1, \vec{R}_2 \right) = \\
		& \quad \exp\!\left( \frac{i}{\hbar} \, \vec{P}_1 \cdot \vec{R}_1 \right)
		\exp\!\left( \frac{i}{\hbar} \, \vec{P}_2 \cdot \vec{R}_2 \right),
	\end{aligned}
\end{equation}   
where $\vec{P}_1$ and $\vec{P}_2$ are the momentum eigenvalues of subsystems $1$ and $2$, respectively, 
$\vec{R}_1$ and $\vec{R}_2$ are the position vectors of their respective centers of mass, 
and $(\vec{P}_1 \cdot \vec{R}_1)$ and $(\vec{P}_2 \cdot \vec{R}_2)$ denote the Euclidean scalar products of the corresponding vectors. 
The expansion coefficients in the basis~\eqref{PsiPP2R1R2} must satisfy the constraint that they vanish unless $\vec{P}_1 + \vec{P}_2 = 0$. 
At the same time, any function belonging to the subspace $\mathcal{H}_{\vec{P}=\vec{0}}$ must be invariant under spatial translations. 
This means that it can only depend on the difference $\vec{R}_2 - \vec{R}_1$. 
Therefore, any linear combination of the form
\begin{equation}\label{PsiP}
	\begin{aligned}
		& \Psi \left( t, \vec{R}_2 - \vec{R}_1 \right) = \int d\vec{p} \; \times \\
		& \quad \times \Psi \left( t, \vec{p} \right)
		\exp\!\left( \frac{i}{\hbar} \, \left( \vec{p} \cdot \left( \vec{R}_2 - \vec{R}_1 \right) \right)\right) ,
	\end{aligned}	
\end{equation}  
where $\int d\vec{p}$ denotes three-dimensional integration over the components of $\vec{p}$, is guaranteed to belong to the subspace $\mathcal{H}_{\vec{P}=\vec{0}}$ for any dependence $\Psi(t,\vec{P})$. 
Thus, the states in the subspace $\mathcal{H}_{\vec{P}=\vec{0}}$ are described by functions of the three independent components of $\vec{p}$ rather than of the six dependent components of $\vec{P}_1$ and $\vec{P}_2$.

An analogy may again be drawn with the description of gauge fields~\cite{faddeev1991gauge}, specifically with the stage prior to quantization where the methods of generalized Hamiltonian dynamics~\cite{Dirac:1958sq} are applied. Such a dynamics arises in cases where constraints are imposed on the generalized coordinates and momenta; the theory of gauge field quantization provides an example of such a situation. The presence of constraints necessitates the identification of the \emph{true dynamical variables}~\cite{faddeev1991gauge}, which are mutually independent and whose evolution is described by the Hamiltonian formalism. The remaining coordinates and momenta are expressed in terms of the true dynamical variables via the constraint equations.

A similar situation arises when considering the states of the system \mbox{$1\otimes 2$} within the subspace $\mathcal{H}_{\vec{P}=\vec{0}}$. In this subspace, the momentum operators ${{\hat{P}}_{1}}$ and ${{\hat{P}}_{2}}$ of subsystems~1 and~2 satisfy the constraint
\begin{equation}\label{Constraint_P}
	{{\hat{P}}_{1}} + {{\hat{P}}_{2}} = \hat{0}.
\end{equation}
Here $\hat{0}$ denotes the operator whose action on any function from $\mathcal{H}_{\vec{P}=\vec{0}}$ yields the function that takes the value zero for all arguments.

The constraint~\eqref{Constraint_P} implies that, within $\mathcal{H}_{\vec{P}=\vec{0}}$, there do not exist two independent true dynamical variables ${{\hat{P}}_{1}}$ and ${{\hat{P}}_{2}}$. The true dynamical variables are the components of a \emph{single} vector rather than two distinct vectors. This single vector could be, for example, the components of ${{\vec{P}}_{1}}$, or of ${{\vec{P}}_{2}}$, or of the relative momentum $\vec{p}$ appearing in Eq.~\eqref{PsiP}, among other choices. Since there is only one true dynamical variable in the form of a three-dimensional vector, both measuring apparatuses can measure only this same quantity. This leads to a measurement scheme that is atypical for quantum mechanics: two distinct apparatuses measure the same dynamical variable in the same state, which is not an eigenstate of that variable. As will be shown later, analogous reasoning applies to measurements of the spin components of the subsystems. Thus, it can be said that the majority of well-known experiments~\cite{ClauserHorneShimonyHoltPhysRevLett.23.880,AspectGrangierRogerPhysRevLett.49.91,A_R_Wilson_1976,V_Paramananda_1987} have been carried out in precisely such an atypical manner. 

In contrast, a typical quantum-mechanical measurement would involve measuring a single vector, such as $\vec{p}$, using only one apparatus. The values of ${{\hat{P}}_{1}}$ and ${{\hat{P}}_{2}}$, being non–true dynamical variables in this case, could then be obtained from the constraint equations, which, as seen from Eq.~\eqref{PsiP}, take the form
\begin{equation}\label{Zvazoc}
	\hat{{\vec{P}}}_{2} = \hat{\vec{p}}, \quad \hat{{\vec{P}}}_{1} = -\hat{\vec{p}}.
\end{equation}
Under such a typical quantum-mechanical measurement, the paradox does not arise: the correlations follow directly from the constraint equations and do not require any propagation of physical quantities from one apparatus to the other.
However, since the experiments~\cite{ClauserHorneShimonyHoltPhysRevLett.23.880,AspectGrangierRogerPhysRevLett.49.91,A_R_Wilson_1976,V_Paramananda_1987} 
are performed with two apparatuses, we should examine them in more detail. 
To this end, let us consider the example of momentum measurement discussed above, but with a certain modification. 
Suppose that two classical particles, $C_1$ and $C_2$, are used as the two apparatuses (Fig.~\ref{fig:twoapparatusmeasurement} below).
The initial classical momenta of both particles are zero in the reference frame of the measurement. 
After the time evolution of the state of the system \mbox{$1\otimes2$}, when the probability amplitude acquires significant weight in the vicinity of each classical particle, 
each particle captures one of the subsystems, either $1$ or $2$. 
In this way, the measurement of the subsystem momenta is carried out as described previously. To describe such a measurement, it is necessary to consider the hybrid system 
\mbox{$1\otimes 2\otimes C_{1}\otimes C_{2}$}, consisting of two classical particles, $C_1$ and $C_2$, 
and two quantum subsystems, $1$ and $2$. 
The state of the system \mbox{$1\otimes 2$} before the measurement can be represented in the form \eqref{PsiP}. 
The state of the two classical particles $C_1$ and $C_2$ prior to their interaction with the quantum subsystems $1$ and $2$ 
can be written in a form analogous to \eqref{PsiP} and \eqref{Clasuchnij_stan_Pcl0}:
\begin{equation}\label{Psi_cl_2_particles}
	\begin{aligned}
		& \Psi_{cl}\left(t,\vec{R}_{1}^{cl},\vec{R}_{2}^{cl},I \right)= \\ 
		& \quad \int d\vec{P}_{cl}\int d\vec{p}_{cl}\,
		C_{cl}\left(t,\vec{P}_{cl},\vec{p}_{cl},I \right) \times \\ 
		& \quad \times \exp\!\left[\frac{i}{\hbar}\,\left( \vec{p}_{cl}\cdot\left(\vec{R}_{2}^{cl}-\vec{R}_{1}^{cl}\right)\right) \right] \times \\ 
		& \quad \times \exp\!\left[\frac{i}{\hbar}\,\left( \vec{P}_{cl}\cdot\frac{\vec{R}_{2}^{cl}+\vec{R}_{1}^{cl}}{2}\right) \right],
	\end{aligned}	
\end{equation}
where $\vec{R}_{1}^{cl}$ and $\vec{R}_{2}^{cl}$ are the position vectors of the centers of mass of the classical particles 
$C_1$ and $C_2$, respectively; $\vec{P}_{cl}$ is the eigenvalue of the total momentum of these particles; 
$\vec{p}_{cl}$ is the eigenvalue of their relative momentum; and $I$ denotes the set of internal degrees of freedom 
of both classical particles. For simplicity, we assume that the two particles have equal masses, so that 
the position vector of their center of mass is $\left(\vec{R}_{1}^{cl}+\vec{R}_{2}^{cl}\right)/2$. 
The function $C_{cl}\left(t,\vec{P}_{cl},\vec{p}_{cl},I\right)$ is significantly nonzero only for values of 
$\vec{P}_{cl}$ and $\vec{p}_{cl}$ that differ from zero within the limits of quantum uncertainty, 
which is negligible compared to the measurement error.

Before the measurement, the probability amplitude of the full system 
\mbox{$1\otimes 2\otimes C_{1}\otimes C_{2}$} is given by the product of 
Eqs.~\eqref{PsiP} and \eqref{Psi_cl_2_particles}. 
This state can therefore be written as a linear combination of states that are 
eigenstates of the total momentum of the system 
\mbox{$1\otimes 2\otimes C_{1}\otimes C_{2}$} with eigenvalue $\vec{P}_{cl}$. 
Importantly, the coefficients of this expansion are significantly nonzero 
only for values of $\vec{P}_{cl}$ that, within the measurement error, 
cannot be distinguished from zero. 

The measurement process, similarly to Eqs.~\eqref{U}--\eqref{U_result}, 
can be described by a unitary time-evolution operator $\hat{\mathcal{U}}$. 
As discussed above, this description also entails a corresponding change in the interpretation of the state, 
in contrast to its meaning during the unitary evolution of the isolated system \mbox{$1\otimes 2$}.

Like the previously considered time-evolution operator $\hat{U}$ for the system of particles 
$C$ and $q$, the operator $\hat{\mathcal{U}}$ is translationally invariant. 
This implies that each total-momentum eigenstate of the hybrid system 
\mbox{$1\otimes 2\otimes C_{1}\otimes C_{2}$} is mapped by $\hat{\mathcal{U}}$ 
onto another eigenstate with the same eigenvalue of the total momentum. 

By repeating the transformations analogous to Eqs.~\eqref{U}--\eqref{U_result}, 
but now with the quantum state \eqref{PsiP} in place of \eqref{Momentum_no_eigenstate} 
and the state of the two classical particles \eqref{Psi_cl_2_particles} 
instead of \eqref{Clasuchnij_stan_Pcl0}, we arrive at the following conclusion. 
In each realization of the hybrid ensemble, after the absorption of the quantum 
subsystems by the classical particles, two composite classical particles are formed. 
Each of these composite objects consists of one original classical particle together 
with the absorbed quantum subsystem. 
The total classical momentum of the two composites is zero. 
Consequently, in every realization of the ensemble, the measured momenta of 
subsystems $1$ and $2$ appear as opposite vectors of equal magnitude. 

Thus, the observed momentum correlations arise directly from the \textbf{translational 
	symmetry} of the time-evolution operator. 
For the case of spin components (to be considered below), the corresponding 
symmetry is the \textbf{rotational invariance} of the entire system. 
Since such symmetries are intrinsic to the \textit{relativistic time-evolution operator} 
(regardless of its specific form or the dynamical quantities to which it is applied), 
as well as to its nonrelativistic approximation, the presence of correlations 
is necessarily consistent with the \textbf{principles of relativity}.

To explain this situation, let us take into account two points. 
The first point is that the presence of the \textit{time-evolution operator} in the 
consideration of the measurement process implies that the process is not \textbf{instantaneous}. 
Instead, the time-evolution operator describes processes occurring over a \textbf{nonzero time interval}, 
which may be very long or even formally infinite (as in scattering theory).

The second point concerns the \textbf{pre-measurement state} of the system, which includes two 
\textbf{complementary potential possibilities}, as illustrated in Fig.~\ref{fig:twoapparatusmeasurement}.
In the present case, the pre-measurement state allows for the observation of particles of subsystem~$1$ 
near apparatus~1 (the particle $C_1$) and particles of subsystem~$2$ near apparatus~2 (the particle $C_2$). 
At the same time, it also includes the possibility that particles of subsystem~$2$ might be observed 
near apparatus~1, and particles of subsystem~$1$ near apparatus~2.

Let us assume that apparatus~1 absorbs subsystem~$1$ as a result of the measurement run under consideration. 
This implies that after the measurement the probability of observing particles of subsystem~$1$ becomes 
negligible everywhere except within a small region near apparatus~1. 
Before the measurement, however, this probability was non-negligible 
not only near apparatus~1 but also, at a considerable distance, in the vicinity of apparatus~2.  
This leads to the conclusion that a probability flow directed from apparatus~2 to apparatus~1 must occur. 
As a consequence of this flow, the probability to observe particles of subsystem~$1$ near apparatus~2 decreases 
from its appreciable pre-measurement values to negligible ones after measurement. 
The principles of relativity place an upper bound on the rate at which probability can spread from apparatus~2 to apparatus~1. 
Consequently, they impose a lower bound on the duration of the measurement process, which is governed 
by the time-evolution operator $\hat{\mathcal{U}}$ mentioned above. 

At the same time, a momentum flow accompanies the probability flow. 
Indeed, during the absorption of subsystem~$1$ by apparatus~1, the apparatus gradually absorbs not only the subsystem itself 
but also its momentum, thereby receiving momentum directed toward it. 
This implies that subsystem~2 must acquire momentum in the opposite direction. 
Such momentum transfer arises from ordinary local interactions, which become possible because 
the potential possibility of observing particles of subsystem~2 near apparatus~1 still exists by the end of the measurement, 
as illustrated in Fig.~\ref{fig:twoapparatusmeasurement}. 
Therefore, when an apparatus captures one subsystem, it simultaneously repels the other. 
The existence of this repulsion explains why an apparatus can capture only one of the subsystems, 
even though it interacts with both. 

The repulsion also implies that, together with the probability flow directed from apparatus~2 to apparatus~1, 
a momentum flow directed from apparatus~1 to apparatus~2 arises. 
Both flows thus cover the same distance between the two apparatuses, 
and the rates at which these quantities spread are bounded by the same relativistic limit. 
Since the measurement time is sufficient for the probability to spread from one apparatus to the other, 
it is also sufficient for the momentum to spread in the opposite direction.
These considerations show that two spatially separated apparatuses can influence each other during the measurement 
without contradicting the principles of relativity. 
The counterflow of momentum between the apparatuses over the sufficiently long duration of the measurement 
establishes long-distance correlations between their outcomes. 
On the other hand, the very fact that the apparatuses mutually affect each other makes this situation unusual in quantum mechanics. 
Indeed, two apparatuses with such mutual influence measure the same dynamical variable, as discussed above.

Note that the discussion of correlations between two distant measurements, given above, addressed the problem only in principle, since it referred to correlations in a single realization of the measurement. In contrast, experiments such as those reported in \cite{KocherPhysRevLett.18.575,ClauserHorneShimonyHoltPhysRevLett.23.880,AspectGrangierRogerPhysRevLett.49.91,A_R_Wilson_1976,V_Paramananda_1987,PhysRevX.13.021031} probed only statistical characteristics of the ensemble as a whole, rather than individual realizations. In that situation, the existence of correlations can be explained by the fact that the two distant apparatuses operate under identical conditions, as illustrated in Fig.~\ref{fig:twoapparatusmeasurement}. Both devices interact in the same way with subsystems $1$ and $2$, and therefore their measurement outcomes must coincide on the statistical level. Such correlations require no transfer of any physical quantity between the apparatuses and hence do not lead to a paradox.

Developing the arguments presented above, we show in the following sections that
the EPR paradox does not arise within quantum mechanics. This conclusion is to a
significant extent based on the fact that quantum dynamics is governed by
\emph{potential possibilities} rather than by actual events, as follows from the
Schr\"odinger equation. Indeed, the Hamiltonian acts on the probability amplitude as a function defined on
the set of potential possibilities, producing a new function on the same set. If
this function is nonzero for certain possibilities, it determines the subsequent
dynamics of the state through the time derivative of the probability amplitude.
This paradoxical feature was discussed in Ref.~\cite{Smolin:2011nn}.
However, in that case the potentialities govern a nonobservable quantity, namely
the time derivative of the probability amplitude. In the second part of the
present work, we draw attention to a different situation, in which the potential
possibilities encoded in a quantum state determine an \emph{observable} quantity.

Specifically, let us consider the electric field generated by a pointlike quantum
particle carrying a nonzero electric charge. As will be justified below, in this
case the gauge principle~\cite{YangMillsPhysRev.96.191}, together with the
corresponding dynamical equations, implies that the observable field strength in
each individual member of the ensemble is determined by \emph{all} potential
positions of the particle present in its quantum state. These potential positions
can coexist within each system of the quantum ensemble only as possibilities.
They are mutually exclusive and, in reality, can be manifested only in different
members of the ensemble. At the same time, the electric field strength, as a
function of the spatial position, exists in each individual system of the
ensemble. Therefore, in each such system, the result of an actual measurement of
the electric field strength is determined exclusively by potential possibilities
that not only do not realize (or manifest) themselves, but in fact cannot manifest
themselves in principle within that system. 

Some experimental support for these considerations can be found in measurements
of the electric field strength inside atoms
\cite{SCHMIDT1993101,MULLERCASPARY201762,Field_in_atom,PhysRevB.98.121408,Shibata2017ElectricFI}.
The result of one such measurement is shown in
Fig.~\ref{fig:atomfield}~\cite{Field_in_atom}. As seen in
Fig.~\ref{fig:atomfield}, there are clearly recognizable field singularities
produced by the pointlike, positively charged classical atomic cores, whereas
\textbf{no singularities are observed that could be attributed to negatively
	charged quantum electrons}. This indicates that quantum electrons manifest
themselves not as pointlike particles, but rather as a continuously distributed
charge, which does not give rise to field singularities, in contrast to a
classical point charge.

Another, albeit indirect, argument supporting the paradox under consideration
is the widespread applicability of the independent-electron approximation in
solid-state theory~\cite{ashcroft1976solid}. Even its refinement within the
mean-field framework~\cite{HartreeDouglasRayner1897_19581957Tcoa,ashcroft1976solid}
relies on representing the effective field as one generated by all potentially
possible positions of the electrons.

The issues outlined above provide the context and motivation for the detailed
analysis presented in the following sections.

\section{Objections to EPR arguments}

\subsection{Objection number 1}

As an example of an entangled state, the paper \cite{EPRPhysRev.47.777} considers the state of a one-dimensional two-particle system, which is an eigenstate of both the relative coordinate of the particles and the total momentum of the system:
\begin{equation}\label{StanEPR}
	\begin{aligned}
		&{{\psi }_{x={{x}_{0}},P=0}}\left( {{x}_{1}},{{x}_{2}} \right)\sim\\
		&\sim\int\limits_{-\infty }^{+\infty }{\exp \left( \frac{i}{\hbar }p\left( {{x}_{1}}-{{x}_{2}}+{{x}_{0}} \right) \right)dp}.
	\end{aligned}
\end{equation} 
Here, $\hbar$ is the Planck constant, ${{x}_{1}}$ and ${{x}_{2}}$ are the coordinates of the particles, $x={{x}_{2}}-{{x}_{1}}$ is the relative coordinate of the particles, and $P$ is the total momentum of the system, canonically conjugate to the center of mass coordinate \mbox{$X={\left( {{m}_{1}}{{x}_{1}}+{{m}_{2}}{{x}_{2}} \right)}/{\left( {{m}_{1}}+{{m}_{2}} \right)}\;$} (${{m}_{1}}$ and ${{m}_{2}}$ are the masses of the particles). 

It is stated in \cite{EPRPhysRev.47.777} that the particles do not interact with each other when measured in this state, and that bringing one of them into interaction with the measuring apparatus will not affect the other.

However, if the particles do not interact, the state cannot remain an eigenstate of the relative coordinate. Indeed, in such a case, the Hamiltonian of the system is
\begin{equation}\label{Hamiltonian_two_particles_without_interaction_1D}
	\hat{H}=-\frac{{{\hbar }^{2}}}{2M}\frac{{{d}^{2}}}{d{{X}^{2}}}-\frac{{{\hbar }^{2}}}{2\mu }\frac{{{d}^{2}}}{d{{x}^{2}}}	
\end{equation} 
Here, $M=m_{1}+m_{2}$ is the total mass of the particles, and $\mu ={\left( {{m}_{1}}{{m}_{2}} \right)}/{\left( {{m}_{1}}+{{m}_{2}} \right)}\;$ is the reduced mass. Using the time evolution operator corresponding to the Hamiltonian \eqref{Hamiltonian_two_particles_without_interaction_1D}, we find that if at the initial time $t=0$ the state of the system is \eqref{StanEPR}, then at any later time $t>0$ it will be
\begin{equation}\label{StanEPRt}
	\begin{aligned}
		& \Psi \left( t,{{x}_{1}},{{x}_{2}} \right)=\\
		&=\exp \left( -\frac{i}{\hbar }\hat{H}t \right){{\psi }_{x={{x}_{0}},P=0}}\left( {{x}_{1}},{{x}_{2}} \right)\sim  \\ 
		& \sim \int\limits_{-\infty }^{+\infty }{\exp \left( -\frac{i}{\hbar }\left( \frac{{{p}^{2}}}{2\mu }t-p\left( {{x}_{1}}-{{x}_{2}}+{{x}_{0}} \right) \right) \right)dp}. \\ 
	\end{aligned}	
\end{equation}   
This state is no longer an eigenstate of the relative coordinate $x$. 

In addition, let us assume that $x>0$. Suppose we measure the coordinates $x_{1}$ and $x_{2}$ in the state \eqref{StanEPR}. According to \eqref{StanEPR}, the probability of observing the particle with coordinate $x_{2}$ to the left of the particle with coordinate $x_{1}$ is zero. If there is no interaction between the particles, then the question arises: what exactly \enquote{prevents} the particle with coordinate $x_{2}$ from being observed to the left of $x_{1}$?

Thus, \textbf{maintaining the eigenstate of the relative coordinate of the particles is impossible without interaction between the particles}.  

It was pointed out in \cite{Selleribook:970920} that the state \eqref{StanEPR} exists only for an instant. However, this was not considered a refutation of the EPR arguments.

Since the state \eqref{StanEPR} describes a system of interacting particles, any measurement on one particle will affect the other. Therefore, whatever the state of the system after the measurement, it will be formed under the condition of influence on both particles. Accordingly, there is no paradoxical formation of the state of a particle without any impact on it.

\subsection{Objection Number  2}
As mentioned earlier in the introduction of this article, where we discussed the situation considered in the works \cite{EPRPhysRev.47.777,schredinger_1935,AharonovBohmPhysRev.108.1070}, an entangled state of a isolated system \mbox{$ 1\otimes2 $}, consisting of two subsystems $ 1 $ and $ 2 $, is considered.  In \cite{EPRPhysRev.47.777,schredinger_1935,AharonovBohmPhysRev.108.1070}, the scenario is examined where subsystems $ 1 $ and $ 2 $ interacted in the past, leading to entanglement. However, as time progresses, these systems cease to interact, and measurements are subsequently performed on the non-interacting systems.

Some of the \enquote{paradoxical} conclusions from the analysis of this situation include the statement that when measuring one of the subsystems, the state of the second is formed \textbf{without any interaction between this subsystem and anything else}. 
%

We would like to draw attention to the fact that when it is stated that there was no interaction between subsystems during the measurement, or there was no interaction between one of them and the apparatus ("without in any way disturbing a system" \cite{EPRPhysRev.47.777}), this is only an \textbf{assumption}. It is not based on experimental confirmation or theoretical considerations.

In the papers \cite{EPRPhysRev.47.777,schredinger_1935,FurryPhysRev.49.393}, it is simply stated that the two subsystems ceased to interact starting from some point in time. However, no reasoning is provided regarding how this is achieved or why it can be assumed that the subsystems have indeed ceased to interact. In the papers \cite{AharonovBohmPhysRev.108.1070,CANTRELL1978499}, the assumption that they have ceased to interact is based on another assumption — that they \enquote{have separated enough so that they cease to interact.} The exact meaning of spatial separation of quantum systems in these works is not provided. One might think that this refers to a state in which particles from different subsystems can be detected with a non-zero probability only at such large distances that they do not interact with each other. But at the same time, the papers \cite{AharonovBohmPhysRev.108.1070,CANTRELL1978499} only consider measurements of the spin projections of the particles, not their spatial coordinates. Furthermore, the theoretical considerations focus solely on the spin dependence of the state. In this case, the question arises as to how it is known that the subsystems are indeed spatially separated and do not interact with each other.

The same applies to the interaction of the subsystems with the apparatus. Now, it concerns the spatial separation between the subsystems and the apparatus. If a certain dynamic variable value was measured for subsystem $1$ in a given system of a quantum ensemble, it means that, in the state, there was a potential possibility of observing subsystem $1$ close to the apparatus, while subsystem $2$ was observed far from it. However, this does not imply that, in the same state, there cannot be a potential possibility of observing subsystem $2$ close to the apparatus and subsystem $1$ far from it.

All of these considerations indicate that before drawing conclusions about whether or not certain interactions influence the formation of the state after the measurement, it is necessary to analyze in more detail the properties of the coordinate representation of the state that existed before the measurement. This state is the result of the time evolution of the initial state, in which the particles of subsystems $1$ and $2$ could have been observed close to each other with high probability and interacted significantly. Therefore, we will begin our analysis with this initial state.

The complete list of interactions that could, in principle, influence the formation of the state of the system $1 \otimes 2$ includes interactions between subsystems $1$ and $2$, as well as interactions of each subsystem with the apparatus. First, we will consider the potential impact of interactions between subsystems on the formation of the state, and then their interactions with the apparatus.

\subsubsection{Reduction postulate insufficiency}

Let $N_1$ be the number of particles in subsystem $1$, and $N_2$ the number of particles in subsystem $2$. Let ${{r}^{\left( 1 \right)}}=\left\{ \vec{r}_{1}^{\left( 1 \right)},\vec{r}_{2}^{\left( 1 \right)},\ldots ,\vec{r}_{{{N}_{1}}}^{\left( 1 \right)} \right\}$ be the set of position vectors of particles in subsystem $1$ that can be observed during a measurement. We will also introduce analogous notation ${{r}^{\left( 2 \right)}}=\left\{ \vec{r}_{1}^{\left( 2 \right)},\vec{r}_{2}^{\left( 2 \right)},\ldots ,\vec{r}_{{{N}_{2}}}^{\left( 2 \right)} \right\}$ for the set of position vectors of particles in subsystem $2$. The state of the system $1 \otimes 2$ is described by the probability amplitude $\Psi(t, r^{(1)}, r^{(2)})$ in the coordinate representation. The influence of the interaction between subsystems $1$ and $2$ on the state of the system $1 \otimes 2$ is determined by the equation
\begin{equation}\label{Shredinger12}
	\begin{aligned}
		& i\hbar \frac{\partial \Psi \left( t,{{r}^{\left( 1 \right)}},{{r}^{\left( 2 \right)}} \right)}{\partial t}=\left( {{{\hat{H}}}^{\left( 1 \right)}}+{{{\hat{H}}}^{\left( 2 \right)}}+ \right. \\ 
		& \left. +\sum\limits_{i=1}^{{{N}_{1}}}{\sum\limits_{j=1}^{{{N}_{2}}}{\hat{H}_{ij}^{\operatorname{int}}\left( \vec{r}_{i}^{\left( 1 \right)}-\vec{r}_{j}^{\left( 2 \right)} \right)}} \right)\Psi \left( t,{{r}^{\left( 1 \right)}},{{r}^{\left( 2 \right)}} \right). \\ 
	\end{aligned}
\end{equation} 
Here, ${{\hat{H}}^{(1)}}$ and ${{\hat{H}}^{(2)}}$ are the Hamiltonians of subsystems $1$ and $2$, respectively, and $\hat{H}_{ij}^{\text{int}}\left(\vec{r}_{i}^{(1)} - \vec{r}_{j}^{(2)}\right)$ is the Hamiltonian of the interaction between the corresponding particles from different subsystems. All these Hamiltonians vanish for large separations, i.e. $\hat{H}_{ij}^{\operatorname{int}}\left( \vec{r}_{i}^{1}-\vec{r}_{j}^{2} \right) \to 0$ as $\left| \vec{r}_{i}^{1}-\vec{r}_{j}^{2} \right| \to +\infty$. 
However, when the state $\Psi$ allows the particles to be observed at short distances, the interaction term contributes to the evolution of the state.
Thus, as is seen from \eqref{Shredinger12}, the change of the state over time is influenced by all potential configurations contained in it, not just those that can be observed experimentally. 

\begin{figure}[!htbp]
	\centering
	\includegraphics[width=1.0\linewidth]{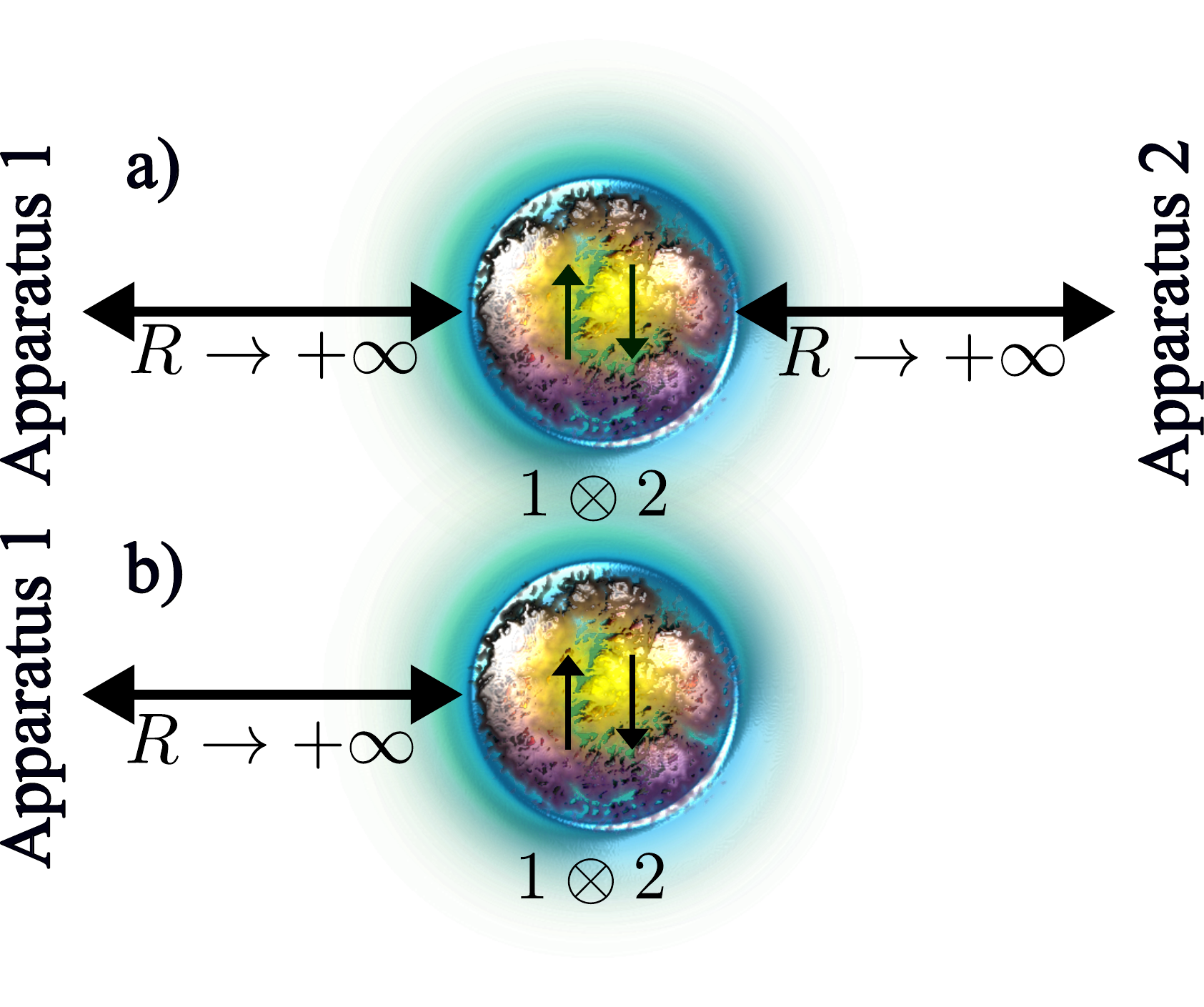}
	\caption{Apparatuses at a large distance from the interaction region.}
	\label{fig:aparatusfar}
\end{figure}

\begin{figure}[!htbp]
	\centering
	\includegraphics[width=0.75\linewidth]{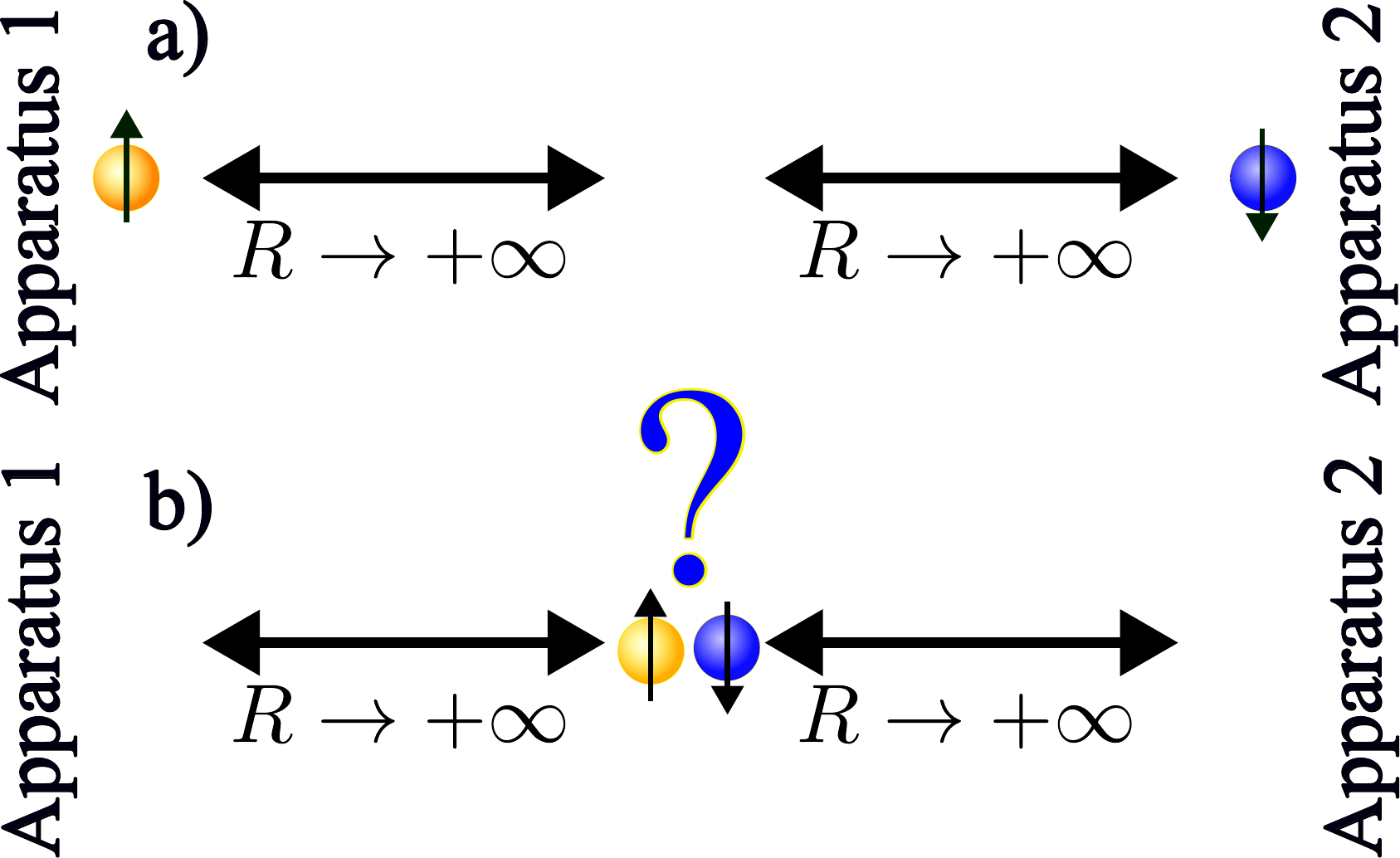}
	\caption{Manifested outcomes are spatially separated, whereas the pre-measurement state might have allowed the particles to be observed in close proximity}
	\label{fig:aparatusfarralisation}
\end{figure}

To illustrate this point, let us consider the \emph{thought-experimental} setups discussed in Refs.~\cite{AharonovBohmPhysRev.108.1070,CANTRELL1978499}, 
where the apparatus measuring the spin component is considered to be at a large distance from the region in which the subsystems have significantly interacted (Fig.~\ref{fig:aparatusfar}). In this way, only a subset of potential outcomes of the measurement is manifested. For these manifested possibilities, spatial separation and absence of interaction are ensured (Fig.~\ref{fig:aparatusfarralisation}(a)). 

At the same time, we do not control whether it is possible to observe the particles of subsystems $1$ and $2$ so close to each other (Fig.~\ref{fig:aparatusfarralisation}(b)), such that the corresponding Hamiltonian $\hat{H}_{ij}^{\operatorname{int}}\left( \vec{r}_{i}^{(1)}-\vec{r}_{j}^{(2)} \right)$ is non-zero. In this case, \eqref{Shredinger12} includes a contribution from the interaction between the subsystems. According to this equation, this contribution affects the state, including during measurement, and the state after measurement is formed under the influence of this interaction, not without any impact on either subsystem.

At least in the considered thought-experimental setup, the possibility of obtaining particles of subsystems close to each other during the measurement will remain unobserved. Thus, to analyze the presence or absence of influences on the formation of the state, we cannot rely solely on the observed results of the experiment and, accordingly, solely on the description of measurement as a reduction in theory. Since there are significant experimental difficulties in realizing the possibility of observing particles of subsystems close to each other, we are left with the theoretical study of the conditions under which the same state simultaneously allows the potential observation of particles both far from and close to each other.

\subsubsection{Intersubsystems interaction non-cessation}

Usually, the probability amplitude $ \Psi \left( t,{{r}^{\left( 1 \right)}},{{r}^{\left( 2 \right)}}\right) $ is defined over a region where each spatial argument ranges from $-\infty$ to $+\infty$. However, since the normalization integral for the state $ \Psi \left( t,{{r}^{\left( 1 \right)}},{{r}^{\left( 2 \right)}} \right) $ converges, there exist finite three-dimensional spatial domains $ D_{1}\left( t \right)$ and $ D_{2}\left( t \right)$ such that, for any time $ t $,
\begin{equation}\label{Integral_normuvanna_eps}	
	1-\varepsilon \le \int\limits_{{{r}^{\left( 1 \right)}}\in {{D}_{1}}\left( t \right)}{d{{r}^{\left( 1 \right)}}}\int\limits_{{{r}^{\left( 2 \right)}}\in {{D}_{2}}\left( t \right)}{d{{r}^{\left( 2 \right)}}} {{\left| \Psi  \right|}^{2}}\le 1. \\ 	
\end{equation}

Here $\Psi=\Psi \left( t,{{r}^{\left( 1 \right)}},{{r}^{\left( 2 \right)}} \right)$ and $\varepsilon$ denotes the probability measurement error. The notation 
$\int\limits_{{{r}^{\left( 1 \right)}}\in {{D}_{1}}\left( t \right)}{d{{r}^{\left( 1 \right)}}}$ means integration over all components of all vectors ${{r}^{\left( 1 \right)}}$ such that each of them lies within the domain $D_{1}\left( t \right)$; the integral 
$\int\limits_{{{r}^{\left( 2 \right)}}\in {{D}_{2}}\left( t \right)}{d{{r}^{\left( 2 \right)}}}$ is defined analogously for vectors in ${{r}^{\left( 2 \right)}}$. 

In other words, the domains $D_{1}\left( t\right)$ and $D_{2}\left( t\right)$ are chosen such that the total probability of finding at least one particle of subsystem~1 outside $D_{1}\left( t\right)$, or at least one particle of subsystem~2 outside $D_{2}\left(t\right)$, is smaller than $\varepsilon$.

If condition~\eqref{Integral_normuvanna_eps} is satisfied for some state with domains \(D_{1}\left( t\right)\) and \(D_{2}\left( t\right)\), we shall say that subsystem~1 is \textit{localized} in \(D_{1}(t)\) and subsystem~2 is \textit{localized} in \(D_{2}(t)\); in what follows we shall use the term “localized” in this sense without further qualification.

\begin{figure}[!htbp]
	\centering
	\includegraphics[width=1.0\linewidth]{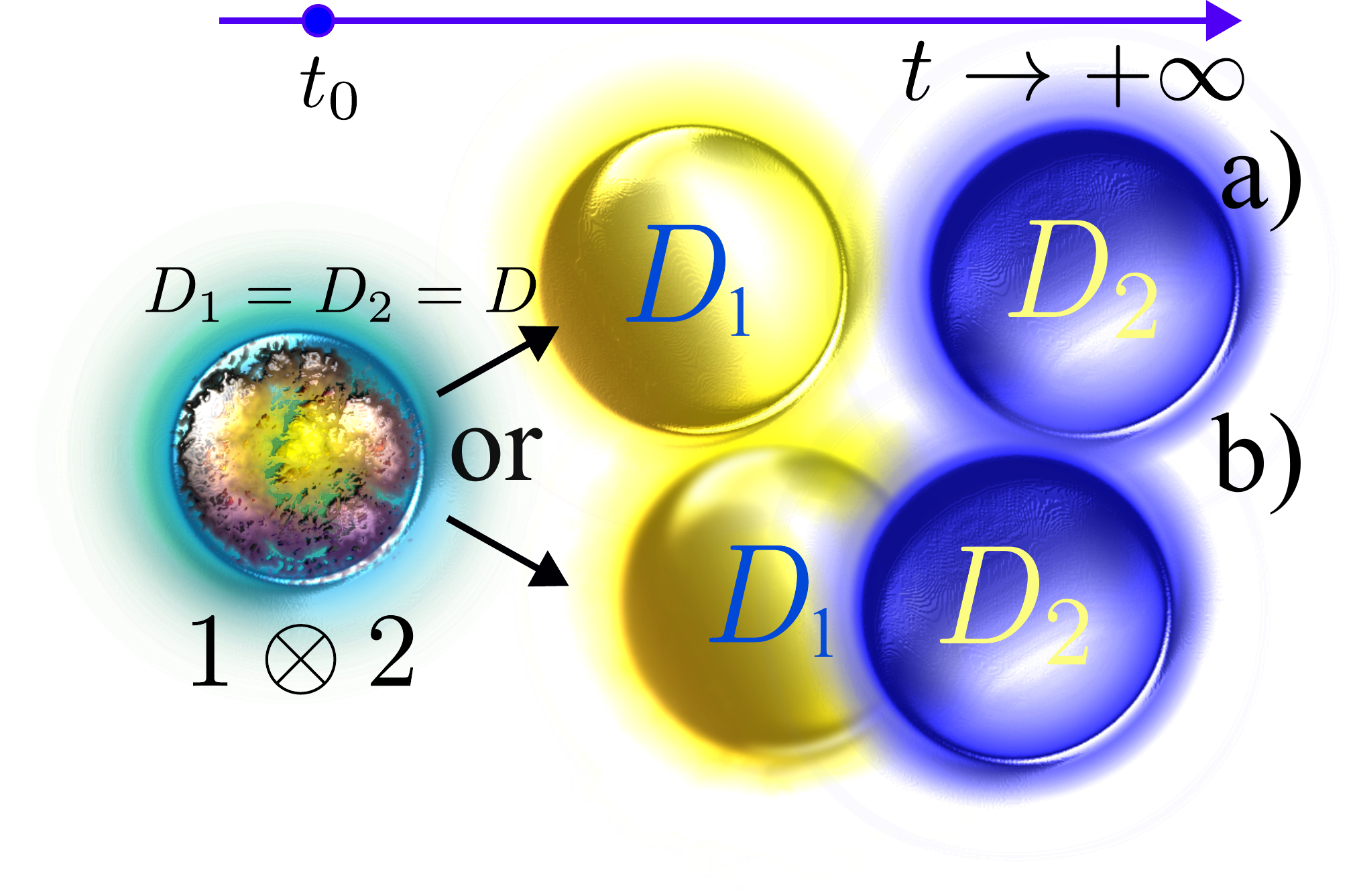}
	\caption{Two possible arrangements of the domains $D_1(t)$ and $D_2(t)$: overlapping or distinct.}
	\label{fig:overlapnotoverlapd1d2}
\end{figure}

At the initial time $t_0$, the state is assumed to satisfy $D_1(t_0) = D_2(t_0) = D$ (Fig.~\ref{fig:overlapnotoverlapd1d2}), corresponding to the scenarios considered in \cite{AharonovBohmPhysRev.108.1070} and \cite{CANTRELL1978499}. In both cases, the subsystems are initially in a bound, entangled state, allowing both particles to be observed within the same domain $D$. An external effect (e.g., photo-disintegration of positronium at $t_0$ in \cite{CANTRELL1978499}) then transforms the system into a linear combination of continuous-spectrum states, which serves as the initial state for our analysis.

After $t_0$, the subsystems can separate spatially, while the domains $D_1(t)$ and $D_2(t)$ also undergo internal expansion due to the uncertainty in their relative momentum.

Figure~\ref{fig:overlapnotoverlapd1d2} illustrates how internal expansion and mutual separation of the domains compete to determine the asymptotic configuration as $t \to +\infty$. The following analysis examines which scenario occurs under the considered conditions.

To address this question, let us introduce Jacobi coordinates for each of the subsystems $1$ and $2$ \cite{Cornilledoi:10.1142/5272,faddeev1993quantum}. Let ${{\vec{R}}_{1}}$ and ${{\vec{R}}_{2}}$ denote the position vectors of the centers of mass of subsystems $1$ and $2$, respectively, and let 
${{y}^{(1)}}=\{\vec{y}_{1}^{(1)}, \vec{y}_{2}^{(1)}, \ldots, \vec{y}_{N_1-1}^{(1)}\}$ and 
${{y}^{(2)}}=\{\vec{y}_{1}^{(2)}, \vec{y}_{2}^{(2)}, \ldots, \vec{y}_{N_2-1}^{(2)}\}$ 
be the sets of relative position vectors for these subsystems.

Then let us move to the Jacobi coordinates for the position vectors ${{\vec{R}}_{1}}$ and ${{\vec{R}}_{2}}$:
\begin{equation}\label{R_i_y}
	\begin{aligned}
&\vec{R}=\frac{{{M}_{1}}{{{\vec{R}}}_{1}}+{{M}_{2}}{{{\vec{R}}}_{2}}}{{{M}_{1}}+{{M}_{2}}}, \\ 
& \vec{Y}={{{\vec{R}}}_{2}}-{{{\vec{R}}}_{1}}. \\ 	
	\end{aligned}
\end{equation}

Here, $ {M}_{1} $ and $ {M}_{2} $ are the total masses of the particles in subsystems $1$ and $2$, respectively. Instead of the probability amplitude $\Psi \left( t, {{r}^{\left( 1 \right)}}, {{r}^{\left( 2 \right)}} \right)$ in Cartesian coordinates, we will obtain the probability amplitude in Jacobi coordinates $\Psi \left( t, \vec{R}, \vec{Y}, {{y}^{\left( 1 \right)}}, {{y}^{\left( 2 \right)}} \right)$. 

Let us suppose that in a certain interval of time $\left(t_1,+\infty\right)$ we can indeed neglect the interaction between particles of subsystem $1$ and particles of subsystem $2$. We will show that this assumption leads to a contradiction, at least in some cases. 


After neglecting the interaction between the particles of subsystem $1$ and the particles of subsystem $2$, the Hamiltonian of the system \mbox{$1\otimes2$} (denoted as \mbox{$\hat{H}^{\left(1\otimes2\right)}$}) can be expressed as a sum of four commuting operators, each depending on a separate group of variables: $\vec{R}$, $\vec{Y}$, $y^{\left(1\right)}$, and $y^{\left(2\right)}$:
\begin{equation}\label{Hamiltonian12bezVzaemodii}
	\begin{aligned}
		&{{\hat{H}}^{\left(1\otimes2\right)}} = -\frac{\hbar^2}{2M} \Delta_{\vec{R}} - \frac{\hbar^2}{2\mu} \Delta_{\vec{Y}} +\\
		&+ \hat{H}_1\left(y^{\left(1\right)}\right) + \hat{H}_2\left(y^{\left(2\right)}\right). 
	\end{aligned}	
\end{equation}
Here, $\Delta_{\vec{R}}$ is the Laplace operator with respect to the components of the center-of-mass position vector $\vec{R}$ of the entire system \mbox{$1\otimes2$}, and $\Delta_{\vec{Y}}$ is the Laplace operator with respect to the components of the relative position vector $\vec{Y}$ between subsystems $1$ and $2$. The parameters $M$ and $\mu$ are the total and reduced masses, respectively:
\begin{equation}\label{Poznachenna_M_mu}
	\begin{aligned}
		M &= M_1 + M_2, \\
		\mu &= \frac{M_1 M_2}{M_1 + M_2}.
	\end{aligned}	
\end{equation}
Additionally, $\hat{H}_1\left(y^{\left(1\right)}\right)$ consists of the operators for the relative kinetic energies of the particles in subsystem $1$ (which include Laplace operators with respect to the components of the vectors in the set $y^{\left(1\right)}$), plus the interaction operator between the particles of subsystem $1$. The operator $\hat{H}_2\left(y^{\left(2\right)}\right)$ has a similar meaning for subsystem $2$.

In contrast to the Hamiltonian, in the expression for the state of the system, even in the absence of interaction in the time interval $\left( {{t}_{1}},+\infty \right)$, we cannot separate the dependencies on $\vec{R}$, $\vec{Y}$, $y^{\left(1\right)}$, and $y^{\left(2\right)}$. The presence of an external influence in the past that transferred the system \mbox{$1\otimes2$} from a bound state to a linear combination of states in the continuous spectrum prevents us from separating the dependence on the center of mass position vector $\vec{R}$ from the dependence on the other variables. During this external influence, the components of this position vector become entangled with the other variables. Therefore, as is usually the case, we cannot expect this entanglement to disappear after the external influence ceases. Similarly, the interaction between the subsystems leads to entanglement of the relative position vector $\vec{Y}$ with the other variables. Thus, the absence of this interaction in the interval $\left( {{t}_{1}},+\infty \right)$ will not result in the separation of the probability amplitude’s dependence on $\vec{Y}$ into a separate factor.
 
Since our goal is to examine the possibility of spatial separation of the subsystems, we are particularly interested in the dependence of the probability amplitude on $\vec{Y}$. To study this dependence, we can express the probability amplitude $\Psi \left( t_1,\vec{R},\vec{Y},{{y}^{\left( 1 \right)}},{{y}^{\left( 2 \right)}} \right)$ as a series expansion:
\begin{equation}\label{Rozclad_Psi_ot_t1n1n2}
	\begin{aligned}
		& \Psi \left( {{t}_{1}},\vec{R},\vec{Y},{{y}^{\left( 1 \right)}},{{y}^{\left( 2 \right)}} \right) = \\ 
		& = \sum_{{n_1}=0}^{{S_1}} \sum_{{n_2}=0}^{{S_2}} \int \frac{d\vec{P}}{{(2\pi \hbar)}^{3/2}} \exp \left( \frac{i}{\hbar} \left( \vec{P} \cdot \vec{R} \right) \right) \, \times \\ 
		& \times \Phi_{{n_1}{n_2}}\left( \vec{P},\vec{Y} \right) \psi_{{n_1}}\left( y^{\left( 1 \right)} \right) \psi_{{n_2}}\left( y^{\left( 2 \right)} \right). \\ 
	\end{aligned}
\end{equation}
Here, $S_1$ and $S_2$ are the numbers of bound internal states of the particles in subsystems $1$ and $2$, respectively, $\psi_{{n_1}}\left( y^{\left( 1 \right)} \right)$ are the eigenstates of the discrete spectrum of the Hamiltonian ${\hat{H}}_1\left( y^{\left( 1 \right)} \right)$ from equation \eqref{Hamiltonian12bezVzaemodii}, and $\psi_{{n_2}}\left( y^{\left( 2 \right)} \right)$ are analogous eigenstates for the Hamiltonian ${\hat{H}}_2\left( y^{\left( 2 \right)} \right)$ from the same equation. The coefficients $\Phi_{{n_1}{n_2}}\left( \vec{P}, \vec{Y} \right)$ are the expansion coefficients that contain the dependence on $\vec{Y}$, which is of interest to us.

Now we can use the commutativity of the terms in the formula \eqref{Hamiltonian12bezVzaemodii} to describe the time evolution of the state \eqref{Rozclad_Psi_ot_t1n1n2}:
\begin{equation}\label{Psi_t_t1}
	\begin{aligned}
		& \Psi \left( t,\vec{R},\vec{Y},{{y}^{\left( 1 \right)}},{{y}^{\left( 2 \right)}} \right) = \\ 
		& = \exp \left( -\frac{i}{\hbar }{{\hat{H}}^{\left( 1\otimes 2 \right)}}\left( t-{{t}_{1}} \right) \right) \times \\ 
		& \times \Psi \left( {{t}_{1}},\vec{R},\vec{Y},{{y}^{\left( 1 \right)}},{{y}^{\left( 2 \right)}} \right). \\ 
	\end{aligned}
\end{equation}  
Using this commutativity, we obtain:
\begin{equation}\label{Time_evolution_expanding}
	\begin{aligned}
		& \Psi \left( t,\vec{R},\vec{Y},{{y}^{\left( 1 \right)}},{{y}^{\left( 2 \right)}} \right) = \\ 
		& = \sum_{{n_1}=0}^{{S_1}} \sum_{{n_2}=0}^{{S_2}} \int \frac{d\vec{P}}{{(2\pi \hbar)}^{3/2}} \exp \left( \frac{i}{\hbar} \left( \vec{P} \cdot \vec{R} \right) \right) \times \\ 
		& \times \exp \left( -\frac{i}{\hbar} \left( \frac{{\vec{P}}^{2}}{2M} + E_{{n_1}} + E_{{n_2}} \right) \left( t - t_1 \right) \right) \times \\ 
		& \times \psi_{{n_1}}\left( y^{\left( 1 \right)} \right) \psi_{{n_2}}\left( y^{\left( 2 \right)} \right) \Phi_{{n_1}{n_2}}\left( t,\vec{P},\vec{Y} \right). \\ 
	\end{aligned}
\end{equation}
Here, we denote $E_{{n_1}}$ and $E_{{n_2}}$ as the eigenvalues corresponding to the eigenfunctions $\psi_{{n_1}}\left( y^{\left( 1 \right)} \right)$ and $\psi_{{n_2}}\left( y^{\left( 2 \right)} \right)$, respectively. We also introduce the notation:
\begin{equation}\label{Fi_ot_t_t1}
	\begin{aligned}
		& \Phi_{{n_1}{n_2}}\left( t,\vec{P},\vec{Y} \right) = \\ 
		& = \exp \left( -\frac{i}{\hbar} \left( -\frac{\hbar^{2}}{2\mu} \Delta_{\vec{Y}} \right) \left( t - t_1 \right) \right) \times \\ 
		& \times \Phi_{{n_1}{n_2}}\left( \vec{P},\vec{Y} \right). \\ 
	\end{aligned}	
\end{equation}
  
Let us define the probability density for the values of the vector $\vec{Y}$ as:
\begin{equation}\label{Ro_Y_t}
	\begin{aligned}
		& {{\rho }_{{\vec{Y}}}}\left( t,\vec{Y} \right)=\int{d\vec{R}d{{y}^{\left( 1 \right)}}d{{y}^{\left( 2 \right)}}}\times  \\ 
		& \times \left| \Psi \left( t,\vec{R},\vec{Y},{{y}^{\left( 1 \right)}},{{y}^{\left( 2 \right)}} \right) \right|^{2}. \\ 
	\end{aligned}	
\end{equation} 
Now, we can use the orthogonality of the basis states in the expansion \eqref{Rozclad_Psi_ot_t1n1n2}. We obtain the expression:
\begin{equation}\label{Ro_Y_t_cherez_Fi}
	\begin{aligned}
		& {{\rho }_{{\vec{Y}}}}\left( t,\vec{Y} \right)= \\ 
		& =\sum\limits_{{{n}_{1}}=0}^{{{S}_{1}}}{\sum\limits_{{{n}_{2}}=0}^{{{S}_{2}}}{\int{d\vec{P}}}} \left| {{\Phi }_{{{n}_{1}}{{n}_{2}}}}\left( t,\vec{P},\vec{Y} \right) \right|^{2}. \\ 
	\end{aligned}
\end{equation}

Hence, to study the properties of the probability density \eqref{Ro_Y_t} and its dependence on $ \vec{Y} $ at different time moments, we need to study the time evolution of the functions ${{\Phi }_{{{n}_{1}}{{n}_{2}}}}\left( t,\vec{P},\vec{Y} \right)$. This is governed by the formula \eqref{Fi_ot_t_t1}. For this purpose, it is convenient to use the well-known result about the time evolution of Gaussian states in the absence of interaction \cite{Griffiths,Flugge2012practical}. It is described by an operator similar to that in the formula \eqref{Fi_ot_t_t1}. To proceed, we represent the function ${{\Phi }_{{{n}_{1}}{{n}_{2}}}}\left( \vec{P},\vec{Y} \right)$ in the formula \eqref{Ro_Y_t} as an expansion in coherent states \cite{klauder2006fundamentals}. We change the relative position vector $\vec{Y}$ to the corresponding dimensionless dynamical variable $ \vec{X} $, by the substitution $\vec{Y}=l\vec{X}$, where $ l $ is the characteristic length scale of the $ 1\otimes2 $ system. Then, \cite{klauder2006fundamentals}
\begin{equation}\label{Rozclad_po_cogerentnim_stanam}
	\begin{aligned}
		\begin{aligned}
			& {{\Phi }_{{{n}_{1}}{{n}_{2}}}}\left( \vec{P},\vec{X} \right)=\int{d\vec{p}}\int{d\vec{q}}{{\Phi }_{{{n}_{1}}{{n}_{2}}}}\left( \vec{P},\vec{q},\vec{p} \right)\times  \\ 
			& \times \exp \left( -\frac{1}{2}{{{\vec{p}}}^{2}} \right)\exp \left( -\frac{1}{2}{{\left( \vec{X}-\left( \vec{q}+ip \right) \right)}^{2}} \right). \\ 
		\end{aligned} 
	\end{aligned}
\end{equation}
Here, the coefficients of the expansion ${{\Phi }_{{{n}_{1}}{{n}_{2}}}}\left( \vec{P},\vec{q},\vec{p} \right)$ represent the initial function ${{\Phi }_{{{n}_{1}}{{n}_{2}}}}\left( \vec{P},\vec{X} \right)$ in the representation of coherent states.

In order to calculate the result of the time evolution operator action in formula \eqref{Fi_ot_t_t1}, it is convenient to switch to dimensionless time values $t={{t}_{0}}\tau$, ${{t}_{1}}={{t}_{0}}{{\tau }_{1}}$, where ${{t}_{0}}$ is a certain magnitude of time dimension.

Since $t_{0}$ serves as a freely chosen time scale, it can be set arbitrarily. In particular, it is convenient to choose $t_{0}$ such that the dimensionless parameter $\hbar t_{0}/(\mu l^{2})$, which enters the time-evolution operator \eqref{Fi_ot_t_t1}, equals unity. 

Thus, we obtain
\begin{equation}\label{Nerozmirnij_operator_na_cogerentnij_stan}
	\begin{aligned}
		& \Phi_{n_{1}n_{2}}(\tau ,\vec{P},\vec{X})
		= \int d\vec{p}\int d\vec{q}\,
		\Phi_{n_{1}n_{2}}(\vec{P},\vec{q},\vec{p}) \\
		& \quad \times \exp\!\left(-\tfrac{1}{2}\vec{p}^{\,2}\right)
		\exp\!\left[-i\left(-\tfrac{1}{2}\Delta_{X}\right)(\tau - \tau_{1})\right] \\
		& \quad \times \exp\!\left[-\tfrac{1}{2}\left(\vec{X}-(\vec{q}+i\vec{p})\right)^{2}\right],
	\end{aligned}
\end{equation}
where $\Delta_{\vec{X}}$ denotes the Laplacian with respect to the components of the dimensionless relative position vector of the subsystems. After transformations similar to those discussed in \cite{Griffiths,Flugge2012practical}, we arrive at
\begin{equation}\label{Fi_rezult}
	\Phi_{n_{1}n_{2}}(\tau ,\vec{P},\vec{X})
	= \frac{\phi_{n_{1}n_{2}}(\tau ,\vec{P},\vec{X})}
	{\left[1+i(\tau - \tau_{1})\right]^{3/2}},
\end{equation}
with
\begin{equation}\label{Fi_rezult_integral}
	\begin{aligned}
		&\phi_{n_{1}n_{2}}(\tau ,\vec{P},\vec{X})
		= \int d\vec{p}\int d\vec{q}\,\\
		&\Phi_{n_{1}n_{2}}(\vec{P},\vec{q},\vec{p}) 
		\exp\!\left(-\tfrac{1}{2}\vec{p}^{\,2}\right) \\
		& \times \exp\!\left[
		-\tfrac{1}{2}
		\frac{\left(\vec{X}-(\vec{q}+i\vec{p})\right)^{2}}
		{1+i(\tau - \tau_{1})}
		\right].
	\end{aligned}
\end{equation}

Let us now consider the asymptotic behavior of Eq.~\eqref{Fi_rezult} as $\tau \to \infty$. 
The prefactor $\left[1+i(\tau - \tau_{1})\right]^{-3/2}$ accounts for the decay of the probability density \eqref{Ro_Y_t_cherez_Fi} with increasing $\tau$, reflecting the expansion of the region where $\vec{X}$ is most likely to be observed. 
This prefactor is independent of $\vec{X}$, whereas our interest lies in comparing the magnitudes of the probability density \eqref{Ro_Y_t_cherez_Fi} at small and large $|\vec{X}|$. 
The relevant dependence is contained in the integral \eqref{Fi_rezult_integral}.

Thus, we aim to analyze the dependence of the integral 
$\phi_{n_{1}n_{2}}(\tau ,\vec{P},\vec{X})$ \eqref{Fi_rezult_integral} on $\vec{X}$ 
in the limit $\tau \to \infty$. 
To this end, we consider three regions: 
$\lvert \vec{X} \rvert \ll \tau$, 
$\lvert \vec{X} \rvert \sim \tau$, 
and $\lvert \vec{X} \rvert \gg \tau$.

We begin with the case $\lvert \vec{X} \rvert \gg \tau$. 
From Eq.~\eqref{Ro_Y_t_cherez_Fi} it follows that
\begin{equation}\label{Integral_menh_za_1}
	\int d\vec{P}\,d\vec{X}\,
	\bigl| \Phi_{n_{1}n_{2}}(\tau ,\vec{P},\vec{X}) \bigr|^{2} \leq 1,
\end{equation}
and, according to Ref.~\cite{klauder2006fundamentals} and the expansion 
\eqref{Rozclad_po_cogerentnim_stanam},
\begin{equation}\label{Integral_vid_kv_mod_Fi_p_q}
	\begin{aligned}
		&\int d\vec{P}\,d\vec{X}\,
		\bigl| \Phi_{n_{1}n_{2}}(\vec{P},\vec{X}) \bigr|^{2} \\
		&= \int d\vec{P}\,d\vec{q}\,d\vec{p}\,
		\bigl| \Phi_{n_{1}n_{2}}(\vec{P},\vec{q},\vec{p}) \bigr|^{2}.
	\end{aligned}
\end{equation}

Hence, $\bigl| \Phi_{n_{1}n_{2}}(\vec{P},\vec{q},\vec{p}) \bigr|$ 
decays rapidly as either $\lvert \vec{q} \rvert$ or $\lvert \vec{p} \rvert \to \infty$. 
Thus, when analyzing the integrand of Eq.~\eqref{Fi_rezult_integral}, 
we may restrict ourselves to finite $\lvert \vec{q} \rvert$ and $\lvert \vec{p} \rvert$, 
while $\tau$ is arbitrarily large and $\lvert \vec{X} \rvert \gg \tau$. 
In this case, the exponent in Eq.~\eqref{Fi_rezult_integral} 
acquires a large negative real part, suppressing the exponential factor 
and rendering the integrand negligible throughout the integration domain. 
Consequently, for $\lvert \vec{X} \rvert \gg \tau$,
\begin{equation}\label{Limit_X_neskonchenist}
	\phi_{n_{1}n_{2}}(\tau ,\vec{P},\vec{X}) \approx 0.
\end{equation}

Next, consider the case $\lvert \vec{X} \rvert \ll \tau$. 
Here, the moduli of all terms in the numerator of the exponent in Eq.~\eqref{Fi_rezult_integral} 
are much smaller than the modulus of the complex denominator. 
Taking the limit $\tau \to +\infty$ yields an expression independent of ${\vec{X}}$:
\begin{equation}\label{Limit_tau_to_infinity}
	\begin{aligned}
		&\left| \lim_{\tau \to +\infty} \phi_{n_{1}n_{2}}(\tau ,\vec{P},\vec{X}) \right|\\
		&= \Biggl| \int d\vec{q}\,d\vec{p}\,  
		\Phi_{n_{1}n_{2}}(\vec{P},\vec{q},\vec{p}) 
		\exp\!\bigl(-\tfrac{1}{2}\vec{p}^{\,2}\bigr) \Biggr|.
	\end{aligned}
\end{equation}

\begin{figure}[!htbp]
	\centering
	\includegraphics[width=0.95\linewidth]{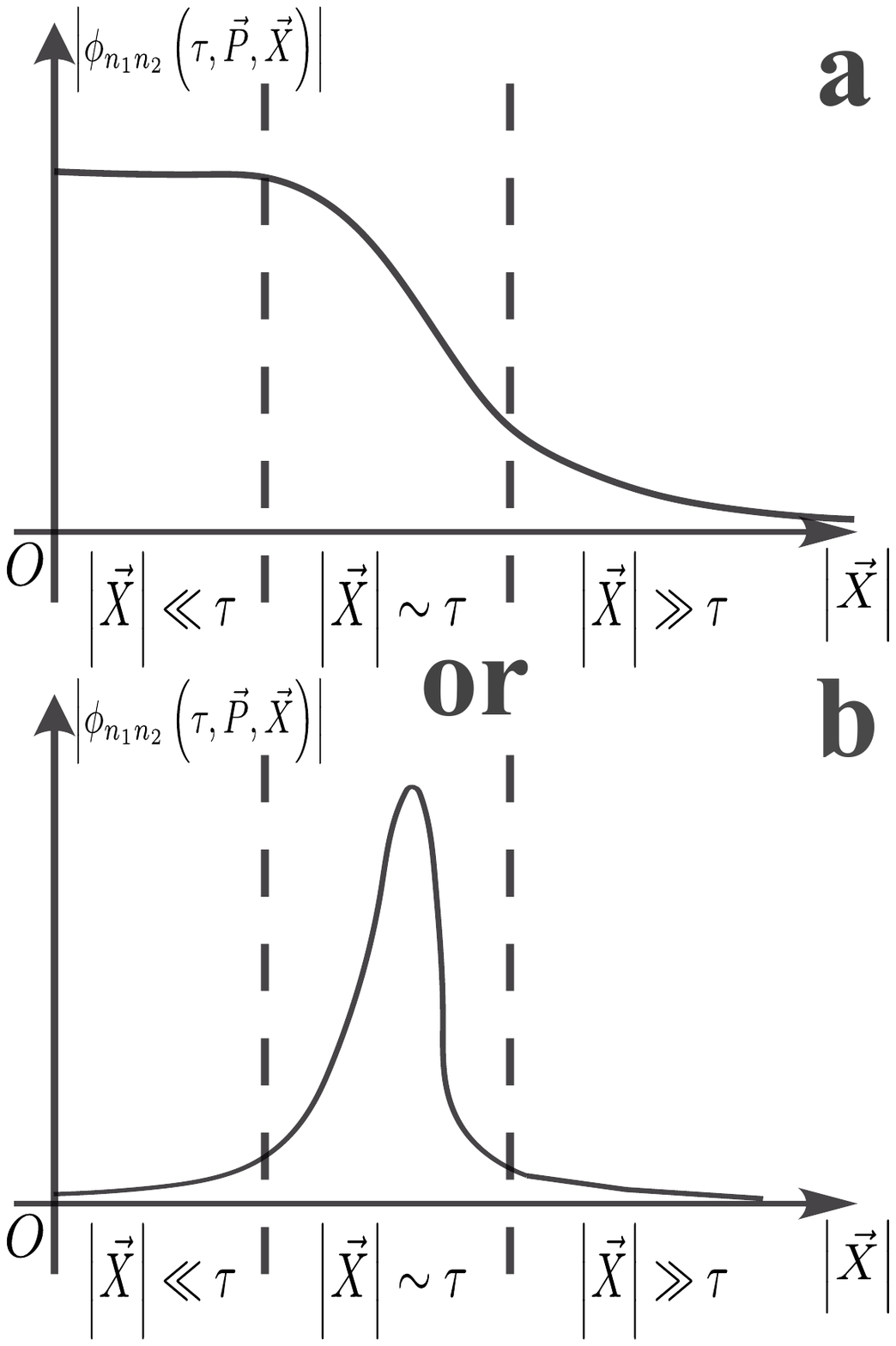}
	\caption{Two possible asymptotic behaviors of 
		$\lvert \phi_{n_{1}n_{2}}(\tau ,\vec{P},\vec{X}) \rvert$ 
		as implied by Eqs.~\eqref{Limit_X_neskonchenist} 
		and \eqref{Limit_tau_to_infinity}.}
	\label{fig:threeregions}
\end{figure}

Taking into account \eqref{Limit_X_neskonchenist} and \eqref{Limit_tau_to_infinity}, 
two possible asymptotic behaviors for $\tau \to \infty$ arise, as illustrated in Fig.~\ref{fig:threeregions}a,b. 

In case (b) of Fig.~\ref{fig:threeregions}, the result corresponds to a spatial separation 
of the subsystems, whereas in case (a) no such separation occurs. 
To determine which behavior is realized, we estimate Eq.~\eqref{Fi_rezult_integral} 
in the region $\lvert \vec{X} \rvert \sim \tau$. 
Under this condition, any vector $\vec{X}$ can be written as  
$\vec{X} = \vec{v}(\tau - \tau_{1})$ for some finite vector $\vec{v}$. 
Substituting this into Eq.~\eqref{Fi_rezult_integral} and taking the limit of large $\tau$, we obtain
\begin{equation}\label{Exp_min_v_kv}
	\begin{aligned}
		&\bigl| \phi_{n_{1}n_{2}}(\tau ,\vec{P},\vec{X} = \vec{v}(\tau - \tau_{1})) \bigr|
		= \exp\!\left( -\tfrac{1}{2}\vec{v}^{\,2} \right) \\
		&\quad\times \left| \int d\vec{p}\,d\vec{q}\,
		\Phi_{n_{1}n_{2}}(\vec{P},\vec{q},\vec{p})
		\exp\!\left( -\tfrac{1}{2}\vec{p}^{\,2} \right) \right|.
	\end{aligned}
\end{equation}

Hence, in the region $\lvert \vec{X} \rvert \sim \tau$, 
the modulus $\left| \phi_{n_{1}n_{2}}(\tau ,\vec{P},\vec{X} = \vec{v}(\tau - \tau_{1}))\right|$ 
cannot exceed that of the limit \eqref{Limit_tau_to_infinity}. 
Consequently, of the two possibilities in Fig.~\ref{fig:threeregions}, only case (a) is realized.

Therefore, neglecting the interaction between the subsystems of the composite system \mbox{$ 1\otimes2 $} leads to a contradiction. 
On the one hand, we assumed free time evolution, i.e., without interactions. 
On the other hand, we found that such evolution yields a state in which particles of different subsystems can still be found with non-negligible probability at small mutual distances. Under these circumstances, the interaction between the subsystems would significantly affect the time evolution of the state.

This contradiction demonstrates that, in an asymptotic state at large times, neither a spatial separation of the subsystems nor the cessation of their interaction can be achieved. 
We therefore regard this contradiction as an objection to the reasoning underlying the EPR paradox.

It is worth noting that the possibility of such an objection was already expressed in Ref.~\cite{Louis_de_Broglie} in a single phrase:  
\enquote{Against the standpoint of Einstein, I think one can argue that the definition of the \enquote{spatial separation} of two systems is not a simple matter when the localizations of the two systems are incomplete, and that one can have localization of the two systems in the same region of space.}

The considerations presented above leave two additional questions to be addressed. 

The first is whether, in the quantum case, spatial separation could arise as a consequence of interactions between subsystems. For instance, might this occur if the composite system $1\otimes 2$ consists of two particles with charges of the same sign, or more generally, when the interaction is repulsive? 

The second is whether, in the quantum case, a finite interval of time may exist during which spatial separation temporarily emerges before disappearing in the asymptotic limit $t\to\infty$ considered above.

Both questions can be analyzed using the analytic properties of the probability amplitude of an isolated quantum system as a function of time. Analyticity with respect to time leads to the conclusion that spatial separation cannot arise through time evolution. If such separation is absent in the initial state, it will not develop under any circumstances during the system’s evolution.

To see this, let us consider the probability amplitude of an arbitrary state of an isolated quantum system. It depends on time $t$ and on a set of dynamical variables, denoted collectively by $K$, and we denote the probability amplitude by $\Psi(t,K)$. The time-evolution operator for an isolated system is defined through its Taylor expansion, valid over the corresponding time interval \cite{Sakurai}. This expansion governs the dependence of $\Psi(t,K)$ over the entire interval from the initial time $t_0$, when $\Psi(t_0,K)$ is prepared, to any later time $t>t_0$. Hence, the interval $\left[t_0,t\right]$ lies entirely within the radius of convergence of this expansion \cite{krantz2002primer}.

Now fix $K$ to definite values $K_0$ and apply the time-evolution operator to $\Psi(t_0,K)$. The resulting function $\Psi(t,K_0)$ is then represented by a convergent Taylor series over the full interval $\left[t_0,t\right]$. Suppose that there exists a subinterval $\left[t_1,t_2\right]\subset\left[t_0,t\right]$ such that
\[
\Psi(t,K_0)=0, \quad \forall t\in\left[t_1,t_2\right], t_2 \ne t_1.
\]  
By the properties of analytic functions \cite{krantz2002primer}, $\Psi(t,K_0)$ must then vanish identically over $\left[t_0,t\right]$, since vanishing on a finite interval forces all coefficients of the Taylor expansion to be zero.  

Thus, either $\Psi(t,K_0)$ vanishes identically on $\left[t_0,t\right]$, or it can vanish only at a discrete set of isolated points within this interval. Consequently, if the probability of observing the configuration $K_0$ is nonzero at any time, it must remain nonzero throughout $\left[t_0,t\right]$, except possibly at isolated instants.

In particular, let us consider an initial state of the isolated system $1\otimes2$ in which particles of subsystem $1$ can be observed at small distances from those of subsystem $2$ with nonzero probability. Then such configurations must also occur with nonzero probability (though possibly small) at all subsequent times, except at isolated instants. In other words, if spatial separation is absent initially, it cannot emerge through time evolution.

This conclusion does not rely on any special property of the Hamiltonian of $1\otimes2$. It therefore holds regardless of the internal structure of the system, including whether interactions are present or absent, or whether they are attractive or repulsive.

The above considerations do not exclude the possibility that, during a finite interval of the time evolution, the modulus of the probability amplitude in the region of small interparticle distances remains within a range of small values without vanishing exactly at every instant. In this case, the terms on the right-hand side of Eq.~\eqref{Shredinger12} describing the interaction between subsystems contribute only small corrections (in absolute magnitude) to the time derivative of the probability amplitude. Under these conditions, neglecting the inter-subsystem interaction may be regarded as an acceptable approximation. This approximation can be described as a \enquote{temporary approximate spatial separation} of the subsystems. However, as shown above, over long times this approximation leads to a contradiction and therefore can only hold within a limited time interval. 

Moreover, over long times even weak interactions may \enquote{accumulate} their effects. By contrast, in the following subsections we consider the interaction of $1\otimes2$ with an apparatus (or multiple apparatuses) and show that state formation in such processes cannot be instantaneous. We will also argue that this formation time exceeds the interval during which the \enquote{temporary approximate spatial separation} remains valid.

Therefore, there is no way to realize a typical EPR measurement on the system $1\otimes 2$ without any interaction between subsystems~$1$ and~$2$.
       
\subsubsection{A model of the state--apparatus interaction}

We now consider the interactions between the particles of system $1 \otimes 2$ and the apparatus $A$. Before doing so, we make two remarks.  

First, due to the locality of interactions, a system particle can interact with the apparatus only if that particle can be observed within a certain region surrounding the apparatus. Let us denote this region by $D_A$. In what follows, we will therefore be concerned with the mutual arrangement of the regions $D_1$ and $D_A$, as well as of $D_2$ and $D_A$. By contrast, the overlap of $D_1$ and $D_2$, discussed above, is not essential for the present considerations. For this reason, in some of the figures below we omit this overlap to simplify the illustrations. These figures should thus be regarded as purely schematic, serving to support the arguments presented rather than as exact representations of the spatial arrangement of the regions.

Second, our aim here is limited to identifying those interactions between the particles of system $1 \otimes 2$ and the apparatus that are relevant for the formation of the post-measurement state. We do not attempt to describe the detailed time evolution of this state by solving the full dynamical problem. Instead, we introduce a simplified model of the system--apparatus interaction. Let us describe its essential features.

The apparatus $A$ measures a certain dynamical variable, which we denote by $\hat{V}$. For subsystem $1$, we denote by $V^{(1)}$ the set of eigenvalues of $\hat{V}$ that can be observed upon measurement. Similarly, for subsystem $2$, the corresponding set is $V^{(2)}$. The set of all possible measurement outcomes for the composite system is then $V = V^{(1)} \cup V^{(2)}$. In each member of the quantum ensemble representing this state, some value $V_k$ ($V_k \in V$) is manifested through the corresponding response of the apparatus.

For the reasons discussed in the Introduction, and in order for the probability of a given outcome to be determined solely by the state of the quantum system prior to the measurement, the apparatus $A$ must be classical \cite{BohrNPhysRev.48.696}. This implies that the interaction of the particles of the $1 \otimes 2$ system with the apparatus reduces to their interaction with a certain number $N_A$ of classical particles that constitute the apparatus, each characterized by a specific position vector
\[
\mathbf{r}^{(A)}(t) = \left\{ \vec{r}_1^{(A)}(t), \vec{r}_2^{(A)}(t), \ldots, \vec{r}_{N_A}^{(A)}(t) \right\}
\]
at any moment in time $t$.

When a particular value $V_k$ is observed in one of the ensemble systems, this corresponds to a definite classical motion of the apparatus particles,
\begin{equation}\label{r_vid_Vk_t}
	\begin{aligned}
		\mathbf{r}^{(A)}(V_k, t) = &\left\{ \vec{r}_1^{(A)}(V_k, t), \vec{r}_2^{(A)}(V_k, t), \ldots, \right.  \\
		& \ldots, \left. \vec{r}_{N_A}^{(A)}(V_k, t) \right\}.
	\end{aligned} 
\end{equation}

A complete treatment of the system--apparatus interaction would require a hybrid ensemble description \cite{Hall:2016oqf}, as discussed in the Introduction. In the present case, however, we do not need the explicit form of the functions \eqref{r_vid_Vk_t}. This is because our aim is narrower: for each particle of the system $1 \otimes 2$, we ask only a yes-or-no question—does its interaction with the apparatus particles affect the time evolution of the state during the measurement? We are not concerned with how this effect occurs, nor with the form of the resulting state.

The existence of such an effect depends solely on the possibility of observing the particle in the vicinity of the apparatus, rather than on the detailed form of the functions \eqref{r_vid_Vk_t}, since all vectors \eqref{r_vid_Vk_t} remain within the spatial region occupied by the apparatus.

Thanks to this observation, we can avoid the full hybrid description and describe the dynamics of the quantum state of $1 \otimes 2$ as that of a system interacting with a set of classical particles undergoing some classical motion \eqref{r_vid_Vk_t}. This constitutes the essential simplification of the system--apparatus model, as outlined above.

With this simplification, the formation of the state of the $1\otimes2$ system as a result of its interaction with the apparatus can be described by
\begin{equation}\label{Classical_apparatus_interaction}
	\begin{aligned}
		& i\hbar \frac{\partial \Psi }{\partial t}=\left( {{{\hat{H}}}^{\left( 1\otimes 2 \right)}} \right.+ \\
		& +\sum\limits_{i=1}^{{{N}_{1}}}{\sum\limits_{j=1}^{{{N}_{A}}}{\hat{H}_{ij}^{1,A}\left( \vec{r}_{i}^{\left( 1 \right)}-\vec{r}_{j}^{\left( A \right)}\left( {{V}_{k}},t \right) \right)}} \\
		& \left. +\sum\limits_{i=1}^{{{N}_{2}}}{\sum\limits_{j=1}^{{{N}_{A}}}{\hat{H}_{ij}^{2,A}\left( \vec{r}_{i}^{\left( 2 \right)}-\vec{r}_{j}^{\left( A \right)}\left( {{V}_{k}},t \right) \right)}} \right)\Psi, \\
	\end{aligned}	
\end{equation}
where $\Psi=\Psi \left( t,{{r}^{\left( 1 \right)}},{{r}^{\left( 2 \right)}},{{V}_{k}} \right)$. 
Here $\hat{H}_{ij}^{1,A}\left( \vec{r}_{i}^{\left( 1 \right)}-\vec{r}_{j}^{\left( A \right)}\left( {{V}_{k}},t \right) \right)$ represents the interaction Hamiltonians between the particles of subsystem $1$ and the apparatus, while $\hat{H}_{ij}^{2,A}\left( \vec{r}_{i}^{\left( 2 \right)}-\vec{r}_{j}^{\left( A \right)}\left( {{V}_{k}},t \right) \right)$ represents those between subsystem $2$ and the apparatus. These Hamiltonians, like those in \eqref{Shredinger12}, satisfy
\begin{equation}\label{No_spooky_action_in_distance}
	\begin{aligned}
		& \hat{H}_{ij}^{1,A}\left( \vec{r}_{i}^{\left( 1 \right)}-\vec{r}_{j}^{\left( A \right)}\left( {{V}_{k}},t \right) \right)\xrightarrow{\left| \vec{r}_{i}^{\left( 1 \right)}-\vec{r}_{j}^{\left( A \right)}\left( {{V}_{k}},t \right) \right|\to +\infty}0, \\ 
		& \hat{H}_{ij}^{2,A}\left( \vec{r}_{i}^{\left( 2 \right)}-\vec{r}_{j}^{\left( A \right)}\left( {{V}_{k}},t \right) \right)\xrightarrow{\left| \vec{r}_{i}^{\left( 2 \right)}-\vec{r}_{j}^{\left( A \right)}\left( {{V}_{k}},t \right) \right|\to +\infty}0. \\ 
	\end{aligned}	
\end{equation}

Lastly, ${{\hat{H}}^{\left( 1\otimes 2 \right)}}$ denotes the Hamiltonian of the $1\otimes2$ system, which also appears on the right-hand side of Eq.~\eqref{Shredinger12}:
\begin{equation}\label{H1otimes2}
	\begin{aligned}
		& {{{\hat{H}}}^{\left( 1\otimes 2 \right)}}={{{\hat{H}}}^{\left( 1 \right)}}+{{{\hat{H}}}^{\left( 2 \right)}}+ \\
		& +\sum\limits_{i=1}^{{{N}_{1}}}{\sum\limits_{j=1}^{{{N}_{2}}}{\hat{H}_{ij}^{\operatorname{int}}\left( \vec{r}_{i}^{\left( 1 \right)}-\vec{r}_{j}^{\left( 2 \right)} \right)}}. \\
	\end{aligned}	
\end{equation}

Using Eq.~\eqref{Classical_apparatus_interaction}, we can now analyze which of the interactions between subsystems~1 and~2 with the apparatus affect the formation of the post-measurement state. To this end, we examine the probabilities of finding the particles of subsystems~1 and~2 near the apparatus in the state that results from the time evolution of the initial state at time \(t_{0}\) (Fig.~\ref{fig:overlapnotoverlapd1d2}). In this initial state, subsystems~1 and~2 are localized in the region \(D_{1}(t_{0}) = D_{2}(t_{0}) = D\), as illustrated in Fig.~\ref{fig:overlapnotoverlapd1d2}. This region is located far from the apparatus, as shown in Fig.~\ref{fig:aparatusfar}. Consequently, the modulus of the probability amplitude near the apparatus is negligible, and so are the contributions of the interaction between the apparatus and the particles of the \(1 \otimes 2\) system in the right-hand side of Eq.~\eqref{Classical_apparatus_interaction}. However, after a sufficiently long time interval, this modulus may increase, and the corresponding interaction terms can become significant. The state obtained as a result of this long-time evolution will be referred to as the \emph{asymptotic state}, i.e., the state in the limit \(t \to +\infty\).

By expanding the asymptotic state in a suitable set of mutually orthogonal basis states, we can analyze the locations of the regions \(D_{1}\) and \(D_{2}\) corresponding to each of these basis states. Our goal is to show that these locations allow the particles of both subsystems to be found near the apparatus and to locally interact with it.

\subsubsection{Possibility of paradox avoidance via \enquote{exchange} terms}

\begin{figure}[!htbp]
	\centering
	\includegraphics[width=0.7\linewidth]{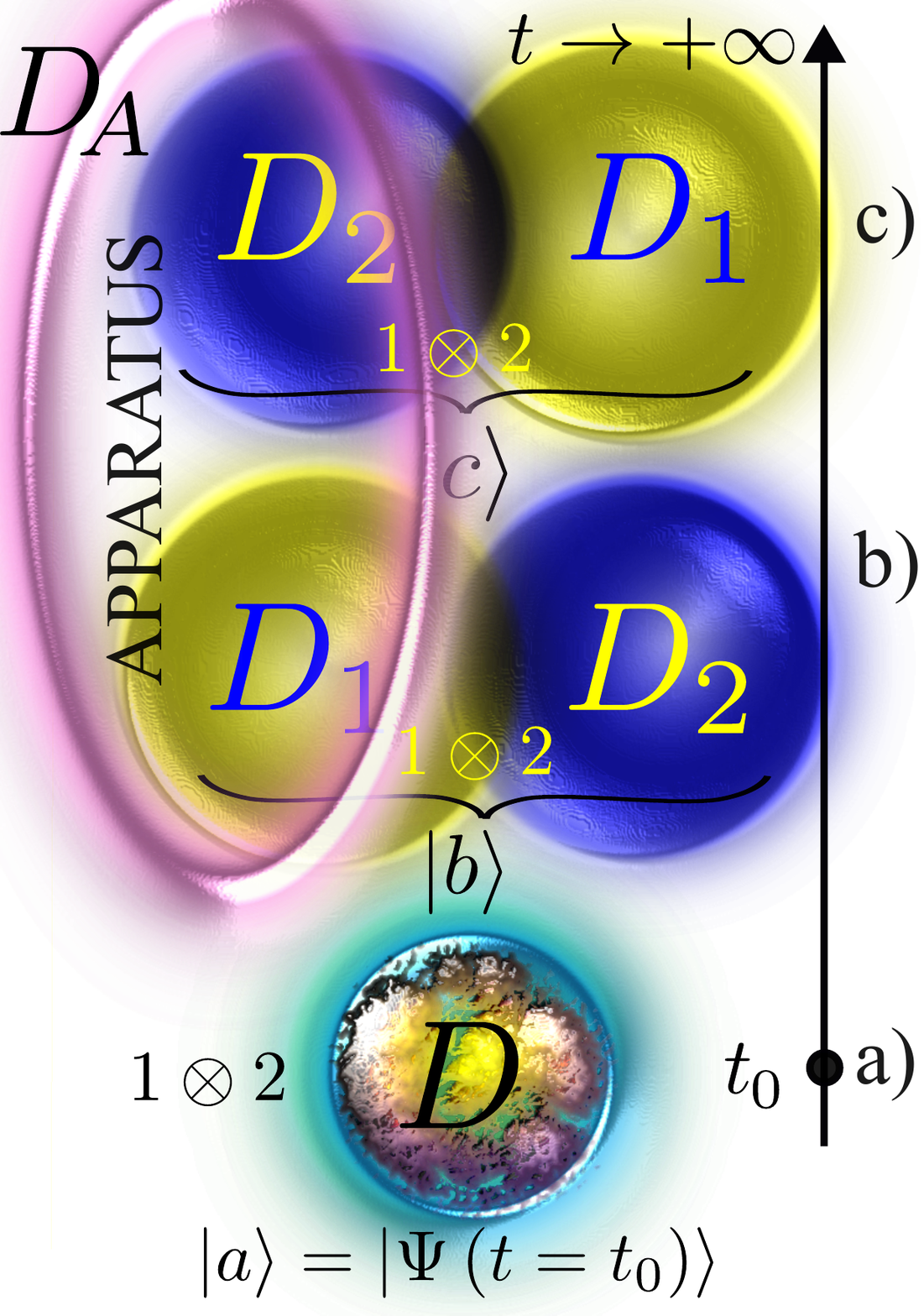}
	\caption{Schematic depiction of the regions $D_{1}$ and $D_{2}$, in which subsystems~1 and~2 of the $1\otimes2$ system are localized, and the region $D_{A}$, where their particles can interact with the apparatus. 
		(a) Configuration associated with the initial state, for which $D_{1}=D_{2}=D$. 
		(b) Configuration associated with one of the basis states, denoted $\left| b \right\rangle$. 
		(c) Configuration associated with the corresponding \enquote{exchange} state $\left| c \right\rangle$.}
	\label{fig:exchangeterms}
\end{figure}

\textbf{Let us provisionally assume} that the expansion of the asymptotic state includes, in addition to the basis state whose configuration of the regions \(D_{1}\) and \(D_{2}\) is shown in Fig.~\ref{fig:exchangeterms}(b), another linearly independent state whose configuration is depicted in Fig.~\ref{fig:exchangeterms}(c).

Because of their superficial resemblance to the exchange states that occur for indistinguishable particles, we will refer to the states $\left| b \right\rangle$ and $\left| c \right\rangle$ shown in Fig.~\ref{fig:exchangeterms} as \enquote{exchange} states. We emphasize, however, that we are dealing with systems composed of generally non-identical particles. As will become clear in the following discussion, these \enquote{exchange} states appear in the state expansion of the \(1 \otimes 2\) system for reasons entirely different from those underlying the true exchange symmetry of identical-particle systems.

The states \(\left| b \right\rangle\) and \(\left| c \right\rangle\) need not be orthogonal. In such a case, they can be orthogonalized, yielding two new orthogonal states that are certain linear combinations of \(\left| b \right\rangle\) and \(\left| c \right\rangle\). These linear combinations approximately reduce to a single dominant term when considered in the region \(D_A\), where the particles of the \(1\otimes2\) system can locally interact with the apparatus. 

If we consider configurations with a high probability of finding the particles of subsystem~1 in the region \(D_A\) but only a negligible probability for subsystem~2, then the probability amplitudes of both orthogonalized states reduce approximately to terms proportional to \(\left| b \right\rangle\). Conversely, for configurations in which the particles of subsystem~2 have a significant probability of appearing in \(D_A\), while those of subsystem~1 are far from it, both orthogonalized states reduce approximately to terms proportional to \(\left| c \right\rangle\).

It follows from the above reasoning that the presence of both \enquote{exchange} states in the expansion of the asymptotic state leads to nonzero probabilities of finding the particles of \emph{both} subsystems in the vicinity of the apparatus. Consequently, the right-hand side of Eq.~\eqref{Classical_apparatus_interaction} contains nonzero terms corresponding to the interaction Hamiltonians with the apparatus for both subsystems. This implies that the interaction with the apparatus \textbf{necessarily affects both subsystems $1$ and $2$}. Thus, the paradoxical conclusion \cite{EPRPhysRev.47.777,schredinger_1935,AharonovBohmPhysRev.108.1070} that the state of one subsystem may be formed after the measurement without any influence on it is avoided.

Hence, the possibility of avoiding the paradox is a direct consequence of the \textbf{assumption} made above. In what follows, we will \textbf{establish} that this \textbf{assumption} is indeed justified.

\subsubsection{Existence of "exchange" states via analytic properties}

We now aim to establish that the probability amplitude of a nonrelativistic quantum system in configuration space can have only a discrete set of isolated zeros within the region determined by the boundary conditions.

This statement will then be applied to the analysis of the probability amplitude of the state of the composite system $1 \otimes 2$. Among other variables, this amplitude is a function of the relative position vector $\vec{Y}$ of subsystems $1$ and $2$, defined in Eq.~\eqref{R_i_y}. As a consequence of the statement above, in the vicinity of both vectors $\vec{Y}$ and $-\vec{Y}$, there must exist an uncountable set of relative position vectors for which the probability amplitude is nonzero. This implies that if there is a possibility to observe subsystem~1 near the apparatus and subsystem~2 far from it, there is likewise a possibility to observe subsystem~2 near the apparatus and subsystem~1 far from it.

We now proceed to a more detailed justification of the statement that the probability amplitude of a quantum system can have only a discrete set of isolated zeros within the region determined by the boundary conditions. To this end, let us consider a quantum system consisting of \(N\) nonrelativistic particles that interact with each other. Its arbitrary state can be represented either in the coordinate representation by the probability amplitude \(\Psi\left(t,J \right)\) or in the momentum representation \(\Psi\left(t,p_J\right)\). Here \(t\) denotes time, \(J\) is the set of all \(3N\) Jacobi coordinates \cite{Cornilledoi:10.1142/5272,faddeev1993quantum} of the \(N\) particles, and \(p_J\) is the set of the corresponding canonically conjugate momenta. The elements of these sets are denoted as \(J_l\) and \(p_{J,l}\), where \(l = 1, 2, \ldots, 3N\).

For \(\Psi\left(t,p_J \right)\), we assume that its modulus decreases sufficiently rapidly and approaches zero as the modulus of any \(p_{J,l}\) becomes large. In particular, we assume that this falloff ensures the convergence of the following integrals over the entire momentum space, i.e., over infinite limits:
\begin{equation}\label{Furrier}
	\begin{aligned}
		&I(n) = \int 
		\left( \prod_{l=1}^{3N} dp_{J,l} \right)\times \\[4pt]
		& \exp\!\left( \frac{i}{\hbar} 
		\sum_{l=1}^{3N} p_{J,l} J_l \right)  
		\left( \prod_{l=1}^{3N} p_{J,l}^{n_l} \right)
		\Psi(t, p_J),
	\end{aligned}
\end{equation}
for any set \(n\) of natural-number exponents \(n_l\). Because arbitrarily large momenta cannot be physically realized in a nonrelativistic system, the assumption \(\left| I(n) \right| < +\infty\) appears physically reasonable.

The condition \(\left| I(n) \right| < +\infty\) implies that the function \(\Psi(t,J)\) possesses all partial derivatives
\begin{align*}
	&\frac{{\partial}^{k_{\mathrm{sum}}}\Psi(t,J)}
	{\partial^{k_1}J_1\,\partial^{k_2}J_2\ldots \partial^{k_{3N}}J_{3N}}, \\[4pt]
	&k_{\mathrm{sum}} = \sum_{l=1}^{3N} k_l,
\end{align*}
of all orders with respect to all elements of the set \(J\), for all values within the region where it is defined. 

We can make use of this property by considering the function $\Psi(t,J)$ not over the entire domain where the Jacobi coordinates may vary, but along some smooth curve within this domain. Such a curve can be defined by expressing each coordinate \(J_l\) as a function \(J_l(q)\) of a single parameter \(q\) defined on an interval \([q_0, q_1]\). We assume that all functions \(J_l(q)\) possess derivatives \({{{d}^{k}}{{J}_{j}}}/{{{ dq }^{k}}}\;\) of all orders \(k\) for all \(q \in [q_0, q_1]\). 

For any fixed moment of time \(t\), let us consider the function \(f(q) = \Psi(t, J(q))\), where \(J(q)\) denotes the set of functions \(J_l(q)\). The argument \(t\) will be omitted hereafter, since we are interested in the dependence of \(f\) on \(q\) for a fixed time. Taking into account the smoothness of both \(\Psi(t,J)\) and \(J_l(q)\), we conclude that the function \(f(q)\) possesses derivatives of all orders with respect to \(q\) for all \(q \in [q_0, q_1]\).

Given that all derivatives of \(f(q)\) exist, we can formally write its Taylor series. If these derivatives are such that the Taylor series centered at any point \(q_c \in [q_0, q_1]\) has a nonzero radius of convergence, then \(f(q)\) is a real analytic function~\cite{krantz2002primer}. As proved in~\cite{krantz2002primer}, a real analytic function either has a discrete set of isolated zeros or vanishes identically. To briefly explain why this statement holds, let us assume the opposite and verify that it leads to a contradiction. The opposite assumption is that there exists an interval \([q_i, q_f] \subset [q_0, q_1]\) of nonzero length such that \(f(q) = 0\) for all \(q \in [q_i, q_f]\), while there also exist points in \([q_0, q_1]\) at which \(f(q) \ne 0\). Since the functions \((q - q_c)^k\) for different values of \(k\) are linearly independent on any interval of nonzero length, all coefficients in the Taylor expansion around any \(q_c \in [q_i, q_f]\) must vanish. Taking into account that, for \(q_c = q_i\) and \(q_c = q_f\), the corresponding Taylor series have nonzero convergence radii \(r_i\) and \(r_f\), we conclude that \(f(q) = 0\) for all \(q \in [q_i - r_i, q_f + r_f]\). Repeating this reasoning iteratively, we find that \(f(q) = 0\) throughout its entire domain of definition.

This result allows us to exclude the case in which the function \(\Psi(t,J)\) has nonisolated zeros. Indeed, if nonisolated zeros existed, we could choose a curve \(J(q)\) passing through the region formed by these zeros. Consequently, there would exist an interval \([q_i, q_f]\) corresponding to points of this region, such that \(f(q) = 0\) for all \(q \in [q_i, q_f]\). Therefore, the function \(f(q)\) would vanish along the entire curve \(J(q)\). On the other hand, since \(\Psi(t,J)\) is normalized in its domain, there must be points where it is nonzero. By choosing the curve \(J(q)\) such that it includes both points where \(f(q) \ne 0\) and points corresponding to the interval \([q_i, q_f]\), we obtain a contradiction. Hence, if the derivatives of \(\Psi(t,J)\) at all points are such that, along any curve \(J(q)\), the function \(f(q) = \Psi(t, J(q))\) is real analytic (i.e., its Taylor series has a nonzero radius of convergence at every point \(q \in [q_0, q_1]\)), we can conclude that \(\Psi(t,J)\) has only isolated zeros.

This conclusion is valid only if the Taylor series of \(f(q)\) has a nonzero
radius of convergence at every point. Yet the mere existence of derivatives of all
orders does not guarantee convergence, since the magnitudes of the derivatives may
grow too rapidly with their order. For example, if at some point \(q\) the derivatives
\({d^n f(q)}/{dq^n}\;\) grow as \((n!)^2\), then the Taylor series has zero radius of
convergence around the point \(q\) ~\cite{fikhtengol1965fundamentals}. Since we have not specified the function
\(f(q) = \Psi(t, J(q))\) and the choice of the curve \(J(q)\), we cannot exclude the
possibility that, despite the existence of derivatives of all orders, the function
\(f(q)\) cannot be represented by its formal Taylor series.

To avoid this issue, we can choose the curve \(J(q)\) such that the point corresponding to \(q = q_0\) coincides with the point corresponding to \(q = q_1\). In other words, we consider a closed smooth curve \(J(q)\). In this case, we obtain a periodic function \(f(q)\), which can be represented by a Fourier series.  

For this series to converge and represent the function \(f(q)\) at each point \(q\), it is sufficient that the first derivative \({df(q)}/{dq}\) exists at every point of the interval \([q_0, q_1]\)~\cite{courant1937differential,fikhtengol1965fundamentals}. As discussed above, in our case the function \(f(q)\) possesses derivatives of all orders. Consequently, it can be represented by its Fourier series.

As in the case of Taylor expansions, the basis functions of a Fourier series are linearly independent on any interval of nonzero length. Therefore, if a function that is represented by a convergent Fourier series vanishes at every point of some interval of nonzero length, then all coefficients of this expansion must be zero. As a result, the function vanishes throughout its entire domain of definition. Thus, similarly to a real analytic function, any smooth periodic function either has only isolated zeros or vanishes identically.

As before, we can exclude the case in which the function $\Psi(t,J)$ has nonisolated zeros. If we assume that there exists a region of nonzero size in which $\Psi(t,J)$ vanishes everywhere, then we can choose a smooth closed curve $J(q)$ that passes both through this region and through a region where $\Psi(t,J)$ is nonzero. Accordingly, the smooth periodic function $f(q) = \Psi\left(t,J(q)\right)$ vanishes on an interval of nonzero length. On the one hand, $f(q)$ must then vanish throughout its entire domain of definition, while on the other hand it must take nonzero values for those values of $q$ that correspond to the part of the curve $J(q)$ where $\Psi(t,J(q)) \ne 0$. This contradiction allows us to conclude that the probability amplitude $\Psi(t,J)$ can vanish only on a discrete set of isolated points.
     
\begin{figure}[!htbp]
	\centering
	\includegraphics[width=1.0\linewidth]{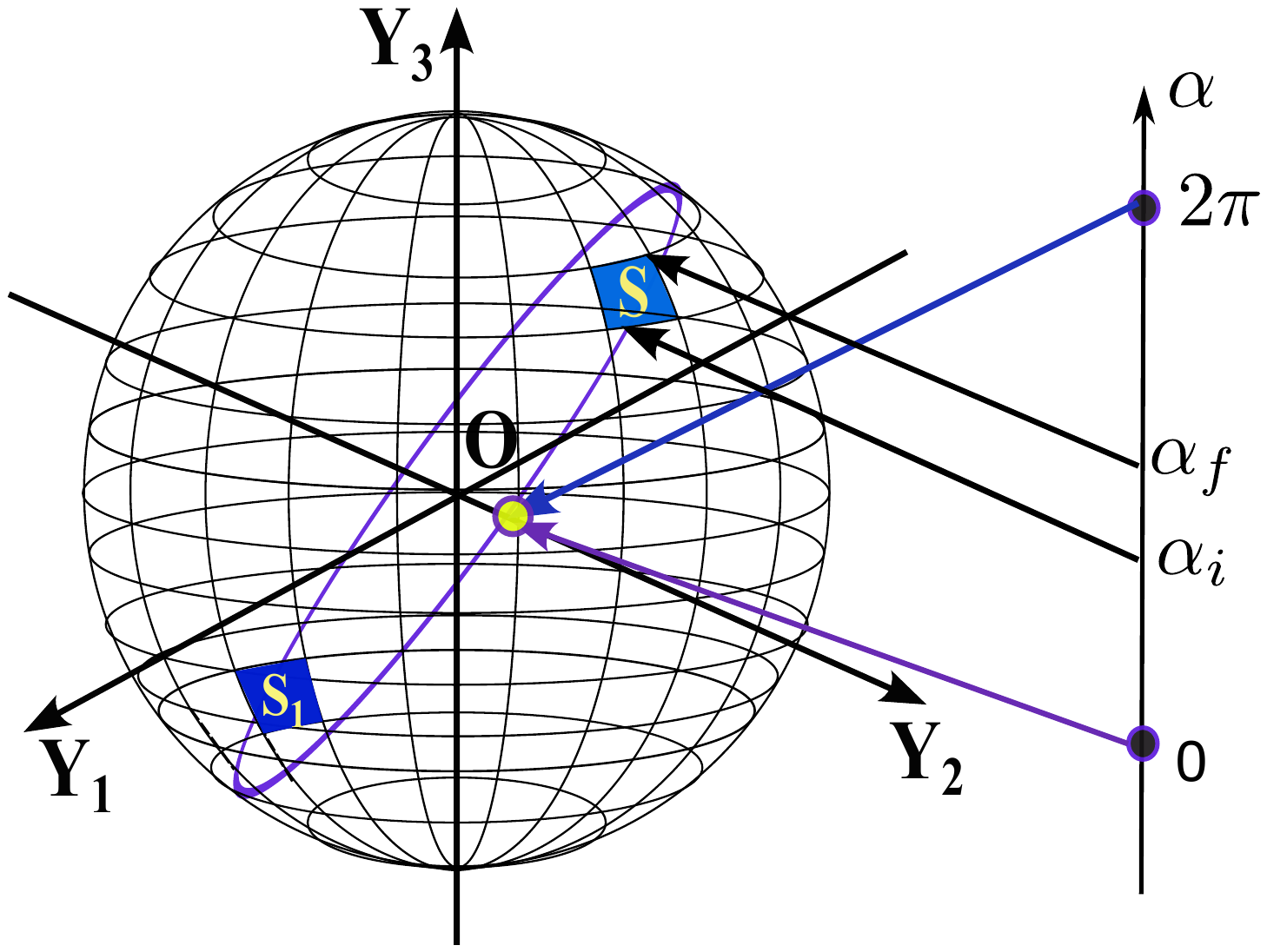}
	\caption{Division of a sphere into rectangular regions by \enquote{meridians and parallels}, with two opposite regions \(S\) and \(S_{1}\) shown.}
	\label{fig:newsfera1}
\end{figure}

For example, let us distinguish the relative position vector \(\vec{Y}\) (Eq.~\eqref{R_i_y}) among the set of Jacobi coordinates: \(J = \{\vec{Y},J_1\}\), where \(J_1\) denotes the set of all remaining Jacobi coordinates except \(\vec{Y}\). By assigning specific values to time \(t\) and to the variables in the set \(J_1\), we can reduce the probability amplitude to a three-dimensional space of vectors \(\vec{Y}\). In this three-dimensional space, we consider a sphere of large radius centered at \(\vec{Y} = \vec{0}\) (Fig.~\ref{fig:newsfera1}). By introducing spherical coordinates on the sphere, we can draw the coordinate lines — \enquote{meridians and parallels} — as shown in Fig.~\ref{fig:newsfera1}. In this way, the sphere is divided into spherical rectangles such as \(S\) and \(S_1\) in Fig.~\ref{fig:newsfera1}. The spherical rectangles \(S\) and \(S_1\) shown in the figure are centrally symmetric with respect to the center of the sphere. In the following, we will refer to such pairs of centrally symmetric spherical rectangles as \emph{opposite regions}.                

Now we consider a circle on the sphere that intersects both opposite spherical rectangles \(S\) and \(S_1\) along arcs of nonzero length, as shown in Fig.~\ref{fig:newsfera1}.

This circle can be defined parametrically by a smooth periodic function \(\vec{Y} = \vec{Y}(\alpha)\) with period \(2\pi\). Accordingly, for any fixed values of \(t\) and \(J_1\), the function \(f(\alpha) = \Psi\bigl(t, \vec{Y}(\alpha), J_1\bigr)\) is also smooth and periodic. 

If we assume that \(\Psi(t,\vec{Y},J_1)\) vanishes within region \(S\), then there must exist an interval \([\alpha_i,\alpha_f]\) such that \(f(\alpha) = 0\) for all \(\alpha \in [\alpha_i,\alpha_f]\) (Fig.~\ref{fig:newsfera1}) . As a consequence, all Fourier coefficients of \(f(\alpha)\) must be zero, and therefore the probability amplitude vanishes along the entire circle. By choosing, in a similar way, other circles that intersect \(S\) over an arc of nonzero length, we conclude that the probability amplitude must vanish over the entire sphere, including region \(S_1\).

However, if \(S_1\) corresponds to values of \(\vec{Y}\) that allow the system \(1 \otimes 2\) to interact with the measuring apparatus, then the probability amplitude cannot vanish at all points of \(S_1\). Consequently, it cannot vanish at all points of \(S\) either. Therefore, if a state contains the potential possibility of observing the position vectors of the centers of mass of subsystems \(\vec{R}_1\) and \(\vec{R}_2\) corresponding to some relative position vector \(\vec{Y}\) (Eq.~\eqref{R_i_y}), then it also contains the \enquote{exchange} possibility corresponding to \(-\vec{Y}\), except possibly for a discrete set of pairs \(\vec{Y}\) and \(-\vec{Y}\).

Thus, if prior to measurement the state of the system \(1 \otimes 2\) contains the potential possibility of local interaction with the apparatus involving one subsystem, then it must contain such a possibility for the other subsystem as well. 

\subsubsection{Existence of "exchange" states via symmetry properties}

We now aim to show that, in addition to the reasons based on analytic properties, the existence of \enquote{exchange} contributions in the asymptotic-state expansion can also follow from the symmetries of the Hamiltonian of the composite system \(1 \otimes 2\). In some cases, symmetry arguments not only imply the presence of such \enquote{exchange} terms but also provide information about their relative magnitudes.

These symmetries can be analyzed by noting that, in a typical situation considered in the context of the EPR paradox, the system \(1 \otimes 2\) remains isolated except for a short interval during which an external perturbation drives it from a bound state of its subsystems to a certain linear combination of continuum-spectrum states~\cite{AharonovBohmPhysRev.108.1070,CANTRELL1978499}. Our analysis here focuses on the time interval \(t \in [t_{0}, +\infty)\) [Fig.~\ref{fig:exchangeterms}], after this external influence has ceased and the system \(1 \otimes 2\) has once again become isolated.

One of the symmetries of an isolated quantum system is spatial inversion invariance, which can be discussed in cases where weak-interaction effects can be neglected. In such cases, the Hamiltonian of the system \(1 \otimes 2\) remains unchanged under inversion with respect to any chosen center. However, the transformation of the system’s state under inversion depends significantly on the choice of this inversion center.

\begin{figure}[!htbp]
	\centering
	\includegraphics[width=1.0\linewidth]{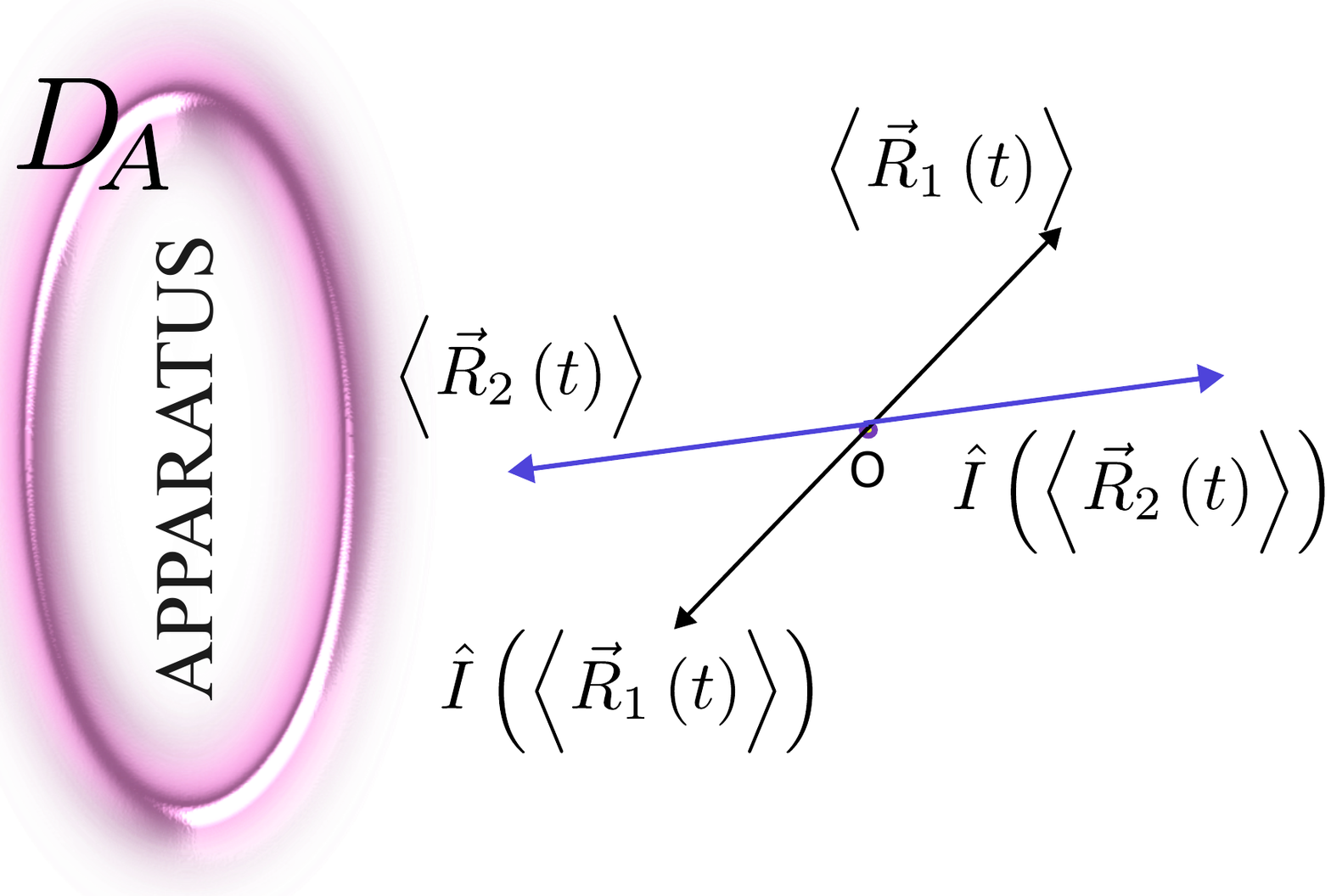}
	\caption{Exchange of the relative positions of the expectation values of the position vectors of the centers of mass of subsystems~1 and~2 with respect to the measuring apparatus under an inversion transformation. Here, \(O\) denotes the center of inversion, and \(\left\langle \vec{R}_{1}(t) \right\rangle\) and \(\left\langle \vec{R}_{2}(t) \right\rangle\) are the expectation values of the position vectors of the centers of mass of subsystems~1 and~2 in the state that results from the time evolution up to the moment \(t\), starting from the initial state at time \(t_0\).}
	\label{fig:exchangetermsinversion}
\end{figure}

We can make use of these properties by considering inversion transformations with respect to different centers. This idea is illustrated in Fig.~\ref{fig:exchangetermsinversion}, which shows an example of a possible choice of the inversion center \(O\). The figure also depicts the possible locations of the expectation values of the position vectors of the centers of mass of subsystems~1 and~2 in the state of the system \(1 \otimes 2\) at a given time \(t\), together with their images under the inversion operator \(\hat{I}\). As illustrated in Fig.~\ref{fig:exchangetermsinversion}, an inversion changes the relative distances between the particles of subsystems~1 and~2 and the measuring apparatus. If, in some state \(\left| \Psi(t) \right\rangle\), the particles of one subsystem are more likely to be found closer to the apparatus than those of the other subsystem, then in the inverted state \(\hat{I}\left|\Psi(t)\right\rangle\) the situation is reversed, and it is the particles of the other subsystem that are more likely to be observed closer to the apparatus.

Given this, we define the \enquote{exchange} state \(\left| c \right\rangle\) shown in Fig.~\ref{fig:exchangeterms} as
\begin{equation}\label{cIb}
	\left| c \right\rangle = \hat{I}\left| b \right\rangle .
\end{equation}

If the center \(O\) of the inversion operator \(\hat{I}\) can be chosen such that both \enquote{exchange} states, \(\left| \Psi(t) \right\rangle\) and \(\hat{I}\left|\Psi(t)\right\rangle\), appear in the asymptotic-state expansion of the system \(1 \otimes 2\), then both subsystems locally interact with the apparatus during the measurement. Our next task, therefore, is to determine the condition for the existence of such a choice of the inversion center \(O\).

To this end, we use the fact that the Hilbert space \(\mathcal{H}^{(1\otimes2)}\) of the system’s \(1 \otimes 2\) states can be decomposed into an orthogonal direct sum of subspaces:
\begin{equation}\label{orthogonal_direct_sum_of}
	\mathcal{H}^{(1\otimes 2)} = \mathcal{H}_{-} \oplus \mathcal{H}_{+}.
\end{equation}
Here, \(\mathcal{H}_{-}\) is the invariant subspace of the inversion operator \(\hat{I}\), consisting of its eigenstates corresponding to the eigenvalue \(-1\), while \(\mathcal{H}_{+}\) is the invariant subspace consisting of eigenstates corresponding to the eigenvalue \(+1\).

Accordingly, the state \(\left| \Psi(t) \right\rangle\) at any time \(t \ge t_0\) can be represented as
\begin{equation}\label{Rozclad_plus_minus}
	\left| \Psi(t) \right\rangle = \left| \Psi_{-}(t) \right\rangle + \left| \Psi_{+}(t) \right\rangle .
\end{equation}
Here, \(\left| \Psi_{-}(t) \right\rangle\) is the projection of the state \(\left| \Psi(t) \right\rangle\) onto the subspace \(\mathcal{H}_{-}\), and \(\left| \Psi_{+}(t) \right\rangle\) is its projection onto the subspace \(\mathcal{H}_{+}\).

Due to the symmetry of the Hamiltonian \(\hat{H}^{(1\otimes 2)}\) with respect to inversion, the subspaces \(\mathcal{H}_{-}\) and \(\mathcal{H}_{+}\) remain invariant under the action of the time-evolution operator. This means that neither \(\left| \Psi_{-}(t_{0}) \right\rangle\) nor \(\left| \Psi_{+}(t_{0}) \right\rangle\) from the decomposition~\eqref{Rozclad_plus_minus} can evolve into a state in which the regions \(D_1\) and \(D_2\) are positioned relative to the apparatus as in either \(\left| b \right\rangle\) [Fig.~\ref{fig:exchangeterms}(b)] or \(\left| c \right\rangle\) [Fig.~\ref{fig:exchangeterms}(c)]. Indeed, regardless of the specific form of the state \(\left| b \right\rangle\), the state obtained as a result of the inversion transformation possesses regions \(D_1\) and \(D_2\) [Fig.~\ref{fig:exchangeterms}] whose locations do not coincide with those of the original state \(\left| b \right\rangle\). Therefore, under this transformation, the state \(\left| b \right\rangle\) is mapped onto a linearly independent state and hence cannot be an eigenstate of the inversion operator, regardless of its other properties. Thus, the state \(\left| b \right\rangle\) belongs to neither of the invariant subspaces \(\mathcal{H}_{-}\) nor \(\mathcal{H}_{+}\). The same considerations apply to the state \(\left| c \right\rangle\).

Hence, each of the initial projections, \(\left| \Psi_{-}(t_{0}) \right\rangle\) and \(\left| \Psi_{+}(t_{0}) \right\rangle\), from the decomposition~\eqref{Rozclad_plus_minus} must evolve in time into some linear combination of the states \(\left| b \right\rangle\) and \(\left| c \right\rangle = \hat{I}\left| b \right\rangle\) [Fig.~\ref{fig:exchangeterms}]. Consequently, in the asymptotic state, each contribution in the decomposition~\eqref{Rozclad_plus_minus} leads to a potential possibility of observing particles from both subsystems~1 and~2 in the vicinity of the apparatus.

The question then arises as to whether one of these two contributions, \(\left| b \right\rangle\) or \(\hat{I}\left| b \right\rangle\), can be canceled as a result of adding the projections \(\left| \Psi_{-}(t) \right\rangle\) and \(\left| \Psi_{+}(t) \right\rangle\) according to Eq.~\eqref{Rozclad_plus_minus}.

To address this question in detail, let us first introduce several notations. 
We denote by \(\mathbf{r}_{1\otimes 2}\) an arbitrary set of position vectors of all the particles of the composite system \(1 \otimes 2\). 
The configuration in which all position vectors of subsystem~1 lie within the region \(D_1\) shown in Fig.~\ref{fig:exchangeterms}(b), 
while those of subsystem~2 simultaneously lie within the region \(D_2\) of the same figure, will be denoted by \(\mathbf{r}_{1\otimes 2}^{b}\). 
Similarly, any configuration in which the position vectors of the particles of subsystems~1 and~2 lie within the regions \(D_1\) and \(D_2\), respectively, 
arranged as in Fig.~\ref{fig:exchangeterms}(c), will be denoted by \(\mathbf{r}_{1\otimes 2}^{c}\).

Using these notations, we can now express the decomposition~\eqref{Rozclad_plus_minus} in the coordinate representation as
\begin{equation}\label{Rozclad_in_coordinate_representation}
	\Psi \left( t, \mathbf{r}_{1\otimes 2} \right)
	= \Psi_{-}\left( t, \mathbf{r}_{1\otimes 2} \right)
	+ \Psi_{+}\left( t, \mathbf{r}_{1\otimes 2} \right).
\end{equation}
Here, \(\Psi \left( t, \mathbf{r}_{1\otimes 2} \right)\), \(\Psi_{-}\left( t, \mathbf{r}_{1\otimes 2} \right)\), and \(\Psi_{+}\left( t, \mathbf{r}_{1\otimes 2} \right)\) 
are the expansion coefficients of each term in Eq.~\eqref{Rozclad_plus_minus} in the basis of eigenstates 
\(\left| \mathbf{r}_{1\otimes 2} \right\rangle\) corresponding to the configurations \(\mathbf{r}_{1\otimes 2}\).

Our question is whether one of the following two cancellations can occur:
\begin{equation}\label{Cancellation_b}
	\Psi_{-}\left( t, \mathbf{r}_{1\otimes 2}^{b} \right)
	+ \Psi_{+}\left( t, \mathbf{r}_{1\otimes 2}^{b} \right)
	= 0,
\end{equation}
for all configurations \(\mathbf{r}_{1\otimes 2}^{b}\), or
\begin{equation}\label{Cancellation_c}
	\Psi_{-}\left( t, \mathbf{r}_{1\otimes 2}^{c} \right)
	+ \Psi_{+}\left( t, \mathbf{r}_{1\otimes 2}^{c} \right)
	= 0,
\end{equation}
for all configurations \(\mathbf{r}_{1\otimes 2}^{c}\).

Let us suppose that one of these conditions holds, for instance, Eq.~\eqref{Cancellation_b}.  
Then, for all configurations \(\mathbf{r}_{1\otimes 2}^{b}\), the following equality must be satisfied:
\begin{equation}\label{Rivnist_moduliv}
	\left| \Psi_{-}\left( t, \mathbf{r}_{1\otimes 2}^{b} \right) \right|
	= \left| \Psi_{+}\left( t, \mathbf{r}_{1\otimes 2}^{b} \right) \right|.
\end{equation} 

To determine the consequences of this equality, we take into account two facts.  
First, by the definition~\eqref{cIb}, an inverted configuration 
\(\mathbf{r}_{1\otimes 2}^{c} = \hat{I} \mathbf{r}_{1\otimes 2}^{b}\) exists for each configuration \(\mathbf{r}_{1\otimes 2}^{b}\).  
Second, \(\Psi_{-}\left( t, \mathbf{r}_{1\otimes 2} \right)\) and \(\Psi_{+}\left( t, \mathbf{r}_{1\otimes 2} \right)\) 
are eigenfunctions of the inversion operator \(\hat{I}\) corresponding to the eigenvalues \(-1\) and \(+1\), respectively.  
Thus, for the configuration \(\mathbf{r}_{1\otimes 2}^{c} = \hat{I}\mathbf{r}_{1\otimes 2}^{b}\), we have
\begin{equation}\label{Rivnist_moduliv_bc}
	\begin{aligned}
		\left| \Psi_{-}\left( t, \mathbf{r}_{1\otimes 2}^{c} \right) \right|
		&= \left| \Psi_{-}\left( t, \mathbf{r}_{1\otimes 2}^{b} \right) \right|,\\
		\left| \Psi_{+}\left( t, \mathbf{r}_{1\otimes 2}^{c} \right) \right|
		&= \left| \Psi_{+}\left( t, \mathbf{r}_{1\otimes 2}^{b} \right) \right|.
	\end{aligned}
\end{equation}

As a consequence of Eqs.~\eqref{Rivnist_moduliv} and~\eqref{Rivnist_moduliv_bc}, we conclude that if the identity~\eqref{Cancellation_b} holds, then for any configuration \(\mathbf{r}_{1\otimes 2}\),
\begin{equation}\label{Consequence}
	\Psi_{-}\left( t, \mathbf{r}_{1\otimes 2} \right)
	= \Psi_{+}\left( t, \hat{I}\mathbf{r}_{1\otimes 2} \right).
\end{equation}
This identity leads to the equality
\begin{equation}\label{Equality_of_norms}
	\left\| \Psi_{-}(t) \right\| = \left\| \Psi_{+}(t) \right\|,
\end{equation}
where \(\left\| \cdot \right\|\) denotes the Hilbert-space norm of the projections in Eq.~\eqref{Rozclad_plus_minus}.

On the other hand, we can consider the expectation value of the inversion operator \(\hat{I}\):
\begin{equation}\label{Seredna_parnist}
	\begin{aligned}
		\left\langle \hat{I} \right\rangle 
		&= \left\langle \Psi(t) \right| \hat{I} \left| \Psi(t) \right\rangle \\[3pt]
		&= \left\| \Psi_{+}(t) \right\|^{2} - \left\| \Psi_{-}(t) \right\|^{2}.
	\end{aligned}
\end{equation}

As a consequence of the inversion symmetry of the Hamiltonian, this expectation value does not depend on time and is fully determined by the initial state \(\left| \Psi(t_{0}) \right\rangle\). If the initial state is such that \(\left\langle \hat{I} \right\rangle \ne 0\), the equality~\eqref{Equality_of_norms} cannot be satisfied. Therefore, the condition~\eqref{Cancellation_b} cannot hold either. Similarly, it can be shown that the condition~\eqref{Cancellation_c} cannot be satisfied.

From the above considerations, we can conclude that if the expectation value of the inversion operator is nonzero, then the asymptotic state must necessarily include \enquote{exchange} contributions. In this case, their appearance can be justified solely on the basis of inversion symmetry arguments.

We can now use these arguments to at least roughly compare the probabilities of observing the particles of subsystems~1 and~2 in the region \(D_A\) [Fig.~\ref{fig:exchangeterms}]. Such a comparison provides insight into the relative magnitudes of the contributions corresponding to the interactions of each subsystem with the apparatus on the right-hand side of Eq.~\eqref{Classical_apparatus_interaction}.

To this end, we note that each term in the decomposition~\eqref{Rozclad_plus_minus} contributes to the probability amplitude with complex values of equal magnitude for mutually inverted configurations. Therefore, any difference between the quantities \(\left| \Psi(t, \mathbf{r}_{1\otimes 2}) \right|\) and \(\left| \Psi(t, \hat{I}\mathbf{r}_{1\otimes 2}) \right|\) can arise only from interference between the terms \(\Psi_{-}(t, \mathbf{r}_{1\otimes 2})\) and \(\Psi_{+}(t, \mathbf{r}_{1\otimes 2})\), as follows from Eq.~\eqref{Rozclad_in_coordinate_representation}.

A detailed description of this interference would require solving the full dynamical problem and therefore cannot be obtained from symmetry considerations alone. Nevertheless, symmetry arguments allow us to identify the cases in which interference is insignificant. Indeed, Eq.~\eqref{Seredna_parnist} shows that if \(\left| \langle \hat{I} \rangle \right|\) approaches its maximal value of unity, then the norm of one of the two projections, \(\left\| \Psi_{-}(t) \right\|\) or \(\left\| \Psi_{+}(t) \right\|\), is much smaller than that of the other.

While this relation between the norms does not, by itself, specify the spatial behavior of the probability amplitudes, in the present situation it is reasonable to expect that the relation between the characteristic magnitudes of \(\Psi_{-}(t, \mathbf{r}_{1\otimes2})\) and \(\Psi_{+}(t, \mathbf{r}_{1\otimes2})\) mirrors that between the norms \(\left\| \Psi_{-}(t) \right\|\) and \(\left\| \Psi_{+}(t) \right\|\).

Indeed, since the state of the system \(1 \otimes 2\) is not a bound state of subsystems~1 and~2, it can be represented as a linear combination of the continuous-spectrum eigenstates of the system Hamiltonian. Consequently, each projection, \(\Psi_{-}(t, \mathbf{r}_{1\otimes 2})\) and \(\Psi_{+}(t, \mathbf{r}_{1\otimes 2})\), is also composed of continuous-spectrum eigenfunctions. As a result, the time evolution of both projections leads to an expansion of the spatial region within which the centers of mass of the subsystems can be observed. This means that, in both cases, the probability distribution of observing the centers of mass is not localized around particular configurations but is instead spread more or less uniformly over that region.

Under such conditions, the vicinities of different configurations contribute approximately equally to the integrals defining the norms. Hence, a larger norm corresponds to a greater characteristic magnitude of the probability amplitude.

Given this relation, we may conclude that when \(\left| \langle \hat{I} \rangle \right|\) is close to unity, the interference between the projections \(\Psi_{+}(t, \mathbf{r}_{1\otimes 2})\) and \(\Psi_{-}(t, \mathbf{r}_{1\otimes 2})\) becomes negligible. In this case, in the sum~\eqref{Rozclad_in_coordinate_representation}, one contribution dominates, while the other represents only a small correction. Accordingly, the moduli of the probability amplitude \(\left| \Psi(t, \mathbf{r}_{1\otimes 2}) \right|\) for mutually inverted configurations are approximately equal. This implies that they are nearly the same for configurations in which the position of the center of mass of subsystem~1 lies near the apparatus and for those in which the center of mass of subsystem~2 occupies that position instead. As a consequence, the terms in Eq.~\eqref{Classical_apparatus_interaction} corresponding to the interactions of the particles of subsystems~1 and~2 with the apparatus are comparable in magnitude. Thus, the interactions between the particles of subsystem~1 and the apparatus, and those of subsystem~2 with the apparatus, affect the formation of the post-measurement state to an equal extent.

Let us now consider the opposite case, when \(\langle \hat{I} \rangle = 0\) or its modulus is close to zero. In this situation, we can make use of the freedom to choose the inversion center in order to maximize \(\left| \langle \hat{I} \rangle \right|\) as a function of the center’s position.
 
To this end, we first select the inversion center \(O\) in such a way as to ensure the largest possible overlap between the region \(D\) and its image \(\hat{I}D\). A reasonably realistic assumption is that the region \(D\) possesses a center of symmetry. In that case, inversion with respect to this center yields \(\hat{I}D = D\). As a result, this choice leads to the coincidence of the regions in which the moduli of the functions \(\Psi(t_{0}, \mathbf{r}_{1\otimes 2})\) and \(\hat{I}\Psi(t_{0}, \mathbf{r}_{1\otimes 2})\) differ significantly from zero.
 
Under these conditions, the integral defining \(\langle \hat{I} \rangle\) can vanish, or become very small, only as a result of the mutual cancellation of contributions from different parts of the integration domain. Such cancellation, in turn, results from the spatial variation of the phase of the complex integrand across that domain.
 
\begin{figure}[!htbp]
 	\centering
 	\includegraphics[width=1.0\linewidth]{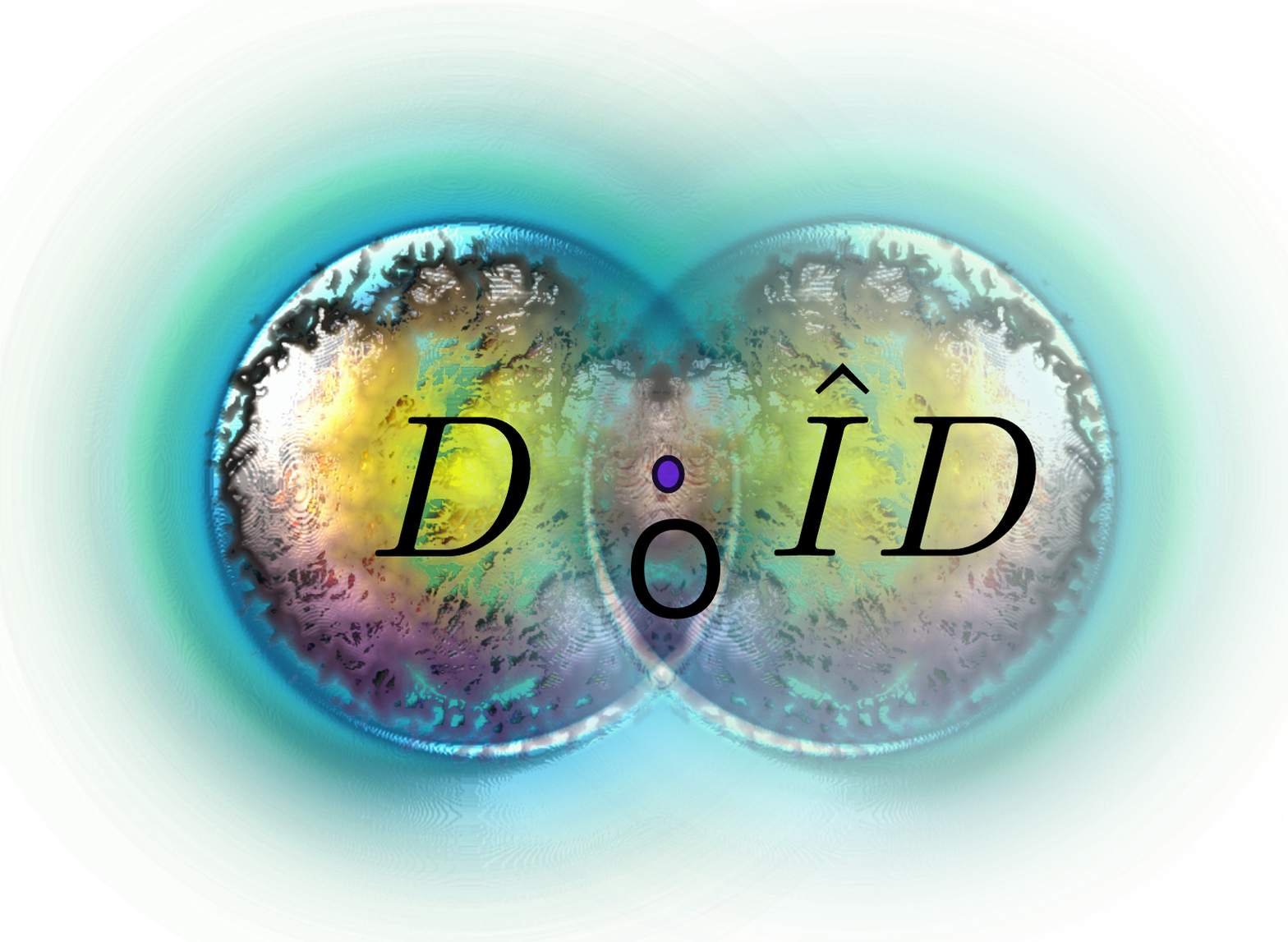}
 	\caption{Reduction of the overlap region \(D \cap \hat{I}D\) by an appropriate choice of the inversion center.}
 	\label{fig:inversionoverlapping}
 \end{figure}

To suppress this variation, one can displace the inversion center, thereby reducing the overlap region \(D \cap \hat{I}D\), as illustrated in Fig.~\ref{fig:inversionoverlapping}. In doing so, we also reduce the portion of the region \(D\) that provides the main contribution to the integral. As this overlap region becomes sufficiently small, the phase variation within it must likewise decrease. Consequently, for a sufficiently small overlap region, the phase variation cannot produce complete cancellation of the integral \(\langle \hat{I} \rangle\). Thus, this integral, regarded as a function of the inversion-center coordinates, is not identically zero, and its modulus must reach a maximal value at some specific position of the inversion center.
 
If the maximum value of \(\left| \langle \hat{I} \rangle \right|\) obtained in this way remains small, it becomes impossible, based on symmetry considerations alone, to compare the extent to which the interaction between each subsystem and the apparatus affects the formation of the post-measurement state. Nevertheless, since this maximal value is nonzero, the above considerations allow us to conclude that this effect, though possibly unequal in magnitude, occurs for both subsystems rather than exclusively for one of them.

Besides, there exist other symmetries that lead to the appearance of \enquote{exchange} contributions in the expansion of the asymptotic state of the composite system \(1 \otimes 2\). To describe the transformation corresponding to the next symmetry that possesses this property, let us introduce several additional notations, again employing the Jacobi coordinates.

In addition to the notations introduced earlier, let us denote
\begin{equation}\label{y}
	y = \{\vec{Y}\} \cup y^{(1)} \cup y^{(2)},
\end{equation}  
as the set of all relative position vectors, and  
\begin{equation}\label{J}
J = \{\vec{R}\} \cup y	
\end{equation}
as the set containing all Jacobi coordinates of the system. Recall that \(\vec{R}\) denotes the position vector of the center of mass of the composite system \(1 \otimes 2\), and \(\vec{Y}\) represents the relative position vector between the centers of mass of subsystems~1 and~2 (Eq.~\eqref{R_i_y}).

We now define a transformation acting on the set \(J\), which we denote by \(\hat{I}_{J}\):
\begin{equation}\label{Peretvorenna_IJ}
	\begin{aligned}
		& \hat{I}_{J} \vec{R} = \vec{R}, \\[2pt]
		& \hat{I}_{J} \vec{Y} = -\vec{Y}, \\[2pt]
		& \hat{I}_{J} \vec{y}_{k}^{(1)} = -\vec{y}_{k}^{(1)}, \quad k = 1, 2, \ldots, N_{1}-1, \\[2pt]
		& \hat{I}_{J} \vec{y}_{k}^{(2)} = -\vec{y}_{k}^{(2)}, \quad k = 1, 2, \ldots, N_{2}-1.
	\end{aligned}
\end{equation}

Here, \(N_{1}\) and \(N_{2}\) denote the numbers of particles in subsystems~1 and~2, respectively.

The representation of the transformation~\eqref{Peretvorenna_IJ} on the state space of the composite system \(1 \otimes 2\) is defined as
\begin{equation}\label{Zobragenna}
	\hat{I}_{J}\Psi(t, \vec{R}, y) = \Psi(t, \vec{R}, -y),
\end{equation}
where \((-y)\) denotes the set consisting of vectors opposite to the corresponding ones in \(y\).

Thus, the transformation~\eqref{Peretvorenna_IJ} inverts all relative coordinates while leaving the center-of-mass position vector \(\vec{R}\) unchanged, in contrast to the inversion transformation \(\hat{I}\) considered earlier. This distinction is essential because of the entanglement between the coordinate \(\vec{R}\) and the remaining Jacobi coordinates. Even if, prior to an external interaction that destroys a bound state, the probability amplitude factorized into a product of a function depending on \(\vec{R}\) and another depending on the relative coordinates, this separability is generally lost during the interaction.

Indeed, the term in the system Hamiltonian corresponding to the interaction with an external field prevents such a separation: it introduces coupling between the center-of-mass coordinate and the other Jacobi coordinates. Consequently, the time-evolution operator cannot be represented as a product of two commuting operators, one of which depends only on \(\vec{R}\), as in the case of an isolated system. As a result, time evolution under this operator leads to entanglement among all position variables, including \(\vec{R}\).

Therefore, the transformations of the entangled state under the inversion operator \(\hat{I}\), which changes the center-of-mass position, and under the operator \(\hat{I}_{J}\) defined by~\eqref{Zobragenna}, which leaves it unchanged, are essentially different.

When the external interaction ceases, the system becomes isolated again, and its Hamiltonian can once again be written as the sum of a term depending only on the center-of-mass coordinate \(\vec{R}\) and a term depending only on the relative coordinates \(y\). This structure of the Hamiltonian ensures symmetry with respect to the transformation~\eqref{Peretvorenna_IJ}.

To see this, let us first note that inversion symmetry is realized not only for the full Hamiltonian of an isolated system but also separately for each of its two parts. Furthermore, observe that the relative coordinates \(y\) transform under~\eqref{Peretvorenna_IJ} in exactly the same way as they do under spatial inversion. This follows from the fact that each relative coordinate is a linear combination of differences of particle position vectors. Under inversion with respect to any center, such differences transform into their opposites, in a way analogous to the transformation illustrated in Fig.~\ref{fig:exchangetermsinversion}.
This transformation law leads to the relations~\eqref{Peretvorenna_IJ} in the special case of an inversion with respect to an arbitrary center.

Consequently, the term in the Hamiltonian that depends only on the relative 
coordinates \(y\) transforms under \(\hat{I}_{J}\) exactly as it does under the inversion 
operator \(\hat{I}\) and therefore remains invariant. The remaining term, which depends 
only on the center-of-mass coordinate, is likewise invariant under \(\hat{I}_{J}\), since 
\(\vec{R}\) itself is left unchanged by this transformation. Hence, the transformation 
\eqref{Peretvorenna_IJ} is indeed a symmetry of the Hamiltonian. The properties of the 
transformation \eqref{Peretvorenna_IJ} are discussed in detail in Appendix~1.

In particular, this transformation interchanges the relative positions of the regions 
\(D_{1}(t)\) and \(D_{2}(t)\) with respect to their distances from the apparatus. If, 
for a given configuration contributing to the state 
\(\Psi(t,\mathbf{r}_{1\otimes 2})\), the region \(D_{1}(t)\) lies closer to the apparatus 
than \(D_{2}(t)\), then in the transformed state 
\(\hat{I}_{J}\Psi(t,\mathbf{r}_{1\otimes 2})\) the situation is reversed: \(D_{2}(t)\) becomes 
closer to the apparatus than \(D_{1}(t)\). Therefore, the symmetry with respect to the 
transformation \(\hat{I}_{J}\) implies the possibility of the appearance of 
\enquote{exchange} contributions in the asymptotic state.

To substantiate that such contributions indeed appear in the asymptotic-state 
decomposition, we may rely on the fact that the algebraic properties of the operator 
\(\hat{I}_{J}\) are the same as those of \(\hat{I}\). In particular, the squares of both 
operators equal the identity; the Hilbert space \(\mathcal{H}^{(1\otimes 2)}\) admits a 
decomposition into an orthogonal direct sum of invariant subspaces for \(\hat{I}_{J}\), 
analogous to the decomposition~\eqref{orthogonal_direct_sum_of} defined with respect 
to \(\hat{I}\); and the operator \(\hat{I}_{J}\) commutes with the time-evolution 
operator, just as \(\hat{I}\) does.

These similarities allow us to repeat, for the operator \(\hat{I}_{J}\), the same line 
of reasoning that was previously applied to \(\hat{I}\). Therefore, if
\begin{equation}\label{Condition_for_I_j}
	\left\langle \hat{I}_{J} \right\rangle
	= \big\langle \Psi(t_{0}) \big| \hat{I}_{J} \big| \Psi(t_{0}) \big\rangle
	\neq 0 ,
\end{equation}
then the asymptotic state necessarily contains a linear combination of the 
\enquote{exchange} states \(\lvert b \rangle\) and \(\hat{I}_{J} \lvert b \rangle\) 
[Fig.~\ref{fig:exchangeterms}]. By analogy with the preceding analysis for \(\hat{I}\), 
we may conclude that if the value \(\big| \langle \hat{I}_{J} \rangle \big|\) is close 
to unity, then the interaction of both subsystems with the apparatus affects the 
formation of the post-measurement state to approximately the same extent.

Concerning these considerations, let us make two remarks.

\textbf{First}, the symmetry with respect to the transformation \(\hat{I}_{J}\) is more 
\enquote{flexible} than the symmetry associated with \(\hat{I}\). As follows from the 
properties of \(\hat{I}_{J}\) discussed in Appendix~1, the image of any configuration 
\(\mathbf{r}_{1\otimes 2}\) under \(\hat{I}_{J}\) is obtained by inverting 
\(\mathbf{r}_{1\otimes 2}\) with respect to its \emph{own} center of mass. Thus, whereas 
the transformation \(\hat{I}\) performs an inversion of all configurations with respect 
to a single fixed point, the transformation \(\hat{I}_{J}\) assigns to each configuration 
its own center of inversion, namely the position of the center of mass of that 
configuration.

This distinction may lead to a substantial difference between the moduli of the 
expectation values \(\big| \langle \Psi \vert \hat{I} \vert \Psi \rangle \big|\) and 
\(\big| \langle \Psi \vert \hat{I}_{J} \vert \Psi \rangle \big|\), making the larger of 
the two more convenient for estimating the extent to which the interaction with the 
apparatus influences the post-measurement state of each subsystem.

\textbf{Second}, as mentioned above and as shown in Appendix~1, in each state belonging 
to a pair of \enquote{exchange} states one of the regions \(D_{1}(t)\) or \(D_{2}(t)\) 
is closer to the apparatus than the other. Appendix~1 demonstrates that, for the state 
\(\Psi(t,\mathbf{r}_{1\otimes 2})\) and its \enquote{exchange} partner 
\(\hat{I}_{J}\Psi(t,\mathbf{r}_{1\otimes 2})\), the distances between the apparatus and 
the nearest region differ. Nevertheless, as shown in Appendix~1, this difference does
not prevent either subsystem~1 or subsystem~2 from interacting with the apparatus during
the measurement process and thereby from influencing the formation of the 
post-measurement state. The arguments presented in Appendix~1 are supported by numerical 
estimates corresponding to the real experimental situation described in 
Ref.~\cite{PhysRevX.13.021031}.

\begin{figure}[!htbp]
	\centering
	\includegraphics[width=0.75\linewidth]{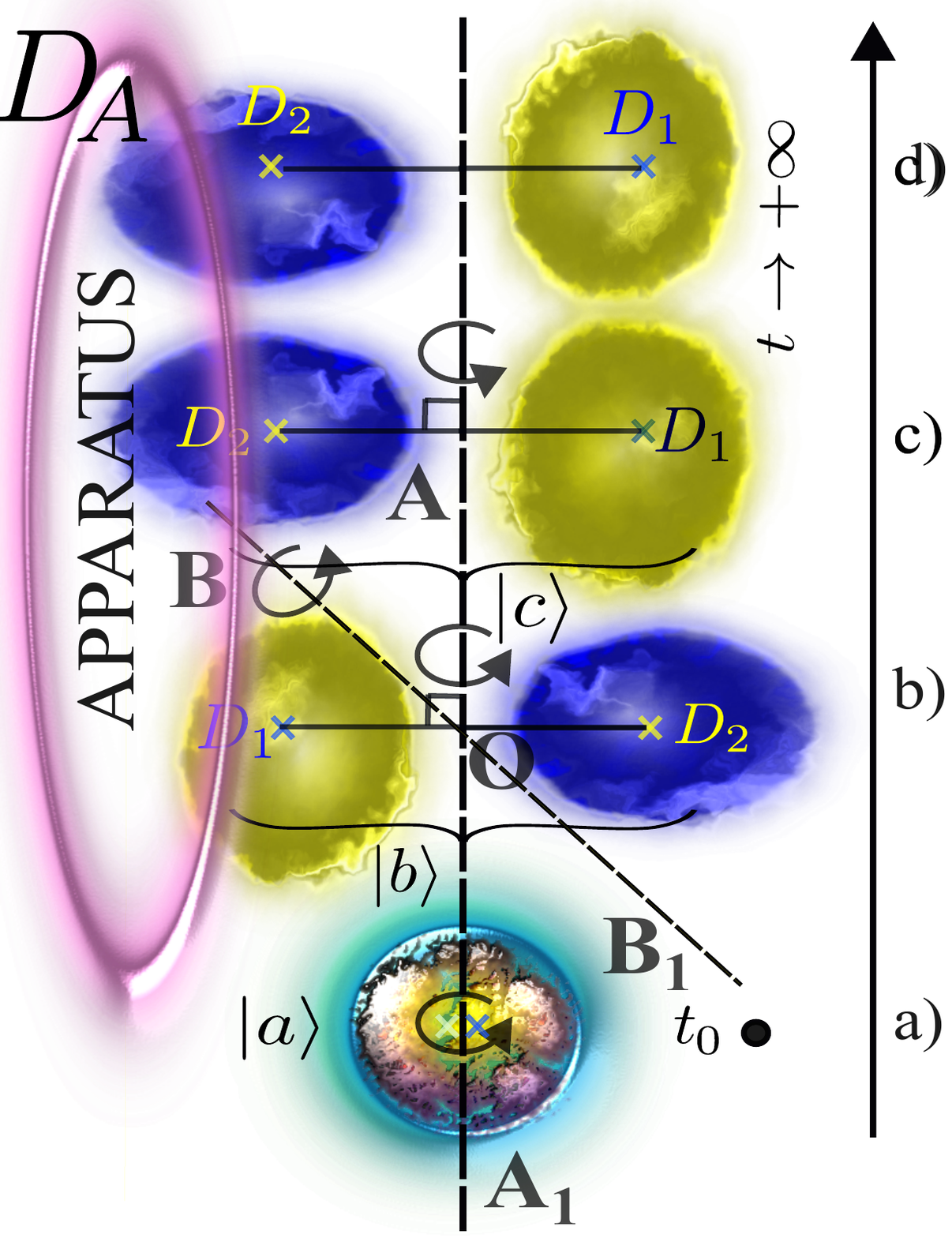}
	\caption{Schematic illustration of the \enquote{exchange} of the regions 
		\(D_{1}\) and \(D_{2}\) with respect to their distances from the apparatus under 
		rotations by the angle \(\pi\) about two different axes. The axis \(AA_{1}\) lies in 
		the plane of the figure, while the axis \(BB_{1}\) is perpendicular to this plane.}
	\label{fig:exchangeterms5}
\end{figure}

Let us now turn to a set of other transformations that \enquote{exchange} the regions
\(D_{1}(t)\) and \(D_{2}(t)\) (Fig.~\ref{fig:exchangeterms5}). These transformations may 
be constructed by exploiting the symmetry of the Hamiltonian of an isolated system with 
respect to spatial rotations.

As in the case of inversion, there exists considerable freedom in choosing the center of 
rotation. For any such choice, the Hamiltonian remains invariant, although the action of 
the rotation operator on the state depends on the chosen center.

For an arbitrary choice of the rotation center, we consider the group of spatial 
rotations around this point. From this group, we may select a subgroup of rotations 
about an axis passing through the chosen center. This subgroup contains, in particular, 
the rotation by an angle \(\pi\). We denote the operator corresponding to this rotation 
by \(\hat{R}(\pi)\). There exist infinitely many choices of the rotation center and the 
rotation axis for which the action of \(\hat{R}(\pi)\) interchanges the regions 
\(D_{1}(t)\) and \(D_{2}(t)\) with respect to their distances from the apparatus, in a 
manner analogous to that produced by spatial inversion.

An example of such an \enquote{exchange} is shown in 
Fig.~\ref{fig:exchangeterms5}. Here, the rotation center \(O\) is chosen at the midpoint 
of the segment connecting the expectation values of the center-of-mass positions of the 
two subsystems at a given moment of time. These expectation values are depicted by crosses. 
In contrast to the earlier schematic figures, where the regions \(D_{1}\) and \(D_{2}\) 
were drawn in a highly symmetric form for clarity, the present example employs irregular, 
asymmetric shapes. The choice of this more general configuration makes it easier to 
illustrate the differences produced by rotations about different axes.

As a consequence, the rotational symmetries generated by operators of the form 
\(\hat{R}(\pi)\), defined for different choices of rotation centers and axes, may also 
lead to the appearance of \enquote{exchange} contributions in the asymptotic state.  
To justify their presence, we again rely on the algebraic properties of the operator 
\(\hat{R}(\pi)\). These properties are the same as those of the inversion operator 
\(\hat{I}\) and of the operator \(\hat{I}_{J}\). Accordingly, the same line of reasoning 
applies here as in the case of inversion symmetry.

In particular, if
\[
\big\langle \Psi(t_{0}) \big| \hat{R}(\pi) \big| \Psi(t_{0}) \big\rangle \neq 0 ,
\]
then the decomposition of the asymptotic state necessarily contains 
\enquote{exchange} contributions arising from rotation symmetry.  
In this setting, the freedom to choose both the rotation center and the rotation axis 
provides substantially more possibilities for selecting a transformation \(\hat{R}(\pi)\) 
for which the modulus of the expectation value 
\(\big| \langle \Psi(t_{0}) | \hat{R}(\pi) | \Psi(t_{0}) \rangle \big|\) 
is maximized.

Thus, to state that the interaction of the apparatus with each subsystem of the 
composite system \(1 \otimes 2\) influences the formation of the post-measurement state 
to approximately the same extent, it is sufficient that there exists at least one 
operator—either \(\hat{I}\) with an appropriately chosen inversion center, or 
\(\hat{I}_{J}\), or \(\hat{R}(\pi)\) with a suitable rotation center and axis—whose 
expectation value has a modulus close to unity.

Here, we have demonstrated the existence of \enquote{exchange} contributions for non-identical systems. These contributions resemble the familiar 
exchange terms for identical particles, but—as shown above—their appearance in the 
state decomposition originates from completely different symmetries, rather than from 
the permutation symmetry characteristic of identical particles.

However, in the well-known experiments 
\cite{KocherPhysRevLett.18.575,AspectGrangierRogerPhysRevLett.49.91,PhysRevLett.28.938},  
the systems under investigation consisted of pairs of identical particles—photons.  
For such particles, the presence of genuine exchange contributions follows directly from 
their bosonic nature and requires no additional justification. Consequently, in these 
experiments each photon could, in principle, be found in the vicinity of either 
apparatus and therefore interact with it locally.

\subsubsection{Experimental confirmation of \enquote{exchange} states}

\begin{figure}[!htbp]
	\centering
	\includegraphics[width=0.75\linewidth]{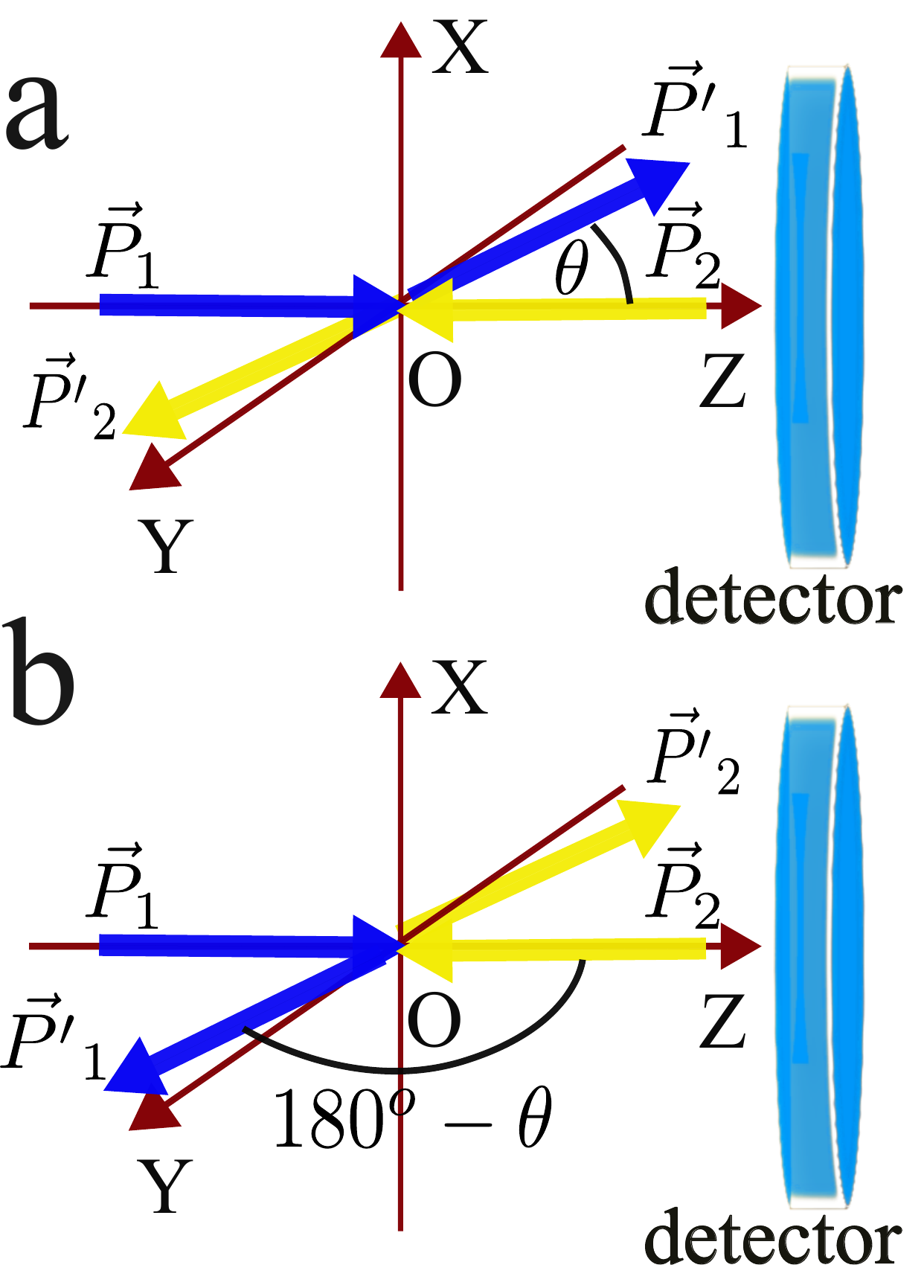}
	\caption{Two example basis states (a and b) that contribute to the expansion 
		of a two-particle state after scattering.}
	\label{fig-collision}
\end{figure}

To confirm the necessity of including the \enquote{exchange} terms in the asymptotic state 
decomposition, we consider experiments on elastic scattering of distinguishable particles. 
Specifically, we analyze the collision of two such particles in their center-of-mass frame, 
with initial momenta $\vec{P}_{1}$ and $\vec{P}_{2} = -\vec{P}_{1}$, as illustrated in 
Fig.~\ref{fig-collision}.

If the momentum configuration shown in Fig.~\ref{fig-collision}a satisfies the conservation 
laws, then the configuration shown in Fig.~\ref{fig-collision}b does so equally well. 
Consequently, the momentum-space representation of the post-collision state could, in principle, 
include \emph{both} configurations.

Experimental data confirm that both configurations indeed appear in the final state. 
Measurements of elastic scattering reported in 
Refs.~\cite{Sakamoto:1996xdz,PhysRevLett.34.233,LEP-2} 
show the differential cross section as a function of the scattering angle $\theta$ 
in the center-of-mass frame. 
These measurements demonstrate that the cross section is nonzero throughout the full angular 
interval from $0^\circ$ to $180^\circ$. 
Although the differential cross sections at angles $\theta$ and $180^\circ-\theta$ generally 
differ in magnitude, both are nonzero. 
Thus, momentum configurations corresponding to angles $\theta$ (as in 
Fig.~\ref{fig-collision}a) and $180^\circ-\theta$ (as in Fig.~\ref{fig-collision}b) occur in 
nonzero fractions of scattering events.

In addition, the measurements reported in 
Refs.~\cite{Newman:1974xa,PhysRevLett.30.507,Perl:1969pg} 
also contain events corresponding to the \enquote{exchange} configurations of 
Fig.~\ref{fig-collision}a,b, although these data do not cover the entire angular interval 
$\theta\in[0^\circ,180^\circ]$.

According to the reduction postulate, the presence of both outcomes in the data implies that 
before measurement the quantum state must have contained both \enquote{exchange} configurations 
of Fig.~\ref{fig-collision}a,b.

As follows from Eq.~\eqref{Classical_apparatus_interaction}, this in turn implies that the 
apparatus interacts with \emph{both} particles, even though only one of these interactions 
produces an observable effect.

This can be understood using the conservation of the expectation value of momentum. 
Before measurement, the expectation value of the total momentum of the combined system---the two 
particles plus the apparatus---is zero. 
Suppose that during the measurement the apparatus captures particle~1 and absorbs its momentum 
$\vec{P}_{1}$. 
After this interaction, the expectation value of momentum of the subsystem consisting of the 
apparatus plus the captured particle becomes nonzero. 
To preserve the total expectation value at zero, particle~2 must acquire an opposite expectation 
value of momentum directed outward from the apparatus. 
As a result, a momentum flow directed outward from the apparatus appears in the system, and, 
accompanying it, a corresponding outward probability flow for observing particle~2 
\cite{Flugge2012practical}. 
Consequently, the probability of finding particle~2 near the apparatus decreases, and the 
particle remains unobserved. 
Nevertheless, the emergence of these flows, together with the momentum exchange, results from 
the local interaction between particle~2 and the apparatus. 
Thus, as concluded above, the analysis of interactions in a quantum system cannot rely solely on 
observable outcomes and the reduction postulate.

Finally, we emphasize that all considerations above concerning the justification of the existence 
of \enquote{exchange} terms apply only to the nonrelativistic case. 
Experiment~\cite{Sakamoto:1996xdz} involves particles with nonrelativistic energies, whereas 
Refs.~\cite{PhysRevLett.34.233,LEP-2,Newman:1974xa,PhysRevLett.30.507,Perl:1969pg} describe relativistic scattering. 
The origin of \enquote{exchange} terms in the relativistic case is different and is briefly 
discussed in Appendix~2.

\subsubsection{Avoiding the EPR paradox in the Aharonov–Bohm setup}

Ref.~\cite{AharonovBohmPhysRev.108.1070} considers only the spin part of the state of two 
spin-${1}/{2}$ particles $A$ and $B$. In that work, the state is written as a superposition of 
two terms,
\begin{equation}\label{Aharonov_Bohm}
	\Psi = \frac{1}{\sqrt{2}}\left( A_{+} B_{-} - A_{-} B_{+} \right),
\end{equation}
where $A_{+}$ and $B_{+}$ denote eigenstates of the spin projection of particles $A$ and $B$ 
along an arbitrarily chosen axis corresponding to the eigenvalue $+1/2$, and $A_{-}$ and $B_{-}$ 
correspond to the eigenvalue $-1/2$.

\begin{figure}[!htbp]
	\centering
	\includegraphics[width=1.0\linewidth]{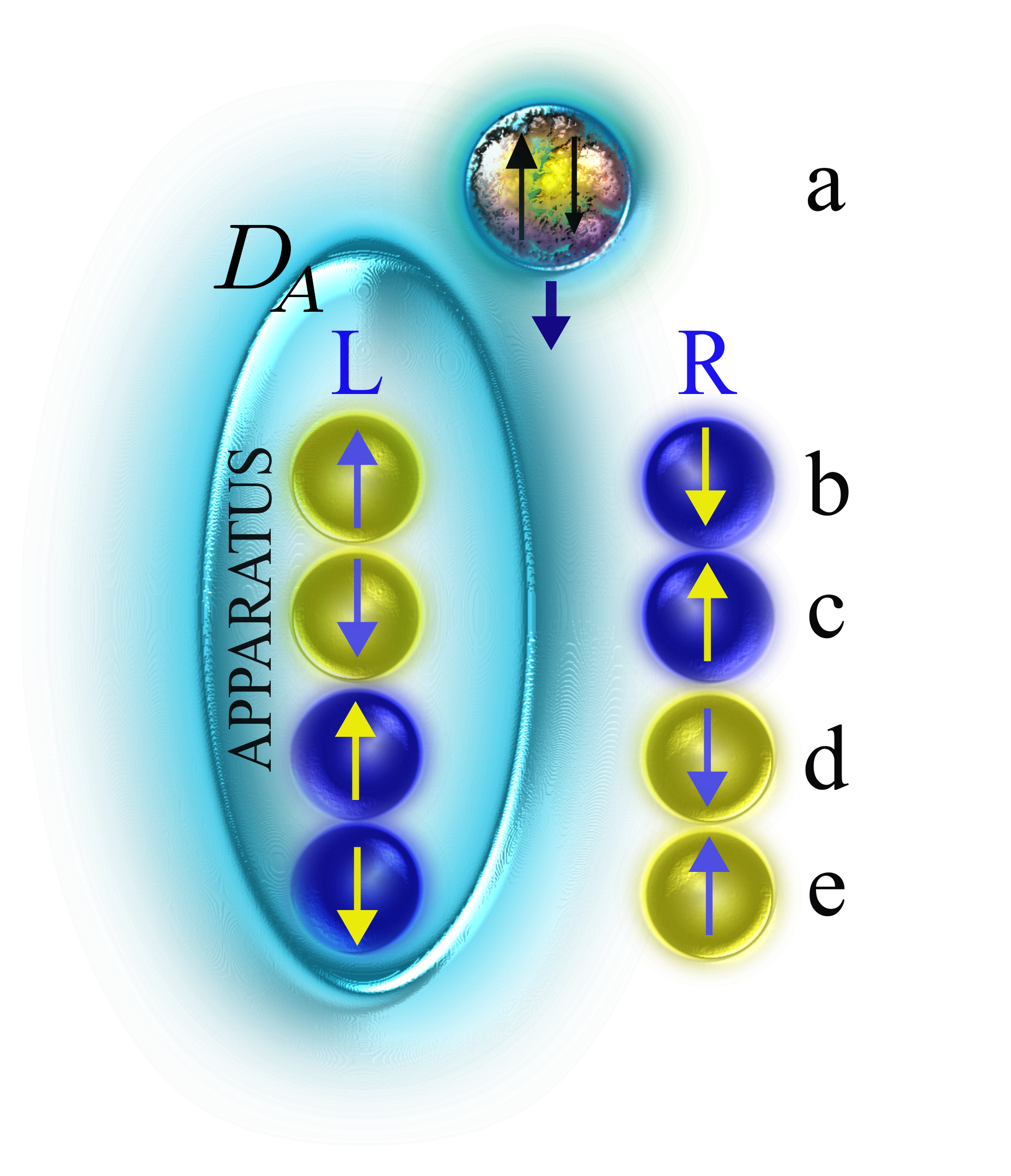}
	\caption{The initial state of the two-particle system $A$ and $B$ (a) evolves into an 
		asymptotic state that is a linear combination of four configurations (b--e).}
	\label{fig:aharonovbohm}
\end{figure}

However, to account for the possibility of interactions between the two particles and their 
individual interactions with the apparatus, one must also include the coordinate dependence 
of the state. In particular, once the \enquote{exchange} terms are taken into account, the asymptotic 
state interacting with the apparatus becomes a linear combination not of two but of four 
distinct configurations (Fig.~\ref{fig:aharonovbohm}(b--e)).

Accordingly, instead of Eq.~\eqref{Aharonov_Bohm}, the asymptotic state takes the form
\begin{equation}\label{Not_Aharonov_Bohm}
	\begin{aligned}
		&\Psi = 
		N\Big[
		\left( A_{+}^{R} B_{-}^{L} - A_{-}^{R} B_{+}^{L} \right)\\
		&+ 
		\left( A_{+}^{L} B_{-}^{R} - A_{-}^{L} B_{+}^{R} \right)
		\Big],
	\end{aligned}
\end{equation}
where $N$ is the normalization factor. As before, $A_{+}$ and $A_{-}$ denote eigenstates of the 
spin projection, but the superscripts $L$ and $R$ now indicate that, in the asymptotic state, the 
region in which particle $A$ is most likely to be detected is shifted to the left or to the right 
relative to its position in the initial state. The quantities 
$B_{+}^{L}$, $B_{-}^{L}$, $B_{+}^{R}$, and $B_{-}^{R}$ have the analogous meaning for particle $B$.
Thus, within the interaction region of the apparatus, both particles $A$ and $B$ may potentially 
be observed. Consequently, the state of the system after measurement is influenced by the 
apparatus acting on \emph{both} particles.

Contrary to the statement in Ref.~\cite{AharonovBohmPhysRev.108.1070}, the coordinate dependence 
of the state \eqref{Aharonov_Bohm} was discussed in Refs.~\cite{Home:1991kk, Selleribook:970920}, 
although without including the \enquote{exchange} terms. Because these terms were omitted, the paradox 
could not be avoided in those analyses.

\subsection{Objection Number 3}

In discussing the measurement of subsystem~1 of the composite system $1\!\otimes\!2$, 
Refs.~\cite{EPRPhysRev.47.777,AharonovBohmPhysRev.108.1070} 
state not only that the post-measurement state of the composite system is formed without any influence on subsystem~2, but also explicitly specify the form of this state.

This line of reasoning is based on a particular part of the reduction postulate~\cite{Von_Neumann_1996}. According to this postulate, if a measurement of a dynamical variable $\hat{V}$ yields an eigenvalue $V_{k}$ for some member of a quantum ensemble, then after the measurement the system is assumed to be projected into the corresponding eigenstate of~$\hat{V}$.

In Ref.~\cite{EPRPhysRev.47.777}, the situation considered is one where the system $1\!\otimes\!2$ is initially in an eigenstate of its total momentum with eigenvalue $\vec{P}$. Invoking the reduction postulate, it is asserted that if measuring the momentum of subsystem~1 yields the eigenvalue $\vec{p}_{1}$, then subsystem~1 is projected into the momentum eigenstate with eigenvalue $\vec{p}_{1}$, while subsystem~2 is projected into a momentum eigenstate with eigenvalue $\vec{P}-\vec{p}_{1}$. An analogous argument, but concerning spin projections, is made in Ref.~\cite{AharonovBohmPhysRev.108.1070}.

As discussed in the Introduction, the reduction postulate can be used for predicting the outcomes of measurements. However, in general, one cannot assert that a quantum system after a measurement is projected into the eigenstate of the measured dynamical variable. Instead, we analyze the evolution of the state during the measurement process on the basis of Eq.~\eqref{Classical_apparatus_interaction}.

This shift in perspective forces us to take into account the fact that the coefficients in the expansion of an entangled state over a chosen basis satisfy an infinite set of linear relations.
These relations ensure that the state belongs to a certain subspace of the full state space of the composite system $1\!\otimes\!2$, such as the subspace of total-momentum eigenstates or that of total-spin eigenstates. 

This feature is irrelevant when the reduction postulate is used alone, but it becomes essential when the process of state formation is analyzed directly from the Schr\"{o}dinger equation~\eqref{Classical_apparatus_interaction}.

To illustrate this point, let us consider the effect of a measurement performed on an eigenstate of the total momentum of the composite system $1\!\otimes\!2$.
We follow the notation of Ref.~\cite{EPRPhysRev.47.777}. 
In addition, $\mathbb{N}$ denotes the set of natural numbers.

Let $x_{1}$ denote the set of all arguments of the probability amplitude associated with subsystem~1, and $x_{2}$ those associated with subsystem~2.
Let $\{u_{i}(x_{1})\}_{i\in\mathbb{N}}$ be an arbitrary basis in the state space of subsystem~1, and $\{v_{j}(x_{2})\}_{j\in\mathbb{N}}$ an arbitrary basis in the state space of subsystem~2.
Then the set of product functions
\[
u_{i}(x_{1})\,v_{j}(x_{2}), \qquad i,j\in\mathbb{N},
\]
forms a basis in the state space of the composite system.

Thus, at time $t_{0}$, before the interaction with the measuring apparatus, the state can be expanded as~\cite{EPRPhysRev.47.777,schredinger_1935}:
\begin{equation}\label{Expansion_u_v}
	\Psi_{\vec{P}}(t_{0},x_{1},x_{2})
	= \sum_{i=1}^{\infty}\sum_{j=1}^{\infty}
	c_{ij}(t_{0})\,u_{i}(x_{1})\,v_{j}(x_{2}) .
\end{equation}

Here $\Psi_{\vec{P}}(t_{0},x_{1},x_{2})$ is the probability amplitude of the eigenstate of the total momentum corresponding to eigenvalue $\vec{P}$.

The linear constraints on the coefficients $c_{ij}(t_{0})$ express the requirement that 
no measurement outcome can yield a pair of subsystem momenta whose sum differs from 
the total momentum of the composite system.

Let $u(\vec{p}_{1},x_{1})$ and $v(\vec{p}_{2},x_{2})$ denote the coordinate representations 
of the momentum eigenstates of subsystems~1 and~2, respectively. 
Then one may write
\begin{equation}\label{Momentum_expansion}
	\begin{aligned}
		u_{i}(x_{1}) &= \sum_{\vec{p}_{1}} k_{i}(\vec{p}_{1})\,u(\vec{p}_{1},x_{1}), \\
		v_{j}(x_{2}) &= \sum_{\vec{p}_{2}} q_{j}(\vec{p}_{2})\,v(\vec{p}_{2},x_{2}),
	\end{aligned}
\end{equation}
where $k_{i}(\vec{p}_{1})$ and $q_{j}(\vec{p}_{2})$ are the corresponding expansion coefficients.

From Eqs.~\eqref{Expansion_u_v} and~\eqref{Momentum_expansion}
it follows that for any pair of eigenvalues $(\vec{p}_{1},\vec{p}_{2})$ satisfying
$\vec{p}_{1}+\vec{p}_{2}\neq \vec{P}$, the coefficients must obey the condition
\begin{equation}\label{Linear_conditions}
	\begin{aligned}
		&\sum_{i=1}^{\infty}\sum_{j=1}^{\infty}
		c_{ij}(t_{0})\,k_{i}(\vec{p}_{1})\,q_{j}(\vec{p}_{2}) = 0,\\
		&if \quad \vec{p}_{1}+\vec{p}_{2}\neq \vec{P}.
	\end{aligned}
\end{equation}

As another example, consider spin instead of momentum, namely states that are eigenstates 
of the spin projection onto a fixed direction~\cite{AharonovBohmPhysRev.108.1070}. 
In this case, the role of the coefficients $c_{ij}(t_{0})$ is played by the 
Clebsch--Gordan coefficients~\cite{Griffiths,Sakurai}, which satisfy linear relations 
ensuring that the state is an eigenstate of the total spin squared and of the total 
spin projection onto the chosen direction.

For the state to remain within the subspace defined by linear relations such as 
\eqref{Linear_conditions}, these relations must be preserved during the time evolution 
of the system in the course of the measurement. However, in the general case, there is 
no reason to expect that the time evolution governed by 
Eq.~\eqref{Classical_apparatus_interaction} preserves these constraints.

In particular, the interaction Hamiltonian between the system $1\otimes 2$ and the 
measuring apparatus depends on the differences between the position vectors of the 
system particles and those of the apparatus particles. Thus, in contrast to internal interactions within an isolated system, which depend only 
on differences of the position vectors of its constituents, the center-of-mass coordinate does not cancel out.

In such setups, the interaction between the system and the apparatus merely determines 
the initial conditions for the functions $\vec{r}_{j}^{(A)}(V_{k},t)$, selecting which 
of several possible stable macroscopic states the apparatus subsequently evolves into. 
The subsequent time dependence of these functions is then governed primarily by the 
internal dynamics of the apparatus. Since these dynamics are, in general, unrelated to 
the linear constraints \eqref{Linear_conditions}, there is no reason to expect that 
these constraints are preserved during the measurement process.

Thus, after a measurement performed on subsystem~1, the composite system will, in general,
cease to be an eigenstate either of its total spin projection or of its total momentum,
depending on which case (spin or momentum) is being considered.
Consequently, the sum of the measured values of the subsystem momenta or of their spin
projections is no longer constrained to take a fixed value, in contrast to the situation
prior to the measurement.

Accordingly, in the state obtained after the measurement on subsystem~1, it becomes 
impossible to make predictions about the outcomes of measurements of the spin projections 
or the momentum of subsystem~2 based solely on the result of that first measurement.

However, 
such predictions may still be possible on the basis of other physical considerations, 
for example, by invoking conservation laws of momentum or angular momentum.

\subsection{Objection Number~4}

Note that the paradoxical conclusions discussed in 
Refs.~\cite{EPRPhysRev.47.777,AharonovBohmPhysRev.108.1070}
rely on the fact that the entangled state is prepared such that
the values of two observables obtained in each run of the measurement
cannot be independent. 
This property exists prior to the measurement and is not a consequence
of the interaction between the system and the measuring apparatus.

In this situation, the following question naturally arises:
Should the quantum dynamics of the system
be described in terms of two independent dynamical variables, 
or in terms of a single dynamical variable?

For instance, for a total momentum eigenstate of the composite system
\(1 \otimes 2\) with eigenvalue \(\vec{P}\), it is sufficient to measure only one
variable. One possible choice is the relative momentum of the subsystems,
\(\vec{P}_{\vec{Y}}\), which is canonically conjugate to the relative position
vector \(\vec{Y}\) [see Eq.~\eqref{R_i_y}]. Once its value is obtained, the
individual momenta of the two subsystems, \(\vec{P}_1\) and \(\vec{P}_2\), can be
determined via their expressions in terms of \(\vec{P}\) and \(\vec{P}_{\vec{Y}}\).

In the following, we provide arguments in favor of the viewpoint that only a single dynamical variable
is physically relevant, rather than two.
This means that, in the situations under consideration, there exists only one quantity
that is actually measured in the quantum sense, namely through the hybrid
quantum--classical dynamics of the composite system including the measuring apparatus
\cite{Hall:2016oqf,PhysRevA.86.042120,Sudarshan:1976bt,PhysRevD.18.4580}.

Accordingly, the quantum \enquote{choice} — understood here as the manifestation
of one potential possibility from a given set of eigenvalues via the hybrid dynamics —
occurs for this single quantity, rather than simultaneously for two independent ones.

In this context, another question arises: How should the values of other quantities,
which can be expressed as functions of the measured variable, be interpreted?

For example, let us choose \(\vec{P}_1\) as the dynamical variable.
Then the measurement should be arranged such that the interaction with the measuring apparatus leads to the manifestation of a particular value of this quantity within the hybrid dynamics. Once this value is obtained, one can calculate \(\vec{P}_2\) from the relation
\(\vec{P}_2 = \vec{P} - \vec{P}_1\).

Thus, together with the statistics of \(\vec{P}_1\), a corresponding statistics
of \(\vec{P}_2\) appears.
The first statistics results from the hybrid dynamics
of the system \(1 \otimes 2\) interacting with the measuring apparatus.
The second statistics, by contrast, arises from calculations rather than from
the hybrid dynamics.
The question, therefore, is how this second statistics should be interpreted physically.

There are two possible interpretations. 
The first is to regard the calculated statistics as a prediction of the outcome
of a quantum measurement that would be obtained if one chose \(\vec{P}_2\),
\emph{instead of} \(\vec{P}_1\), as the dynamical variable.
In this case, the values of \(\vec{P}_2\) would arise from the hybrid dynamics,
whereas the values of \(\vec{P}_1\) would be obtained via calculations.

The second interpretation is to regard the calculated statistics as a prediction
of the outcome of a measurement of \(\vec{P}_2\), performed \emph{along with}
the measurement of \(\vec{P}_1\).
In this case, the values of both \(\vec{P}_1\) and \(\vec{P}_2\)
would be produced by the hybrid dynamics of the composite system
\(1 \otimes 2\) interacting with the corresponding measuring apparatuses.

It is this second interpretation that gives rise to the paradox.
Indeed, in this case the quantum \enquote{choice}, realized via the hybrid dynamics,
is performed twice by two spatially separated measuring devices, 
and the corresponding \enquote{choices} cannot be independent.

In Refs.~\cite{EPRPhysRev.47.777,
	AharonovBohmPhysRev.108.1070,barut1990heisenberg}, this interpretation is introduced
implicitly through the application of the state-reduction postulate to predict
measurement outcomes for both subsystems.

In Refs.~\cite{PhysicsPhysiqueFizika.1.195,ColloquiumRevModPhys.81.1727}, such a
measurement scheme involving two spatially separated and mutually independent
apparatuses is discussed explicitly.
Moreover, this scheme has not only been analyzed theoretically, but has also
been implemented in several widely cited experiments~\cite{KocherPhysRevLett.18.575,
	ClauserHorneShimonyHoltPhysRevLett.23.880,
	AspectGrangierRogerPhysRevLett.49.91,
	PhysRevX.13.021031,A_R_Wilson_1976,V_Paramananda_1987}.

If one adopts the viewpoint that only a single dynamical variable is physically
relevant, then both apparatuses measure this same variable, either
directly or via functions thereof. However, this leads to a situation that is
atypical in standard quantum mechanics: two distinct apparatuses measure
the same dynamical variable in the same state, which is not an eigenstate of that
variable.

At this point, two possible approaches to the further analysis arise.
The first is to consider only a \enquote{proper} quantum measurement, performed
by a single apparatus measuring a single dynamical variable of the composite
system \(1 \otimes 2\). In this case, the quantum \enquote{choice} among the set
of possible outcomes occurs only once and is made by only one apparatus.
As a result, no paradoxical correlation between two spatially separated
\enquote{choices} arises.
 
The second approach is to nonetheless consider such a measurement scheme,
despite its nonstandard character within conventional quantum mechanics.
This is motivated by the fact that it admits experimental realization and
therefore calls for a theoretical understanding.

This measurement involving two apparatuses will be discussed in the next
subsection. Here we aim to substantiate the statement that, in the situation
under consideration, there exists only a single dynamical variable rather than
two. We treat separately the case of a total momentum eigenstate of the
composite system \(1 \otimes 2\) and, independently, the case of a total spin
eigenstate.

\subsubsection{Measuring subsystems’ momenta}

Consider the subspace \(\mathcal{H}_{\vec{P}}\) consisting of states of the
composite system \(1 \otimes 2\) that are eigenstates of the total momentum
operator with eigenvalue \(\vec{P}\).
For any state belonging to this subspace, the eigenvalues of the subsystem
momenta \(\vec{P}_{1}\) and \(\vec{P}_{2}\) necessarily satisfy the relation
\[
\vec{P} = \vec{P}_{1} + \vec{P}_{2}.
\]

As a consequence, the decomposition of a state
\(\lvert \Psi \rangle \in \mathcal{H}_{\vec{P}}\) in terms of the momentum
eigenstates of the subsystems involves a Dirac delta function.
Accordingly, the corresponding coefficients cannot be normalized to unity and,
therefore, cannot be interpreted as probability amplitudes for the state
\(\lvert \Psi \rangle\).
We regard this property as an indication that the subsystem momenta
\(\vec{P}_{1}\) and \(\vec{P}_{2}\) are not \enquote{proper} dynamical
variables for a total momentum eigenstate.

To eliminate the delta function, one must integrate either over
\(\vec{P}_{1}\) or over \(\vec{P}_{2}\).
Equivalently, one may pass to the Jacobi variables \(\vec{Q}\) and
\(\vec{P}_{\vec{Y}}\), defined in terms of the subsystem masses
\(M_{1}\) and \(M_{2}\) as
\begin{equation}\label{Jacoby_Q_PY}
	\begin{aligned}
		\vec{P}_{1} &= \frac{M_{1}}{M_{1} + M_{2}}\,\vec{Q} - \vec{P}_{\vec{Y}}, \\
		\vec{P}_{2} &= \frac{M_{2}}{M_{1} + M_{2}}\,\vec{Q} + \vec{P}_{\vec{Y}}.
	\end{aligned}
\end{equation}
Integrating over \(\vec{Q}\) then ensures that the remaining dynamical
variables correspond to the components of a single vector rather than two
independent ones. Consequently, only the components of this single vector can
be regarded as measurable dynamical variables.

To clarify this point, let us pass to the center-of-mass frame, in which the
total-momentum eigenvalue is \(\vec{P}=\vec{0}\), and choose the components
of the relative momentum of the subsystems,
\(\vec{P}_{\vec{Y}}=\vec{P}_{2}-\vec{P}_{1}\), as the dynamical variable.
The probability amplitude of the state of the composite system
\(1 \otimes 2\) can then be represented in the form
\begin{equation}\label{Rozclad_po_p_malomu}
	\begin{aligned}
			&{\Psi }_{\vec{P}=0}
		\bigl(t,\vec{R}_{1},\vec{R}_{2},y^{(1)},y^{(2)}\bigr)\\
		&= \int d\vec{P}_{\vec{Y}}\,
		{\Psi }_{\vec{P}=0}\bigl(t,\vec{P}_{\vec{Y}},y^{(1)},y^{(2)}\bigr) \\
		&\quad \times
		\exp\!\left[
		\frac{i}{\hbar}\,
		\left( \vec{P}_{\vec{Y}}\cdot\bigl(\vec{R}_{2}-\vec{R}_{1}\bigr)\right) 
		\right].
	\end{aligned}
\end{equation}
Here \(\vec{R}_{1}\) and \(\vec{R}_{2}\) are the position vectors of the
centers of mass of subsystems \(1\) and \(2\), respectively, while
\(y^{(1)}\) and \(y^{(2)}\) denote the previously introduced sets of relative
Jacobi coordinates.

Suppose that the two apparatuses measure the tracks of subsystems \(1\) and
\(2\). These tracks are determined by the probability current density, the
momentum current density, or the energy current density. In all cases, the
corresponding density vectors are expressed in terms of spatial gradients
with respect to \(\vec{R}_{1}\) and \(\vec{R}_{2}\) \cite{Flugge2012practical}.
As is evident from Eq.~\eqref{Rozclad_po_p_malomu}, both gradients are
governed by a single vector \(\vec{P}_{\vec{Y}}\).

Specifically, the apparatus interacting with subsystem \(2\) measures the
value \(\vec{P}_{\vec{Y}}\), whereas the apparatus interacting with
subsystem \(1\) measures the same dynamical variable, but via the function
\(-\vec{P}_{\vec{Y}}\). Thus, in a two-apparatus experiment, both observers measure
a single dynamical variable in the same non-eigenstate, although using two
independent apparatuses.

Let us note that restricting ourselves to \enquote{proper} measurements, in which
a single dynamical variable is measured by a single apparatus, not only removes
the paradox associated with the interdependence of the \enquote{choices} made by distant apparatuses. It also eliminates the paradoxical possibility of simultaneously measuring the momentum and position of a particle, originally formulated in \cite{EPRPhysRev.47.777}.

Indeed, in \cite{EPRPhysRev.47.777}, a two-particle one-dimensional system is
considered in a state that is an eigenstate of both the total momentum and the
relative coordinate. A \enquote{proper} measurement of the particles' coordinates
should correspond to measuring the center-of-mass coordinate in such a way that
the state remains within the same subspace of eigenstates of the relative
coordinate. Knowing the center-of-mass coordinate and the relative coordinate,
one can then calculate the coordinates of each particle.

However, a measurement of the center-of-mass coordinate inevitably changes the
state of the two-particle system so that it no longer belongs to the subspace of
eigenstates of the total momentum. As a result, the individual particle momenta
are no longer constrained and thus cannot be inferred from a measurement of a
single dynamical variable.

Alternatively, a \enquote{proper} measurement of the particles' momenta, performed
via the relative momentum, leads to a situation in which the state of the
two-particle system ceases to be an eigenstate of the relative coordinate. As a
consequence, the particles' coordinates can no longer be calculated from the
measurement of a single dynamical variable.

Therefore, by restricting ourselves to \enquote{proper} measurements, we return
to the usual situation governed by the uncertainty principle.
                 
\subsubsection{Measuring subsystems’ spin projections}

Let us show that, in the case of measuring the spin projections of subsystems using two apparatuses, one encounters a situation analogous to that arising in the case of momentum measurements.

Consider measurements performed on a bipartite system $\mbox{$A\otimes B$}$ consisting of two spin-$\frac{1}{2}$ particles prepared in a state with total spin zero. This situation has been discussed, e.g., in Refs.~\cite{AharonovBohmPhysRev.108.1070,PhysicsPhysiqueFizika.1.195}. Following Ref.~\cite{PhysicsPhysiqueFizika.1.195}, we assume that the spin projections of both particles are measured using the Stern--Gerlach method~\cite{Gerlach:1922dv}.

Due to the interaction with the magnetic fields of the Stern--Gerlach apparatuses, the $\mbox{$A\otimes B$}$ system cannot remain in its initial total-spin-zero state. This follows from the fact that not all components of the total spin operator of the system commute with the Hamiltonian describing its interaction with the external magnetic fields. The Hamiltonian has the form
\begin{equation}\label{Stern_Gerlach_Hamiltonian}
	\begin{aligned}
		&\hat{H} = -\frac{\hbar^{2}}{2m_{A}} \Delta_{A}
		-\frac{\hbar^{2}}{2m_{B}} \Delta_{B}\\
		&+ \mu_{A}\sum_{k=1}^{3} \hat{s}_{k}^{(A)} H_{k}(\vec{r}_{A}) 
		+ \mu_{B}\sum_{k=1}^{3} \hat{s}_{k}^{(B)} H_{k}(\vec{r}_{B})\\
		&+ \hat{H}^{\mathrm{int}}_{A,B}.
	\end{aligned}
\end{equation}

\begin{figure}[!htbp]
	\centering
	\includegraphics[width=1.0\linewidth]{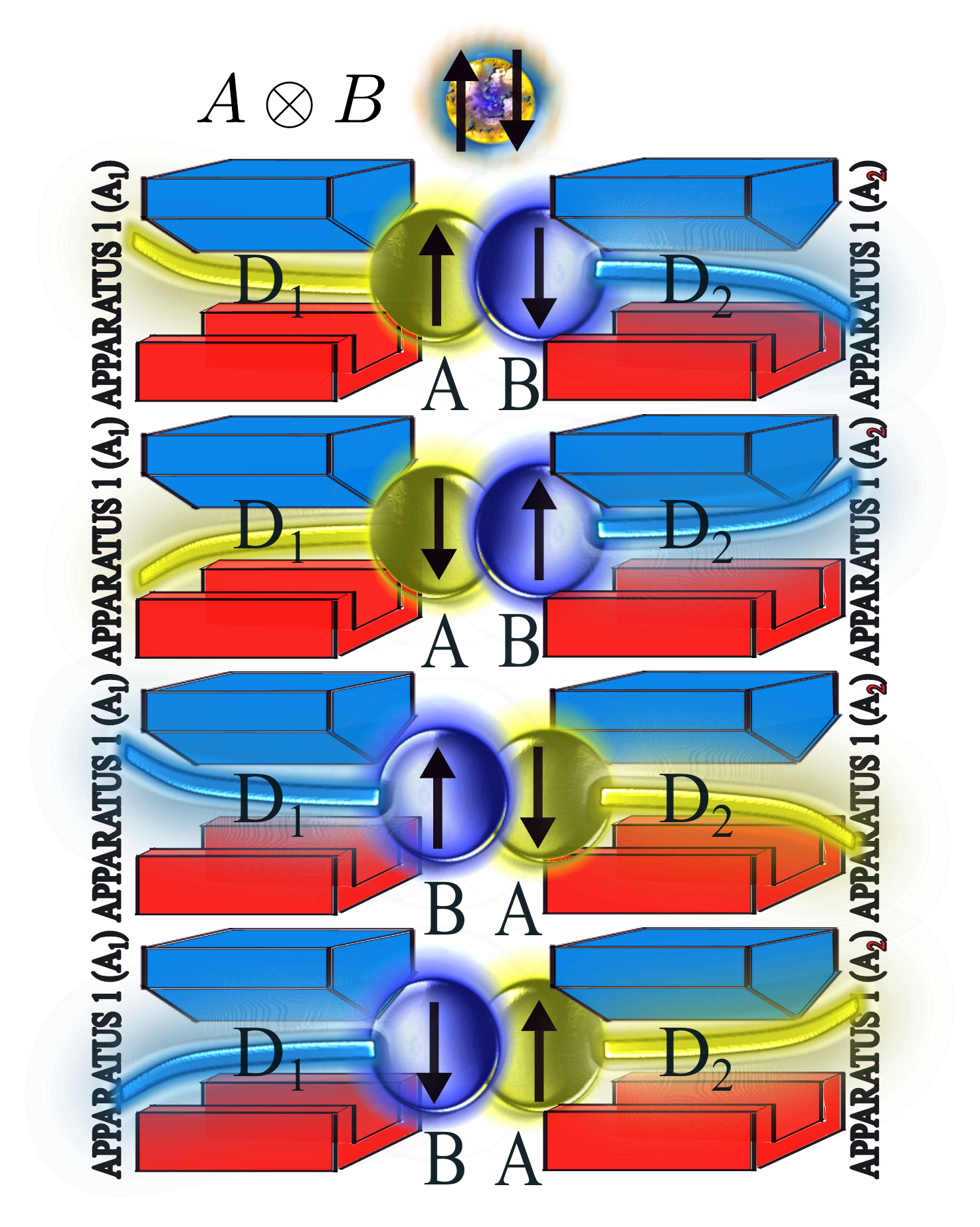}
	\caption{The regions $D_{1}$ and $D_{2}$ in which the magnetic field is nonzero during the measurement by two Stern--Gerlach apparatuses. 
		Due to the presence of \enquote{exchange} terms, both particle $A$ and particle $B$ can be detected in each of these regions.}
	\label{fig:twoapparatusmeauirementspin}
\end{figure}

Here $\vec{r}_{A}$ and $\vec{r}_{B}$ denote the position vectors of particles $A$ and $B$, $m_{A}$ and $m_{B}$ are their masses, and $\Delta_{A}$ and $\Delta_{B}$ are the Laplacians with respect to $\vec{r}_{A}$ and $\vec{r}_{B}$, respectively. The constants $\mu_{A}$ and $\mu_{B}$ characterize the magnetic moments of the particles, while $\hat{s}_{k}^{(A)}$ and $\hat{s}_{k}^{(B)}$ ($k=1,2,3$) are their spin component operators. The term $\hat{H}^{\mathrm{int}}_{A,B}$ represents the interaction Hamiltonian of particles $A$ and $B$, which accounts for the possible overlap of their spatial regions during the measurement (see Fig.~\ref{fig:twoapparatusmeauirementspin}).

The functions $H_{k}(\vec{r}_{A})$ and $H_{k}(\vec{r}_{B})$ are the components of the magnetic field in the Stern--Gerlach apparatuses acting on particles $A$ and $B$, respectively. The magnetic field is assumed to be nonzero only within two spatially separated regions $D_{1}$ and $D_{2}$ (Fig.~\ref{fig:twoapparatusmeauirementspin}). Within $D_{1}$ it is described by the function $\vec{H}^{(1)}(\vec{r})$, while within $D_{2}$ it is described by $\vec{H}^{(2)}(\vec{r})$, with $\vec{r}$ denoting the position vector of a point in the corresponding region.

Thus, the magnetic field entering the Hamiltonian \eqref{Stern_Gerlach_Hamiltonian} is defined as
\begin{equation}\label{Magnitne_pole}
	\vec{H}(\vec{r}) =
	\begin{cases}
		\vec{H}^{(1)}(\vec{r}), & \vec{r} \in D_{1}, \\
		\vec{H}^{(2)}(\vec{r}), & \vec{r} \in D_{2}.
	\end{cases}
\end{equation}

The Hamiltonian \eqref{Stern_Gerlach_Hamiltonian} determines the time evolution of the initial state \eqref{Not_Aharonov_Bohm} considered above.

If the coefficients of all six operators $\hat{s}_{k}^{(A)}$ and $\hat{s}_{k}^{(B)}$, $k=1,2,3$, entering the Hamiltonian \eqref{Stern_Gerlach_Hamiltonian} are not identically zero, then this Hamiltonian  does not commute with any of the operators
\begin{equation}\label{Total_spin_operator}
\hat{S}_{k} = \hat{s}_{k}^{(A)} + \hat{s}_{k}^{(B)}, \quad k=1,2,3,
\end{equation}
corresponding to the total spin of the system $A \otimes B$.
Therefore, as a result of the time evolution in the magnetic field \eqref{Magnitne_pole}, the state \eqref{Not_Aharonov_Bohm} becomes a non-eigenstate of all three components $\hat{S}_{k}$.

Moreover, if the Hamiltonian \eqref{Stern_Gerlach_Hamiltonian} contains 
nonzero coefficients multiplying the spin projection operators along at 
least two different axes — either 
$\hat{s}_{k}^{(A)}$ and $\hat{s}_{j}^{(A)}$, or 
$\hat{s}_{k}^{(A)}$ and $\hat{s}_{j}^{(B)}$, or 
$\hat{s}_{k}^{(B)}$ and $\hat{s}_{j}^{(B)}$, with $k \neq j$ — 
then for each total spin operator $\hat{S}_{k}$ there exists at least one 
term in the sum \eqref{Total_spin_operator} that does not commute with the 
Hamiltonian \eqref{Stern_Gerlach_Hamiltonian}.

Hence, if one wishes to preserve a given component of the total spin of the system 
$A \otimes B$, the only possible option is to generate magnetic fields in regions 
$D_{1}$ and $D_{2}$ (see Fig.~\ref{fig:twoapparatusmeauirementspin}) oriented along 
the same direction.

Let us choose this direction to coincide with the $OZ$ axis (axis $3$ in our notation).
Then, in expression \eqref{Magnitne_pole}, one has
\begin{equation}
	\label{Umova_zberegenna_Sz}
	\begin{aligned}
		& \vec{H}^{(1)}(\vec{r}) = \bigl(0,\, 0,\, H_{3}^{(1)}(\vec{r}) \bigr), \quad \vec{r} \in D_{1}, \\
		& \vec{H}^{(2)}(\vec{r}) = \bigl(0,\, 0,\, H_{3}^{(2)}(\vec{r}) \bigr), \quad \vec{r} \in D_{2}.
	\end{aligned}
\end{equation}

In this case, the terms in the Hamiltonian \eqref{Stern_Gerlach_Hamiltonian} describing 
the interaction with the magnetic field contain only the operators 
$\hat{s}_{3}^{(A)}$ and $\hat{s}_{3}^{(B)}$. Consequently, the total spin component operator \(\hat{S}_{3}\) \eqref{Total_spin_operator} commutes with these terms.

The interaction Hamiltonian $\hat{H}_{A,B}^{\mathrm{int}}$ can contain the spin operators 
of both particles only in the form of the scalar product
\[
\left( \hat{\vec{s}}^{(A)} \cdot \hat{\vec{s}}^{(B)} \right) .
\]
This operator commutes with all components $\hat{S}_{k}$, $k=1,2,3$, of the total spin.

Hence, under condition \eqref{Umova_zberegenna_Sz}, the operator $\hat{S}_{3}$ 
commutes with the full Hamiltonian \eqref{Stern_Gerlach_Hamiltonian}. Therefore, 
the eigenstate of $\hat{S}_{3}$ given by \eqref{Not_Aharonov_Bohm} evolves in time, 
according to the Hamiltonian \eqref{Stern_Gerlach_Hamiltonian}, into another state 
that is also an eigenstate of $\hat{S}_{3}$ with the same eigenvalue $S_{3}=0$ as the initial state.

Since the time evolution of the initial state \eqref{Not_Aharonov_Bohm} takes place 
entirely within the subspace defined by the condition
\[
s_{3}^{(A)} + s_{3}^{(B)} = 0,
\]
there are no longer two independent dynamical variables 
$\hat{s}_{3}^{(A)}$ and $\hat{s}_{3}^{(B)}$ within this subspace.

Indeed, in the Hamiltonian \eqref{Stern_Gerlach_Hamiltonian}, one may replace the operator 
$\hat{s}_{3}^{(B)}$ by the expression $\hat{S}_{3} - \hat{s}_{3}^{(A)}$. Since the time 
evolution is restricted to the subspace defined by $S_{3}=0$, the operator $\hat{S}_{3}$ 
acts as zero on all states in this subspace. The Hamiltonian \eqref{Stern_Gerlach_Hamiltonian} 
can therefore be projected onto this subspace, resulting in the effective Hamiltonian
\begin{equation}
	\label{Zvugenna_Hamiltoniana_na_Sz0}
	\begin{aligned}
		\hat{H} ={}& -\frac{\hbar^{2}}{2 m_{1}} \Delta_{A}
		-\frac{\hbar^{2}}{2 m_{2}} \Delta_{B} \\
		& + \mu_{A} \, \hat{s}_{3}^{(A)} \, H_{3}\!\left( \vec{r}_{A} \right)
		- \mu_{B} \, \hat{s}_{3}^{(A)} \, H_{3}\!\left( \vec{r}_{B} \right) \\
		& + \hat{H}_{A,B}^{\mathrm{int}} .
	\end{aligned}
\end{equation}

As can be seen from Eq.~\eqref{Zvugenna_Hamiltoniana_na_Sz0}, the time evolution of the probability amplitude of the state under consideration, with respect to both $\vec{r}_{A}$ and $\vec{r}_{B}$, is governed by the same dynamical variable $\hat{s}_{3}^{(A)}$. 
In other words, the spatial deflections of the detection probability distributions in the magnetic field are determined by the same operator $\hat{s}_{3}^{(A)}$.

This implies that both observers measure the same dynamical variable. More precisely, as follows from Eq.~\eqref{Zvugenna_Hamiltoniana_na_Sz0}, the outcomes registered by the observer detecting particle $A$ correspond to the eigenvalues of $\hat{s}_{3}^{(A)}$, whereas those registered by the observer detecting particle $B$ correspond to the eigenvalues of $-\hat{s}_{3}^{(A)}$.

Thus, as before, we arrive at a situation in which two observers determine the 
same dynamical variable in the same quantum state using two different and 
independent apparatuses.

Instead of $\hat{s}_{3}^{(A)}$, one may introduce another spin dynamical variable. 
For this purpose, the two operators $\hat{s}_{3}^{(A)}$ and $\hat{s}_{3}^{(B)}$ 
can be expressed in terms of a different pair of dynamical variables.

Since we consider the subspace of states invariant under the total spin component 
operator $\hat{S}_{3}$, it is natural to choose
\[
\hat{S}_{3} = \hat{s}_{3}^{(A)} + \hat{s}_{3}^{(B)}
\]
as one of these variables. As the other variable, one may take a linear combination
\begin{equation}\label{relative_spin}
\hat{s}_{3} = k_{A}\,\hat{s}_{3}^{(A)} + k_{B}\,\hat{s}_{3}^{(B)}.
\end{equation}

Here the coefficients $k_{A}$ and $k_{B}$ should be chosen such that the system
\begin{equation}\label{Sistema_rivnan_spin}
	\left\{
	\begin{aligned}
		& \hat{S}_{3} = \hat{s}_{3}^{(A)} + \hat{s}_{3}^{(B)}, \\
		& \hat{s}_{3} = k_{A}\,\hat{s}_{3}^{(A)} + k_{B}\,\hat{s}_{3}^{(B)}
	\end{aligned}
	\right.
\end{equation}
has a unique solution with respect to $\hat{s}_{3}^{(A)}$ and $\hat{s}_{3}^{(B)}$.

For example, one may introduce
\begin{equation}\label{relative_spin1}
\hat{s}_{3} = \hat{s}_{3}^{(B)} - \hat{s}_{3}^{(A)},
\end{equation}
in analogy with the relative Jacobi coordinate.

For any such choice of the variable $\hat{s}_{3}$, within the invariant subspace 
of eigenstates of the operator $\hat{S}_{3}$, the Hamiltonian of the system becomes 
a function of $\hat{s}_{3}$ only.

Note that arguments concerning the existence of paradoxical correlations between the 
results of distant measurements are usually based on the assumption that the 
measurement is performed in an eigenstate corresponding to zero eigenvalues of 
all three projections of the total spin of the system $A \otimes B$ 
\cite{bohm1951quantum,AharonovBohmPhysRev.108.1070}. 

In Ref.~\cite{bohm1951quantum,AharonovBohmPhysRev.108.1070}, it is also emphasized that, in such a state, 
different spin projections may be measured on different particles. 
Measurements of different spin projections for different particles were also 
analyzed in Ref.~\cite{PhysicsPhysiqueFizika.1.195}.

\begin{figure}[!htbp]
	\centering
	\includegraphics[width=1.0\linewidth]{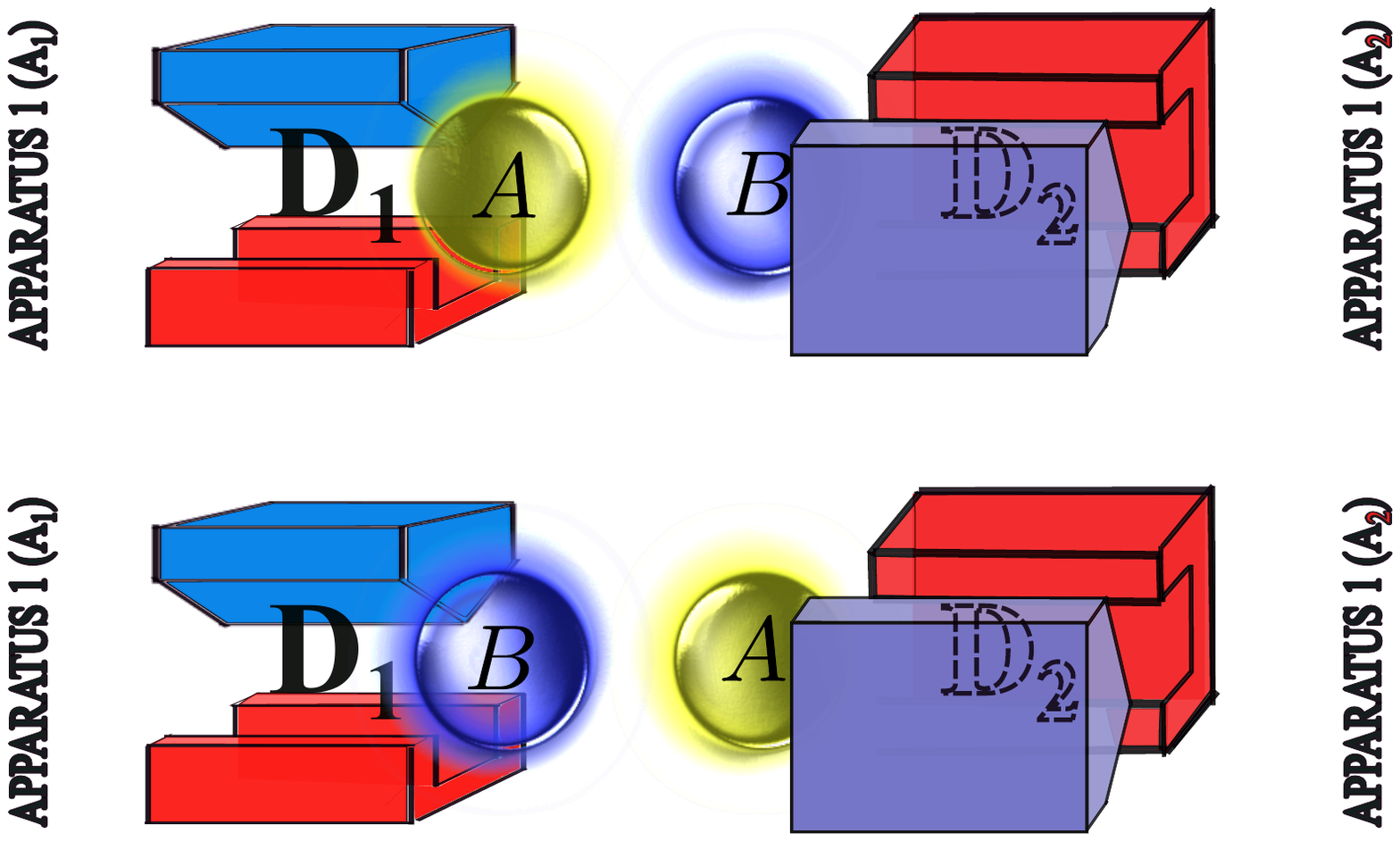}
	\caption{Example of an arrangement of two Stern--Gerlach apparatuses in which the 
		magnetic field directions in the regions $D_{1}$ and $D_{2}$ are not aligned. 
		Owing to the presence of an \enquote{exchange} term in the asymptotic state, both 
		particles $A$ and $B$ interact with magnetic fields that do not satisfy the 
		condition \eqref{Umova_zberegenna_Sz}. As a result, no projection of the total 
		spin of the system $A \otimes B$ is conserved in time under the evolution governed 
		by the Hamiltonian \eqref{Stern_Gerlach_Hamiltonian}.}
	\label{fig:tryrotate}
\end{figure}

However, from the considerations presented here, it follows that if one 
attempts to measure spin projections of different particles along different 
directions (at least within the Stern--Gerlach scheme, since real experiments 
\cite{AspectGrangierRogerPhysRevLett.49.91,PhysRevLett.28.938,KocherPhysRevLett.18.575} 
were performed with photons and did not involve a magnetic field), then such a 
measurement cannot be realized in a state that remains an eigenstate of at least one 
projection of the total spin. Indeed, this requires an experimental arrangement in 
which the magnetic field directions in the regions $D_{1}$ and $D_{2}$ are different 
(see Fig.~\ref{fig:tryrotate}).

As discussed above, in this case the operator of any total spin projection of the 
system $A \otimes B$ does not commute with the Hamiltonian 
\eqref{Stern_Gerlach_Hamiltonian}. This implies that, after the interaction with the 
magnetic field, the state of the system $A \otimes B$ is no longer an eigenstate of 
any of these projections.

Therefore, when particles $A$ and $B$ subsequently interact with detectors (for 
example, photographic plates) after passing through the magnetic field, the 
measurement does not take place in the same quantum state as the one prepared before 
the interaction with the magnet. In particular, it does not occur in the total 
spin-zero state. Hence, the arguments presented in Ref.~\cite{bohm1951quantum} are 
not applicable to this situation.

\subsubsection{Generalized Hamiltonian dynamics and the EPR Paradox}

Let us support the previous considerations by invoking an analogy with the
methods of generalized Hamiltonian dynamics
\cite{Dirac_1950,Dirac:1958sq,dirac2001lectures,Slavnov:904821,chaichian1984introduction}.

In particular, the conditions~\eqref{Linear_conditions} single out a subspace
\(\mathcal{H}_{\vec{P}}\) of the full state space \(\mathcal{H}\) of the composite quantum system
\(1 \otimes 2\). Since this subspace is invariant under the Hamiltonian of the system
\(1 \otimes 2\), the time evolution of any initial eigenstate belonging to
\(\mathcal{H}_{\vec{P}}\) remains confined to this subspace. This observation allows us to
restrict our analysis to \(\mathcal{H}_{\vec{P}}\) rather than to the full space
\(\mathcal{H}\).

Such a restriction is also a standard feature of generalized Hamiltonian dynamics
\cite{Dirac_1950,dirac2001lectures}. This type of dynamics arises when some of the
Euler--Lagrange equations do not serve as dynamical equations but instead impose
constraints on the dynamical variables. A typical example is provided by gauge
field theories, where the Lagrangian does not contain time derivatives of certain
field components \cite{Slavnov:904821,chaichian1984introduction}. As a result, the
variation with respect to these components produces equations that also lack time
derivatives, such as the Gauss law. Consequently, upon transition to the
Hamiltonian formulation, these equations do not determine the time derivatives of
the corresponding generalized coordinates or momenta. They are constraints on the
allowed dynamical variables.

The quantization of such systems
\cite{Dirac_1950,dirac2001lectures,Gupta:1949rh,Bleuler1950-od,Bleuler1950ANM,Slavnov:904821,chaichian1984introduction,Christ:1980ku}
proceeds by imposing the constraints on the states rather than on the operators
associated with the constrained dynamical variables. As a result, the constraint
equations single out a subspace of the full state space of a quantum system
governed by generalized Hamiltonian dynamics. Only the states belonging to this
subspace can be regarded as physically admissible states of the system.

Thus, the restriction to a subspace of the full state space is a common feature in
both situations: the case of entangled eigenstates of the total momentum or total
spin of a composite system, and the case of quantum systems governed by
generalized Hamiltonian dynamics.

However, the analogy is not limited to this similarity. Another common feature of
the two cases under consideration is the existence of a set of \enquote{proper} or
\enquote{physical} variables that form a subset of all quantities of the system.
Indeed, since the constraint equations must be satisfied, it is natural to
separate all dynamical quantities into two subsets. One subset contains mutually
independent variables whose values are determined by the Hamiltonian dynamics. The
members of this first subset are the \enquote{proper} or \enquote{physical}
variables. The second subset is formed by quantities whose values are determined
by the requirement that the constraint equations be satisfied for the given values
of the \enquote{physical} variables. For example, in the electromagnetic field,
the \enquote{physical} variables are the transverse polarization components of the
field configuration, while the members of the second subset are the
\enquote{longitudinal} and \enquote{scalar} polarizations
\cite{Gupta:1949rh,Slavnov:904821,Nelipa}.

In the case of gauge fields, there is an infinite number of degrees of freedom
and, correspondingly, an infinite number of constraints. This situation is
analogous to the infinite set of constraints~\eqref{Linear_conditions} in our
case. To satisfy this infinite set of constraints, one may employ the method
proposed for non-Abelian gauge fields in Ref.~\cite{Christ:1980ku}. Following
\cite{Christ:1980ku}, we decompose the state
\(\Psi_{\vec{P}}(t_{0},x_{1},x_{2})\) in~\eqref{Expansion_u_v}, whose expansion
coefficients must satisfy the infinite set of constraints~\eqref{Linear_conditions},
in terms of a complete set of basis functions, each of which individually
satisfies the constraints.
   
To construct a suitable system of basis functions, we employ the Jacobi
coordinates for each subsystem. In this representation, the variables associated
with subsystem~1 are
\( x_{1} = \{\vec{R}_{1},\, y^{(1)}\} \), where \(\vec{R}_{1}\) is the center-of-mass
position of subsystem~1 and \(y^{(1)}\) denotes the set of relative Jacobi
coordinates. Analogously, \( x_{2} = \{\vec{R}_{2},\, y^{(2)}\} \). In what follows,
the dependence of the probability amplitude on \(y^{(1)}\) and \(y^{(2)}\) plays no
essential role. Therefore, we omit these variables and denote the probability
amplitude simply as \(\Psi_{\vec{P}}(t_{0}, \vec{R}_{1}, \vec{R}_{2})\).
     
To satisfy the constraints~\eqref{Linear_conditions}, the basis functions must be
eigenfunctions of the total momentum operator \(\hat{\vec{P}}\) of the composite
system \(1 \otimes 2\). Moreover, the eigenvalue of the total momentum determines
how these eigenfunctions transform under spatial translations
\cite{Sakurai,LandauLifshitz}. In particular, an eigenfunction corresponding to
the eigenvalue \(\vec{P}=\vec{0}\) must be invariant under translations. Therefore,
any function of the form \(f(\vec{R}_{2} - \vec{R}_{1})\), which is invariant under
spatial translations, is an eigenfunction of the total momentum operator with
eigenvalue \(\vec{P}=\vec{0}\). Such functions automatically satisfy the infinite
set of constraints~\eqref{Linear_conditions} in the case \(\vec{P}=\vec{0}\).

Thus, for \(\vec{P}=\vec{0}\), we may choose any complete set of functions
\(f_{n}(\vec{R}_{2} - \vec{R}_{1})\), \(n = 1,2,\ldots\), as a basis for the
decomposition of the state \(\Psi_{\vec{P}=0}(t_{0}, \vec{R}_{1}, \vec{R}_{2})\).

For a generic eigenvalue \(\vec{P}\), which may be nonzero, the corresponding
decomposition takes the form
\begin{equation}\label{Alternative_decomposition}
	\Psi_{\vec{P}}\!\left(t_{0}, \vec{R}_{1}, \vec{R}_{2}\right)
	= \exp\!\left( \frac{i}{\hbar}\,\vec{P}\cdot\vec{R} \right)
	\sum_{n=0}^{\infty} a_{n}\, f_{n}(\vec{Y}),
\end{equation}
where the Jacobi coordinates \(\vec{R}\) and \(\vec{Y}\) are defined by
Eqs.~\eqref{R_i_y} in terms of \(\vec{R}_{1}\) and \(\vec{R}_{2}\).

Hence, for the same state 
\(\Psi_{\vec{P}}(t_{0}, \vec{R}_{1}, \vec{R}_{2})\), 
there exist two decompositions, \eqref{Expansion_u_v} and 
\eqref{Alternative_decomposition}.  
The essential difference between them is that in 
Eq.~\eqref{Expansion_u_v} the expansion coefficients \(c_{ij}\) are constrained by 
the conditions~\eqref{Linear_conditions}, whereas the coefficients \(a_{n}\) in 
Eq.~\eqref{Alternative_decomposition} are independent.

\begin{figure}[!htbp]
	\centering
	\includegraphics[width=1.0\linewidth]{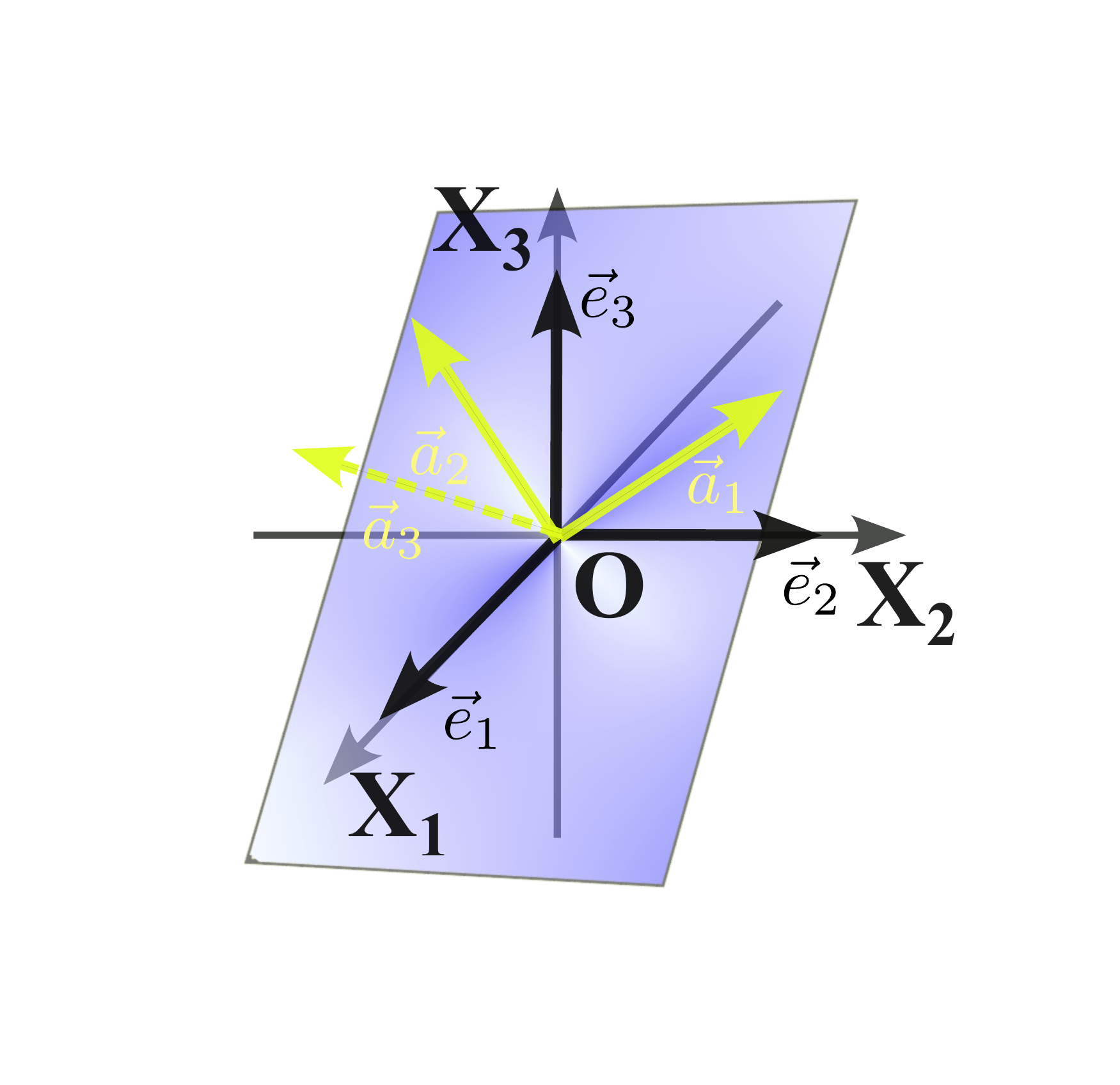}
	\caption{An analogy between defining a linear subspace of the state space
		via linear conditions such as \eqref{Linear_conditions} and defining a plane
		in three-dimensional Euclidean space.}
	\label{fig:pidprostir}
\end{figure}

Let us note that the analysis in Refs.~\cite{EPRPhysRev.47.777,schredinger_1935} 
is entirely based on the decomposition \eqref{Expansion_u_v}, to which the 
reduction postulate is applied. However, the presence of the constraints 
\eqref{Linear_conditions} implies that certain projections appearing in 
\eqref{Expansion_u_v} are canceled. 
This effect can be illustrated by a simple geometrical analogy shown in 
Fig.~\ref{fig:pidprostir}.

We can represent the position vector of any point lying on the plane shown in 
Fig.~\ref{fig:pidprostir} by expanding it in the basis 
\(\vec{e}_{1}, \vec{e}_{2}, \vec{e}_{3}\).  
The coefficients of such an expansion must satisfy constraints analogous to 
\eqref{Expansion_u_v} in order to ensure that the point indeed lies within the plane.  
The same vector, however, may also be expanded in the basis 
\(\vec{a}_{1}, \vec{a}_{2}\), which spans the plane itself.  
In this second representation the expansion coefficients are independent, 
just as in \eqref{Alternative_decomposition}.  

In the first case, the decomposition necessarily contains projections onto
directions orthogonal to the plane.  
Since the resulting vector must lie in the plane, these projections must cancel as
a consequence of the constraints.  
Because the reduction postulate~\cite{Von_Neumann_1996} relies on the presence
of such projections in a state decomposition, its application to
\eqref{Expansion_u_v} appears questionable.

To apply the reduction postulate to a decomposition of the form
\eqref{Alternative_decomposition}, one may compute the matrix representation of a
dynamical variable (i.e., an observable) in the basis used in that
decomposition.  
Diagonalizing this matrix yields eigenvectors that are specific linear
combinations of the basis functions in \eqref{Alternative_decomposition}.  
Consequently, these eigenvectors necessarily belong to the subspace selected by
the constraints~\eqref{Linear_conditions} and contain no nonzero projections
outside this subspace.

However, for a general matrix obtained in this way, it is not always possible to
realize the canonical commutation relations.
This feature is analogous to generalized Hamiltonian dynamics, where the
canonical commutation relations can be realized only for the independent
\enquote{physical} dynamical variables, but not for those whose values are fixed by
constraints rather than determined by Hamiltonian dynamics.

As an example, let us consider the quantities \( \vec{R}_{1} \) and
\( \vec{R}_{2} \).  
As the basis functions \( f_{n}(\vec{R}_{2} - \vec{R}_{1}) \), we choose the
eigenfunctions of the relative momentum \( \vec{P}_{\vec{Y}} \), as in
Eq.~\eqref{Rozclad_po_p_malomu}.  
To simplify the analysis, we replace the continuous spectrum with a discrete one
in the standard way, by confining the system to a large cubic box of edge
length \( L \) and volume \( V = L^{3} \) \cite{Flugge2012practical}.

Imposing periodic boundary conditions yields a discrete spectrum for the
momenta of the subsystems, \( \vec{P}_{1} \) and \( \vec{P}_{2} \).  
Since for any eigenstate of the total momentum one can choose an inertial
reference frame in which the corresponding eigenvalue is zero, we restrict our
consideration to this case.  
In the subspace \( \mathcal{H}_{\vec{P}=\vec{0}} \), the spectrum of the
relative momentum \( \vec{P}_{\vec{Y}} \) coincides with the spectra of
\( \vec{P}_{1} \) and \( \vec{P}_{2} \), and is therefore discrete as well.

Direct calculation of the matrix elements of the operators
\( \hat{\vec{R}}_{1} \) and \( \hat{\vec{R}}_{2} \) in the basis of relative-momentum
eigenfunctions leads to a trivial result:
\begin{equation}\label{Matrix_R_1_R2}
	\begin{aligned}
		& \bigl(\vec{R}_{j}\bigr)_{\vec{P}_{\vec{Y},1}\vec{P}_{\vec{Y},2}}
		= \int d\vec{R}_{1} \int d\vec{R}_{2}\;
		\vec{R}_{j}\,
		\\[2mm]
		& \quad \times
		\exp\!\left(
		-\frac{i}{\hbar}\,
		\vec{P}_{\vec{Y},2}\!\cdot\!\left(\vec{R}_{2}-\vec{R}_{1}\right)
		\right)
		\\[1mm]
		& \quad \times
		\exp\!\left(
		\frac{i}{\hbar}\,
		\vec{P}_{\vec{Y},1}\!\cdot\!\left(\vec{R}_{2}-\vec{R}_{1}\right)
		\right)
		= 0, \qquad j = 1,2 .
	\end{aligned}
\end{equation}

As a consequence, the canonical commutation relations cannot be realized for
the matrices representing \( \hat{\vec{R}}_{1} \) and \( \hat{\vec{R}}_{2} \) on the
subspace \( \mathcal{H}_{\vec{P}=\vec{0}} \).  
Moreover, the matrix of the total momentum operator \( \hat{\vec{P}} \)
on any subspace \( \mathcal{H}_{\vec{P}} \) is proportional to the identity
matrix; therefore, the canonical commutation relations cannot be realized for
this operator on such subspaces either. 

Thus, the canonical commutation relations on the subspace
\( \mathcal{H}_{\vec{P}=\vec{0}} \) can be realized only for the pair consisting
of the relative position vector of the subsystems' centers of mass,
\( \vec{Y} \), and the corresponding relative momentum \( \vec{P}_{\vec{Y}} \),
up to a canonical transformation.  
These quantities are the \enquote{physical} dynamical variables of the system
on the subspace \( \mathcal{H}_{\vec{P}=\vec{0}} \).

Consequently, the decomposition of any state belonging to
\( \mathcal{H}_{\vec{P}=\vec{0}} \) can be performed only in terms of a basis
unitarily equivalent to the one formed by the eigenfunctions of one of these
dynamical variables.  
In other words, the decomposition must be of the form
\eqref{Alternative_decomposition}.  
Given the reduction postulate, this restricts the set of observables whose
measurement may be regarded as a \enquote{proper} measurement.
  
Let us now examine the analogy between the dynamics of an entangled state and
generalized Hamiltonian dynamics from a different perspective.
To this end, we reconsider the expression~\eqref{Alternative_decomposition}.    
As follows from~\eqref{Alternative_decomposition}, the state
$\Psi_{\vec{P}}\!\left(t_{0}, \vec{R}_{1}, \vec{R}_{2}\right)$ is entangled with
respect to the variables $\vec{R}_{1}$ and $\vec{R}_{2}$, but nonentangled with
respect to $\vec{R}$ and $\vec{Y}$. These two dependences exhibit different
properties.  
The dependence on $\vec{R}$ can be modified by applying a transformation
belonging to a unitary representation of the Galilei group.
Such a transformation can remove this dependence entirely.
In contrast, the dependence on the relative position vector $\vec{Y}$ is
determined solely by the quantum Hamiltonian dynamics.

This situation has a close analogue in the description of non-Abelian gauge
fields~\cite{Christ:1980ku}. In that case, the local gauge group plays a role
similar to that of the Galilei group in the present context: it modifies the
form of the dependence of field configurations and of the states of a system of
interacting gauge and fermionic fields without changing the system’s physical
properties. As a consequence of this symmetry, the state is entangled with respect to
both the \enquote{physical} dynamical variables and the parameters of the local gauge
group~\cite{Christ:1980ku}.  As shown in~\cite{Christ:1980ku}, in the Coulomb gauge the state functional
assumes a form analogous to~\eqref{Alternative_decomposition}: it can be written
as a unitary operator, depending only on the local gauge group parameters,
acting on a functional of the \enquote{physical} dynamical variables.  
This unitary operator has been removed by an appropriate unitary transformation
on the state space, together with the unwanted dependence on the gauge-group
parameters.  
This is analogous to the removal of the $\vec{R}$-dependence by a transformation
from the unitary representation of the Galilei group in our case.

This consideration suggests that the total momentum $\vec{P}$ of states in the
subspace $\mathcal{H}_{\vec{P}}$ should be regarded as a parameter of the
Galilei group rather than a dynamical variable. Its value in a given inertial
reference frame is fixed prior to the measurement and is not selected through
the hybrid quantum--classical dynamics during the measurement process.
Consequently, the momentum conservation law must be imposed already at the
operator level, before it appears at the level of observables. Formally, this
means that the conservation law should be written as
\begin{equation}\label{Conservation_law_momentum}
	\hat{\vec{p}}_{1} + \hat{\vec{p}}_{2} = \vec{P}\,\hat{E},
\end{equation}
rather than as
\[
\hat{\vec{p}}_{1} + \hat{\vec{p}}_{2} = \hat{\vec{P}} .
\]
Here $\hat{E}$ denotes the identity operator.  
Equation~\eqref{Conservation_law_momentum} implies that the operators
$\hat{\vec{p}}_{1}$ and $\hat{\vec{p}}_{2}$ are not independent.

Thus, the analogy with generalized Hamiltonian dynamics supports our previous
considerations.

\subsubsection{An example of a "proper" measurement setup}

Some experimental illustration for the above considerations can be found in
elastic scattering experiments.  
In the center-of-mass reference frame, one deals with a total-momentum
eigenstate whose eigenvalue is zero.  
The momenta of the particles in the initial state are fixed by the
preparation procedure.  
In the final state, the momenta ${{\vec{P}}'_{1}}$ and ${{\vec{P}}'_{2}}$
have six components in total (Fig.~\ref{fig-collision}); however, these
components are not independent.  
They are constrained by four conservation laws: energy conservation and the
three components of total momentum conservation.

The remaining two independent variables are usually chosen to be the
spherical angles $\theta$ and $\varphi$, which specify the direction of the
line along which the vectors ${{\vec{P}}'_{1}}$ and ${{\vec{P}}'_{2}}$
lie (Fig.~\ref{fig-collision}).  
The angle $\theta$ is defined as shown in Fig.~\ref{fig-collision}.  
The coordinate $\varphi$ is the angle by which the $XOZ$ plane in
Fig.~\ref{fig-collision} must be rotated around the $OZ$ axis so that it
coincides with the plane containing all four momentum vectors 
${{\vec{P}}_{1}}$, ${{\vec{P}}_{2}}$, ${{\vec{P}}'_{1}}$, and ${{\vec{P}}'_{2}}$.

In most experimental situations, the system exhibits rotational symmetry
around the $OZ$ axis, so all values of $\varphi$ are equally probable.
Consequently, experiments typically measure only the angle~$\theta$.  
This can be seen, in particular, from the experimental results presented in
Refs.~\cite{Sakamoto:1996xdz,PhysRevLett.34.233,LEP-2}.

Thus, in the majority of scattering experiments the measured dynamical
variables are precisely the independent, \enquote{proper} variables.
A well-known example that deviates from this pattern is provided in
Ref.~\cite{ComptonPhysRev.26.289}.  
In that experiment, the directions of both the electron and photon
momenta in the final state were recorded.  

However, this type of measurement was specifically chosen to address
particular physical questions.  
One goal was to demonstrate that, in each run of the experiment, the
energy flow of the electromagnetic field has a definite direction and is
not spherically symmetric.  
Another goal was to show that the direction of the photon momentum is
governed by the total-momentum conservation law.  
Such experimental tests of conservation laws require the measurement of
quantities that are not independent and are therefore \enquote{nonproper} 
when regarded as quantum measurements. These measurements will be discussed later.

To illustrate the above theoretical considerations, let us examine the
experimental setup described in Ref.~\cite{Perl:1969pg}.  
This work presents a rare example of a scattering experiment in which both
the scattering angle and the modulus of the particle momentum are measured.
Nevertheless, given the structure of the setup, the measurement reported in
Ref.~\cite{Perl:1969pg} qualifies as a \enquote{proper} measurement.

The setup employs two detectors placed sequentially, one behind the other.
The first detector measured only the scattering angle.  
This was a standard \enquote{proper} measurement, during which the value of
the angle was \enquote{chosen} through the interaction with the detector,
and no other interaction influenced this \enquote{choice}.  
\emph{After} this value had been determined, the particle passed through the
second detector, positioned so that it could interact with the particle only
after it had passed the first one.  
This second detector measured the modulus of the particle momentum.

The second detection was not a measurement in the quantum-mechanical sense.  
It did not involve a \enquote{choice} among several possible outcomes.  
The \enquote{choice} had already been made by the first \enquote{proper}
measurement as a consequence of the constraint between the scattering angle
and the momentum modulus imposed by the conservation laws.  
Therefore, the value of the momentum modulus revealed in the second
detection had been fixed---together with the value of the angle---prior to
that detection, and not through the interaction with the second detector.

\subsection{Objection number~5}

Another paradox commonly discussed in EPR considerations arises in connection with 
experiments that involve two measurement apparatuses separated by a large distance 
\cite{article,Piccioni1989,1982FoPh...12.1171H,Ghirardi:1983ff,DIEKS1982271,Stapp1988,Selleribook:970920,Kupczynski:2016ysv,A_R_Wilson_1976,V_Paramananda_1987}.  
The paradox typically appears as an alleged exceeding of the relativistic speed limit \(c\).  

The reasoning that leads to this paradox usually proceeds by estimating the time interval \( \tau \) 
between the appearance of an observable outcome at one apparatus and the earliest observable outcome at the other \cite{A_R_Wilson_1976,V_Paramananda_1987}.
 
For sufficiently large spatial separation \(l\) between the apparatuses, it is possible to achieve the condition  
\(
l/c > \tau
\)
\cite{A_R_Wilson_1976,V_Paramananda_1987}.  
Under such circumstances, any exchange of physical quantities between the apparatuses during a single experimental run is impossible.  
Nevertheless, conservation laws require interdependence between the outcomes recorded by the two apparatuses.

For instance, let us consider the interaction between two apparatuses and the system \(1 \otimes 2\). We assume that the classical momentum of each apparatus in its initial state is zero, and that the system is in a total momentum eigenstate corresponding to the eigenvalue \(\vec{P}=\vec{0}\). Therefore, the expectation value of the total momentum of the two apparatuses together with subsystems~1 and~2 in the initial state is the zero vector.

As discussed in the introduction, we consider the situation in which each apparatus captures one of the subsystems during the measurement. Regardless of the distance between the apparatuses, if the classical momentum of one apparatus together with its captured subsystem is \(\vec{p}\), then the classical momentum of the other apparatus together with its captured subsystem must be \(-\vec{p}\). An analogous dependence of spin projections arises as a consequence of angular momentum conservation.

A similar situation was observed in real experiments \cite{A_R_Wilson_1976,V_Paramananda_1987}. In these works, the authors succeeded in achieving space-like separation between the two measurements. Under such conditions, any mutual influence between the measurements within a single experimental run is excluded. Nevertheless, correlations between these space-like separated measurements were observed. Moreover, the magnitude of these correlations did not diminish as the distance between the detectors was increased over a wide range.

Let us note that in most known experiments
\cite{ClauserHorneShimonyHoltPhysRevLett.23.880,AspectGrangierRogerPhysRevLett.49.91,A_R_Wilson_1976,V_Paramananda_1987,PhysRevX.13.021031},
the reported correlations were obtained as ensemble-averaged quantities rather than within a single run of a measurement on an individual system from the ensemble.

The descriptions in \cite{A_R_Wilson_1976,V_Paramananda_1987} emphasize this explicitly, which is crucial for assessing the alleged conflict with relativity:
\begin{quote}
	\enquote{Polarimeters based on this effect suffer from the major disadvantage
		that they do not give a clear yes/no answer for each pair of quanta, but only a statistical one. Thus the behaviour of individual photon pairs no longer has any significance.} \cite{A_R_Wilson_1976}.
\end{quote}

Keeping this in mind, we now once again consider the interaction of the composite system \(1 \otimes 2\) with both apparatuses, taking into account the existence of the \enquote{exchange} contributions discussed above.

\begin{figure}[!htbp]
	\centering
	\includegraphics[width=0.7\linewidth]{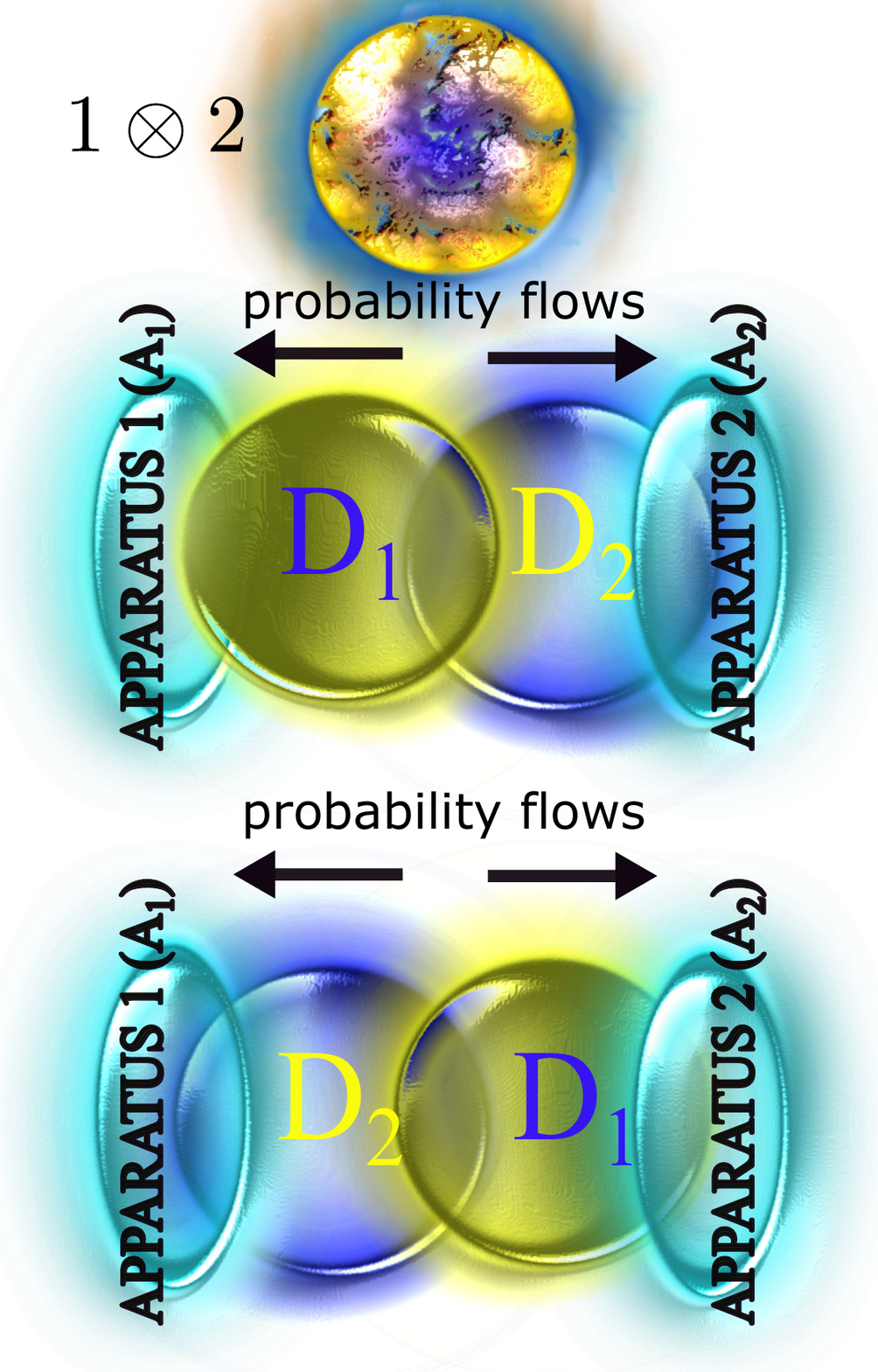}
	\caption{Symmetry of the arrangement of the regions in which particles of subsystems~1 and~2 are most likely to be found with respect to the measurement apparatuses.}
	\label{fig:twoapparatusmeasurement}
\end{figure}

As illustrated in Fig.~\ref{fig:twoapparatusmeasurement}, the measurement statistics recorded by each apparatus are determined by its interaction with both subsystems. If the magnitudes of the \enquote{exchange} contributions are equal, then the interactions between each apparatus and the system \(1 \otimes 2\) are identical for both apparatuses. Consequently, they statistically register the same outcomes.
However, this resulting full correlation does not require the transfer of any physical quantity from one apparatus to the other. 

If the magnitudes of the \enquote{exchange} contributions are unequal, the recorded results differ accordingly, yet they remain correlated. This correlation arises from the correlation between the measurement conditions experienced by the two apparatuses (see Fig.~\ref{fig:twoapparatusmeasurement}), rather than from any interaction between the apparatuses themselves. As follows from the above reasoning, the origin of these statistical correlations---as a consequence of the existence of \enquote{exchange} states---does not depend on the distance between the apparatuses. This may explain the experimental observations reported in Ref.~\cite{A_R_Wilson_1976,V_Paramananda_1987}, which found no dependence of the correlations on the distance between the detectors.

Therefore, the existence of a contradiction with relativity cannot be established without experimental data describing how the correlations are formed in time in each individual run of the measurement. However, such correlations were indeed observed, for example, in the experiment discussed above~\cite{ComptonPhysRev.26.289}, although their temporal formation was not analyzed. In addition, the very existence of such correlations must follow from conservation laws, as mentioned earlier.

For these reasons, we now consider the possible formation of dependencies between the outcomes recorded by distant apparatuses in a single run of the measurement. Note that the analysis of whether or not there exists any interchange of physical quantities between the apparatuses cannot rely solely on the observable properties of the experiment, as is done in Refs.~\cite{A_R_Wilson_1976,V_Paramananda_1987}. 

To clarify this point, let us return to the discussion of the measurement in which each apparatus (Fig.~\ref{fig:twoapparatusmeasurement}) captures one of the subsystems. As discussed in the Introduction, the presence of \enquote{exchange} contributions implies that the formation of the post-measurement state requires a sufficiently long time interval. This is due to the nonzero probability of finding particles belonging to the subsystem captured by one apparatus in the vicinity of the other apparatus. This probability is provided by the \enquote{exchange} terms in the asymptotic state of the system under measurement. The capture of either subsystem by a given apparatus requires this probability to decrease via its flow directed from one apparatus toward the other. Therefore, there exists a time interval required for this flow to \enquote{collect} the probability within the region associated with one of the two apparatuses. This interval determines the duration of the formation of the post-measurement state. As discussed in the Introduction, this duration is sufficient for flows of physical quantities to allow interchange between the apparatuses.

However, these processes remain unobserved, as does any process of time evolution of a state. Nevertheless, they may lead to observable outcomes that appear simultaneously to an observer. If the observer relies solely on these observable outcomes, they will interpret them as space-like separated, and hence independent and causally disconnected. Yet at the unobserved level, an interchange between the apparatuses may have occurred prior to the emergence of the observable results. This unobservable interchange can thus generate a dependence between the outcomes of both apparatuses in a single run of the measurement without violating the relativistic speed limit.

\section{A paradox of potentiality-determined observables}

In our previous considerations, an important role was played by the fact that
the dynamics of a quantum state is governed by all potential possibilities contained within that state. This follows directly from the dynamical equation~\eqref{Shredinger12}.
For instance, the \enquote{exchange} contributions in the asymptotic state provide the potential possibility to observe particles of both subsystems in the vicinity of the apparatus (Fig.~\ref{fig:exchangeterms}), as well as the possibility to observe particles of each subsystem near either of the two distant apparatuses (Fig.~\ref{fig:twoapparatusmeasurement}). Among these possibilities---to observe a particle near apparatus~1 (Fig.~\ref{fig:twoapparatusmeasurement}) or near apparatus~2---a real experiment may realize only one of them, or neither, if the particle is detected far from both apparatuses. However, irrespective of whether a particular possibility is realized or not, the mere \emph{existence} of that possibility influences the time evolution of the state through the dynamical equation~\eqref{Shredinger12}.

In other words, the time evolution of a quantum state is determined not by what actually occurs in the experiment but by what \emph{could} occur~\cite{Smolin:2011nn}:

\begin{quote}
	\enquote{Neither is it acceptable to imagine that there is a spooky way in which \enquote{potentialities affect realities}.}~\cite{Smolin:2011nn}
\end{quote}

However, in the present case, this influence of potentialities concerns a nonobservable object—the quantum state. A paradox arises when potential possibilities affect an \emph{observable} quantity. 
This situation occurs in the description of gauge fields, in particular for the electromagnetic field, which is unique among gauge fields in that its field strengths are directly observable.

Let us consider a single nonrelativistic, electrically charged quantum particle in a state described by the probability amplitude $\Psi(t,\vec{r},s_{z})$, where $t$ is time, $\vec{r}$ is the position vector, and $s_{z}$ is the spin projection along a chosen direction. We examine how the electromagnetic field created by this particle can be described.

We assume that this field is sufficiently strong to be treated as classical rather than quantum. For example, one may consider the field created by the electron in a hydrogen atom. As is well known, the quantum description of the properties of this atom does not require taking into account any quantum effects of the field inside it.

The contribution of the electron to this field at each point in space and at each moment in time is an observable quantity. Indeed, the total field created by the electron and the proton together is observable, and the proton can be treated as a pointlike, immovable classical particle that generates a Coulomb field. As classical dynamical variables, the electric and magnetic field strengths exist prior to measurement. The only characteristic of a quantum particle that exists prior to measurement is its state. Therefore, the field must be determined by the state. Moreover, the measurement of the field strength is not equivalent to a measurement of the particle's position vector. That is, as a result of measuring the field strength, no potential possibility for the particle's position is manifested.

For these reasons, we arrive at the paradoxical conclusion that the classical field in each member of an ensemble representing the state is determined by all potential possibilities contained in the state, rather than by just one of them. That is, the potential possibilities, which can manifest only across different members of the ensemble, determine the field in each single member.

Let us now support the above considerations by appealing to the most fundamental level of the description of the electromagnetic field, namely the gauge principle~\cite{YangMillsPhysRev.96.191}.

We consider Dirac’s quantized bispinor field $\hat{\Psi}_{s}(x)$ and its Dirac-conjugated field $\hat{\bar{\Psi}}_{s}(x)$, where $s=1,2,3,4$. Here, $x=(t,\vec{r})$ denotes a space–time point (we use a system of units in which the relativistic speed limit $c$ and Planck’s constant $\hbar$ are dimensionless and equal to $1$), and $\hat{\bar{\Psi}}_{s}$ is the Dirac conjugate of $\hat{\Psi}_{s}$. After introducing the electromagnetic field and imposing local $U(1)$ symmetry, one obtains the interaction Lagrangian of the bispinor field with the electromagnetic field.

From this form of the interaction Lagrangian, the Euler--Lagrange equations yield the constraint
\begin{equation}\label{Maxwell_operators}
	\operatorname{div}\!\bigl(\hat{\vec{E}}(x)\bigr)
	= e\,\hat{\bar{\Psi}}_{s_{1}}(x)\,\gamma^{0}_{s_{1}s_{2}}\,\hat{\Psi}_{s_{2}}(x),
\end{equation}
where $\hat{\vec{E}}(x)$ is the operator of the electric field strength,  
$\operatorname{div}$ denotes the spatial divergence, \(e\) is the electron charge,  
and $\gamma^{0}_{s_{1}s_{2}}$ are the matrix elements of the time-like Dirac matrix.

Let us now suppose that we can take the quantum nonrelativistic limit of the bispinor field. As an example of such a situation, we may continue considering the hydrogen atom. Most of its properties can be described within the nonrelativistic approximation for the electron state. In this limit, the bispinor field operator reduces to the probability amplitude, and the zeroth component of the four-current on the right-hand side of Eq.~\eqref{Maxwell_operators} becomes the probability density
\begin{equation}\label{Ro}
	\rho(x)
	= \sum_{s_{z}=-1/2}^{1/2}
	\Psi^{*}(x,s_{z})\,\Psi(x,s_{z}).
\end{equation}

Even when the fine structure of the hydrogen spectrum is taken into account and the relativistic Dirac equation~\cite{BjorkenDrell1965} is employed, the bispinor is still treated as a relativistic analogue of the probability amplitude rather than as a field operator. In this case, expression~\eqref{Ro} therefore remains valid.

As noted above, in addition to the nonrelativistic limit for the bispinor field, the electromagnetic field can often be treated in the classical (i.e., non-quantized) approximation. In this case, the field operator \(\hat{\vec{E}}(x)\) is replaced by the classical electric-field strength \(\vec{E}(x)\). Substituting this into Eq.~\eqref{Maxwell_operators} yields an equation formally identical to the usual Maxwell equation:
\begin{equation}\label{Maxwell}
	\operatorname{div}\!\bigl( \vec{E}(x) \bigr) - e\,\rho(x) = 0.
\end{equation}
However, unlike the standard Maxwell source term, the source term here is not a charge density but the probability density multiplied by the charge \(e\). While the charge density in Maxwell’s equation describes the actual distribution of charge in space, the probability density refers only to potential possibilities, not to events that have actually occurred. Consequently, the real \textbf{observable quantity} \(\vec{E}(x)\) is determined by the probabilities of the potential outcomes of a potential measurement.

Moreover, the field \(\vec{E}(x)\) exists in each individual system of the quantum ensemble, and, as can be seen from the previous considerations, it is determined by the state of the charged particle. We consider this as supporting the viewpoint~\cite{omnes2018interpretation} that, prior to measurement, the state corresponds to an individual system of the quantum ensemble, as discussed in the introduction.

Because the field \(\vec{E}(x)\) exists in each individual system of the quantum ensemble, the constraint~\eqref{Maxwell} implies that it depends on potential possibilities that may manifest only in different members of the ensemble.
In this way, the electric field in each system of an ensemble reflects properties of the ensemble as a whole.  
Let us note that a similar situation was discussed in Ref.~\cite{Smolin:2011nn}. In that work, a new interaction was proposed to account for these effects:

\begin{quote}
\enquote{These would be a new kind of interaction among spatially separated but identical systems. This may seem odd, but it brings with it an opportunity: perhaps the apparent influence of the wavefunction on the individual entities could be replaced and explained by interactions between the elements of the ensemble.}~\cite{Smolin:2011nn}
\end{quote}

However, as follows from the present considerations, the introduction of a new interaction is not required. It is sufficient to take into account the familiar gauge interactions.

\begin{figure}[!htbp]
	\centering
	\includegraphics[width=\linewidth]{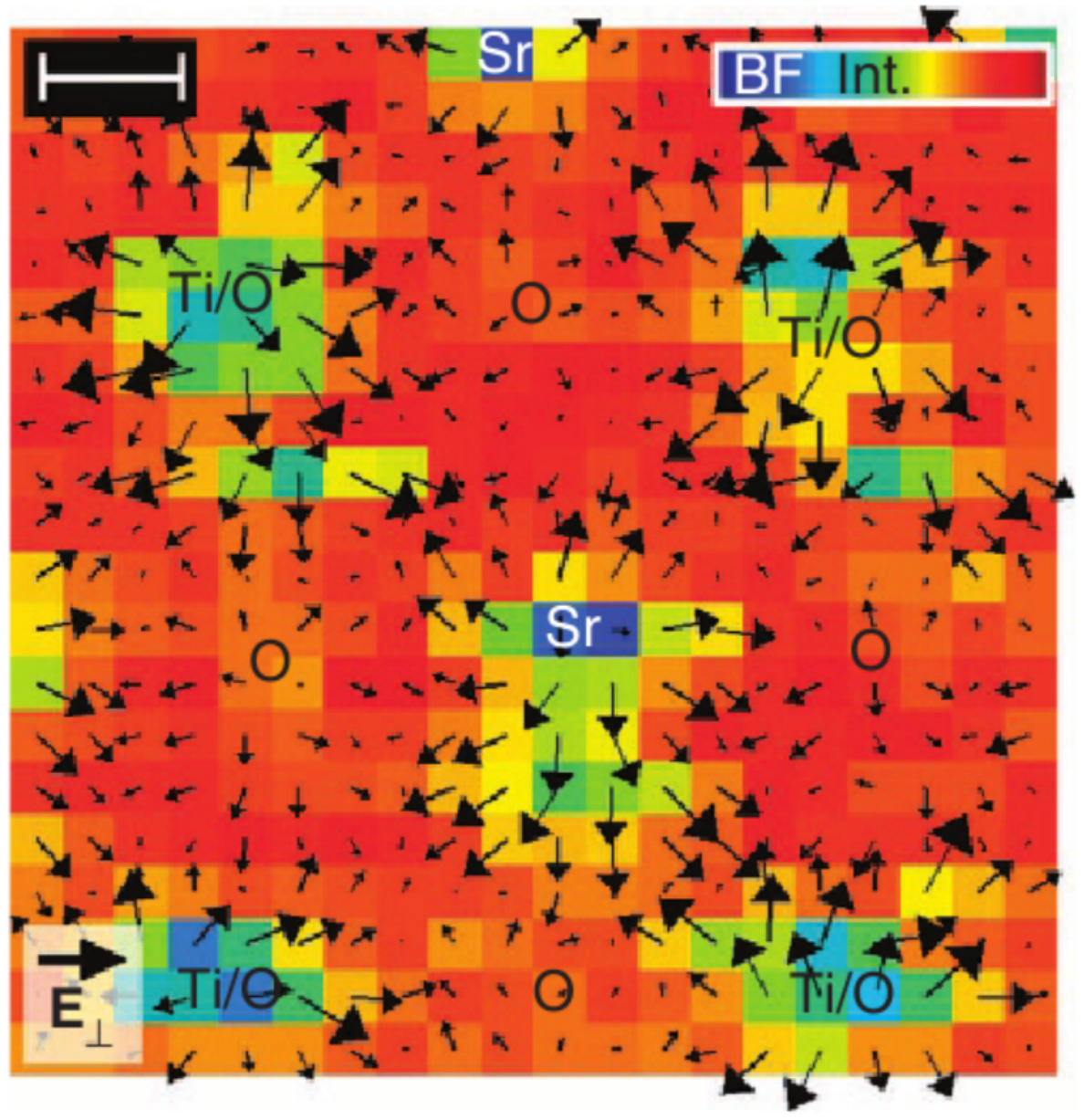}
	\caption{Intrinsic electric field measured in Ref.~\cite{Field_in_atom} (reproduced under CC BY license).  
		Singularities corresponding to the Coulomb field of classical, point-like nuclei are clearly visible.  
		In contrast, no analogous singularities are observed for the field generated by quantum electrons.}
	\label{fig:atomfield}
\end{figure}

We can find partial experimental support for the above considerations in measurements of the electric-field strength inside atoms~\cite{SCHMIDT1993101,MULLERCASPARY201762,Field_in_atom,PhysRevB.98.121408,Shibata2017ElectricFI}. Let us examine the result (Fig.~\ref{fig:atomfield}) reproduced from~\cite{Field_in_atom} and used here under the terms of the Creative Commons CC BY license. This figure shows the measured electric-field strength within a crystal lattice of \( \mathrm{SrTiO_3} \). The field is created both by atomic cores and by valence electrons. Since atomic cores contain massive nuclei, they can be treated as classical point-like particles. In contrast, valence electrons must be regarded as quantum particles.

Thus, the electric field shown in Fig.~\ref{fig:atomfield} is a superposition of fields generated by two distinct types of point-like charges: positive classical charges and negative quantum ones. As seen from the experimental data, Fig.~\ref{fig:atomfield} clearly exhibits Coulomb singularities associated with the positive point-like charges of the cores. However, no corresponding singularities appear for the negative quantum charges. Hence, the quantum electrons do not manifest themselves as point-like sources but instead behave as if their charge were distributed continuously, in accordance with the constraint~\eqref{Maxwell}.
     
To clarify this point, let us first consider the electric field created by a single charged particle (for example, an electron) governed by Eq.~\eqref{Maxwell} when it is in the state
$\Psi(t,\vec{r},s_{z})$, which is an energy eigenstate.
In this case, the probability density $\rho(\vec{r})$ in Eq.~\eqref{Ro} depends only on the spatial coordinates and is independent of time.
Accordingly, the electric field $\vec{E}(\vec{r})$ also depends only on $\vec{r}$.
From Eq.~\eqref{Maxwell} we obtain  

\begin{equation}\label{Naprugennist_Culomb}
	\vec{E}(\vec{r})
	= \frac{e}{4\pi}
	\int d\vec{r}\,'\,
	\frac{\rho(\vec{r}-\vec{r}\,')}{|\vec{r}\,'|^{2}}\,
	\vec{e}_{\vec{r}\,'},
\end{equation}
where $\vec{e}_{\vec{r}\,'}$ is the unit vector in the direction of $\vec{r}\,'$, and  
$\int d\vec{r}\,'$ denotes integration over the Cartesian components of $\vec{r}\,'$.

Next, let us transform this integral from Cartesian to spherical coordinates  
$r' = |\vec{r}\,'|$, $\theta$, and $\varphi$.  
The Jacobian of this transformation introduces a factor $r'^{2}$, which cancels the denominator in the integrand.  
As a result, the Coulomb singularity is removed.

Thus, measuring the electric-field strength inside a hydrogen atom should not reveal such singularities, in agreement with the results of Ref.~\cite{Field_in_atom}, as illustrated in Fig.~\ref{fig:atomfield}.  
However, the measurements and simulations reported in Refs.~\cite{SCHMIDT1993101,MULLERCASPARY201762,Field_in_atom,PhysRevB.98.121408,Shibata2017ElectricFI} concern multi-electron atoms.

To address the case of the electric field generated by a system of non-relativistic electrons, we adopt (with certain modifications) the approach proposed in Ref.~\cite{Christ:1980ku}.  
This approach also frees us from the restriction of treating the electric field classically: following Ref.~\cite{Christ:1980ku}, we will describe the field as a quantum object.  
In this setting, the constraint \eqref{Maxwell_operators} for operator-valued functions cannot be directly replaced by its analogue for number-valued functions, 
since the previous approximations are no longer applicable. Nevertheless, we shall show that the constraint \eqref{Maxwell_operators} for number-valued functions emerges, now as a consequence of more rigorous arguments than before.

To this end, we use the fact that, according to the quantization method in generalized Hamiltonian dynamics \cite{Dirac:1958sq,dirac2001lectures,Slavnov:904821,chaichian1984introduction,Christ:1980ku}, constraints like \eqref{Maxwell_operators} are imposed on elements of the space of states, rather than on operators.  
This allows us to choose a representation in which the basis states are eigenstates of the operators appearing in the constraint \eqref{Maxwell_operators}.  
Applying the constraint to the decomposition in this basis, for each basis state we obtain a numerical constraint analogous to \eqref{Maxwell_operators}.  
We can then apply the same considerations as in the single-particle case. They lead to a similar conclusion — for a system of multiple non-relativistic quantum particles, when the electric field is treated as a quantum object, the field realizations do not contain Coulomb singularities. This conclusion is in accordance with experimental results~\cite{Field_in_atom}.

\medskip

To implement this plan, let us start by introducing some notation.
Let $N_e$ be the number of non-relativistic electrons.  
The system under consideration consists of two subsystems: the electromagnetic field and the set of $N_e$ non-relativistic electrons.  
Their combined system, including the interaction between the subsystems, will be denoted by $E \otimes N_e$.

\medskip

To impose the constraint \eqref{Maxwell_operators} on the elements of the state space of the system $E \otimes N_e$, we must first describe this space.  
Following Ref.~\cite{Christ:1980ku}, it is represented by functionals.  
To introduce their arguments, we now specify the dynamical variables of each subsystem in a given reference frame.  
We use the Schr\"{o}dinger picture; therefore, the dynamical variables are time-independent, whereas the state of the full system $E \otimes N_e$ evolves in time.

\medskip

For the electromagnetic field, we choose the Hamiltonian gauge \cite{Slavnov:904821,Christ:1980ku}, in which the time-like component of the four-vector potential is set to zero.
\[
A_0(\vec{r}) = 0
\]
at all spatial points $\vec{r}$.

In this gauge, as is well known \cite{Slavnov:904821,Christ:1980ku}, the spatial components $A_k(\vec{r})$ for $k=1,2,3$ (collectively denoted $\vec{A}(\vec{r})$) serve as generalized coordinates.  
Their canonically conjugate momenta are the components of the electric field $E_k(\vec{r})$ (or $\vec{E}(\vec{r})$) at the same spatial points.

\medskip

In both cases, $A_k(\vec{r})$ and $E_k(\vec{r})$, the vector $\vec{r}$ is considered as a continuous index which, together with the discrete index $k$, uniquely labels each generalized coordinate and its conjugate momentum.

\medskip

For the system of $N_e$ non-relativistic electrons, we choose as dynamical variables their position vectors $\vec{r}_l$ ($l=1,2,\dots,N_e$) and their spin projections $s_{zl}$ along the $OZ$ axis of the chosen reference frame.  
Let the full set of these variables be denoted as
\[
\vec{r}, s_z \equiv 
\{\vec{r}_1, s_{z1},\, \vec{r}_2, s_{z2},\, \dots, \vec{r}_{N_e}, s_{zN_e}\}.
\]

\medskip

As mentioned earlier, following Ref.~\cite{Christ:1980ku}, we describe the state of the system $E \otimes N_e$ by a functional of the field configuration.  
However, unlike Ref.~\cite{Christ:1980ku}, where the functional depended on the generalized coordinates $\vec{A}(\vec{r})$, here we use the momentum representation and consider a functional of the generalized momenta $\vec{E}(\vec{r})$ instead.

\medskip

Moreover, this functional also depends on $\vec{r}$ and $s_z$.  
Thus, the state of the system $E \otimes N_e$ is described by
\[
\Psi\!\left[t,\,\vec{E}(\vec{r}),\,\vec{r},\,s_z\right].
\]

Its physical meaning is analogous to a probability amplitude.  
In particular, consider
\begin{equation}
	\label{Physical_sense_of_functional}
	\begin{aligned}
		dP = {} & \left|\Psi\!\left[t,\,\vec{E}(\vec{r}),\,\vec{r},\,s_z\right]\right|^2
		\\[2mm]
		& \times \prod_{\vec{r}}\prod_{k=1}^3 dE_k(\vec{r}) \;
		\prod_{l=1}^{N_e}\prod_{j=1}^3 d(\vec{r}_l)_j.
	\end{aligned}
\end{equation}

Here, $\prod_{\vec{r}}\prod_{k=1}^3 dE_k(\vec{r})$ is the measure of the functional (Feynman) integral \cite{feynman2010quantum,Slavnov:904821}, $(\vec{r}_l)_j$ denotes the $j$-th Cartesian component of $\vec{r}_l$, and $\prod_{l=1}^{N_e}\prod_{j=1}^3 d(\vec{r}_l)_j$ is the measure in the $3N_e$-dimensional Euclidean space of all possible electron positions.

\medskip

Suppose that at a given time $t$, we simultaneously measure the electric field $\vec{E}(\vec{r})$ at all points $\vec{r}$, the positions $\vec{r}_l$, and the spin projections $s_{zl}$ of all electrons.  

Then $dP$ is interpreted as the probability that all measured components of the electric field lie in
\[
[E_k(\vec{r}),\,E_k(\vec{r})+dE_k(\vec{r})],
\]
all position components lie in
\[
[(\vec{r}_l)_j,\,(\vec{r}_l)_j + d(\vec{r}_l)_j],
\]
and all spin projections simultaneously take the values $s_{zl}$.

\medskip

Considering that the electrons are non-relativistic, we may apply the non-relativistic approximation to the operators of the bispinor field and its Dirac-conjugate.  
In this approximation, they are replaced by the annihilation operator $\hat{\Psi}(\vec{r},s_z)$ and the creation operator $\hat{\Psi}^\dagger(\vec{r},s_z)$ in the occupation-number (second-quantized) representation \cite{LandauLifshitz}, where $\dagger$ denotes Hermitian conjugation.

\medskip

Equation~\eqref{Maxwell_operators} applies to time-dependent (Heisenberg-picture) operators.  
Operators in the Schr\"{o}dinger picture specify the initial values of the Heisenberg operators.  
Thus, the constraint~\eqref{Maxwell_operators} also applies to the Schr\"{o}dinger-picture operators $\hat{\vec{E}}(\vec{r})$ and
\begin{equation}
	\label{Particles_number_density}
	\hat{\rho}(\vec{r}) = \sum_{s_z=-1/2}^{1/2} \hat{\Psi}^\dagger(\vec{r},s_z) \, \hat{\Psi}(\vec{r},s_z).
\end{equation}

\medskip

From Eq.~\eqref{Particles_number_density}, $\hat{\rho}(\vec{r})$ represents the particle number density. In contrast, in the single-particle case, one obtains the probability density. 
Nevertheless, in both cases, $e\,\rho(\vec{r})$ plays the role of an effective \enquote{charge density}. This makes it appear that the charge is not concentrated at $N_e$ point particles but is continuously distributed according to this effective \enquote{charge density}, which remains free of singularities.  
The reason is that $\hat{\Psi}(\vec{r},s_z)$, $\hat{\Psi}^\dagger(\vec{r},s_z)$, and their derivatives appear in the Hamiltonian of the system $E \otimes N_e$.

As discussed above, within the quantization procedure of generalized Hamiltonian dynamics
\cite{Dirac:1958sq,dirac2001lectures,Slavnov:904821,chaichian1984introduction,Christ:1980ku},
the generalized momenta are treated as independent operator variables.
Instead, an additional condition is imposed on the states:
\begin{equation}\label{Umova_na_stani_zobragenna_Shredingera}
	\Bigl(\operatorname{div} \hat{\vec{E}}(\vec{q}) - e\,\hat{\rho}(\vec{q}) \Bigr)\,
	\Psi\bigl[t,\vec{E}(\vec{r}),\vec{r},s_{z}\bigr] = 0,
\end{equation}
where \(\vec{q}\) denotes the position vector of an arbitrary point in 3D Euclidean space.
The constraint \eqref{Umova_na_stani_zobragenna_Shredingera} selects a subspace of the full state space
on which the standard Hamiltonian dynamics is realized.
This condition must hold at every point $\vec{q}$ within the spatial domain
where the state of the combined system $E \otimes N_{e}$ is defined.

\medskip

Since the operator $\hat{\rho}(\vec{q})$ represents a physical observable at each spatial point,  
its eigenstates form a basis in the state space of $E \otimes N_{e}$.  
Therefore, an arbitrary state functional  
$\Psi\bigl[t,\vec{E}(\vec{r}),\vec{r},s_{z}\bigr]$  
can be expanded in this basis:
\begin{equation}\label{Rozclad_po_ro}
	\begin{aligned}
		&\Psi\bigl[t,\vec{E}(\vec{r}),\vec{r},s_{z}\bigr]
		= \int \prod_{\vec{q}} d\rho(\vec{q})\,\\
		&\times c\bigl(t,\rho(\vec{q})\bigr)\,
		\psi\bigl[\rho(\vec{q}),\vec{E}(\vec{r}),\vec{r},s_{z}\bigr],
	\end{aligned}
\end{equation}
where $\prod_{\vec{q}} d\rho(\vec{q})$ is the functional integration measure,  
$c\bigl(t,\rho(\vec{q})\bigr)$ is the expansion coefficient, and  
$\psi\bigl[\rho(\vec{q}),\vec{E}(\vec{r}),\vec{r},s_{z}\bigr]$ is the eigenfunctional of  
$\hat{\rho}(\vec{q})$ corresponding to the eigenvalue $\rho(\vec{q})$:
\begin{equation}\label{Eigenfunctional}
	\begin{aligned}
		&\hat{\rho}(\vec{q})\,
		\psi\bigl[\rho(\vec{q}),\vec{E}(\vec{r}),\vec{r},s_{z}\bigr]
		=\\
		&\rho(\vec{q})\,\psi\bigl[\rho(\vec{q}),\vec{E}(\vec{r}),\vec{r},s_{z}\bigr].
	\end{aligned}
\end{equation}

Let us substitute the expansion \eqref{Rozclad_po_ro} into the constraint equation \eqref{Umova_na_stani_zobragenna_Shredingera}.  
We also take into account that we have chosen the generalized momentum representation $\vec{E}(\vec{r})$ for the state of the system $E \otimes N_e$.  
In this representation, the operators of generalized momenta and generalized coordinates, which realize the canonical commutation relations, are defined as \cite{Christ:1980ku}:

\begin{equation}\label{Uzagalneni_impulsi_i_coordinati}
	\begin{aligned}
		& \hat{E}_k(\vec{q}) \,\Psi[t, \vec{E}(\vec{r}), \vec{r}, s_z] \\
		&	= E_k(\vec{q}) \,\Psi[t, \vec{E}(\vec{r}), \vec{r}, s_z], \\ 
		& \hat{A}_k(\vec{q}) \,\Psi[t, \vec{E}(\vec{r}), \vec{r}, s_z]	=  \\
		&	-i\hbar \frac{\delta}{\delta E_k(\vec{q})} \Psi[t, \vec{E}(\vec{r}), \vec{r}, s_z].
	\end{aligned}
\end{equation}

For each eigenstate of the \enquote{charge density} operator $e \hat{\rho}(\vec{q})$, we have
\begin{equation}\label{Dla_valsnogo_stanu_ro}
	\left( \operatorname{div}\vec{E}(\vec{q}) - e \rho(\vec{q}) \right) 
	\psi[\rho(\vec{q}), \vec{E}(\vec{r}), \vec{r}, s_z] = 0.
\end{equation}

Equation \eqref{Dla_valsnogo_stanu_ro} shows that only those field configurations that satisfy the numerical analog of the constraint \eqref{Maxwell} can be observed with nonzero probability in the state 
$\psi[\rho(\vec{q}), \vec{E}(\vec{r}), \vec{r}, s_z]$.

Since this analog  leads to \eqref{Naprugennist_Culomb}, the same reasoning as in the single-particle case can be applied.  
This implies that, when measuring the field strength in an eigenstate of the \enquote{charge density} $e \rho(\vec{q})$, no Coulomb singularities are observed.  
Thus, this result holds not only for a single particle but also for a system of an arbitrary number of non-relativistic charged particles.

For a non-eigenstate $\Psi[t, \vec{E}(\vec{r}), \vec{r}, s_z]$, we can apply the reduction postulate.  
According to this postulate, each outcome of a measurement of the electric field corresponds to one of the possible outcomes defined by \eqref{Dla_valsnogo_stanu_ro}.  
Therefore, in this case as well, no Coulomb singularities are observed, as illustrated in Fig.~\ref{fig:atomfield}.

Our conclusion that, in a system of charged point-like quantum particles, the field is determined by a continuously distributed \enquote{charge density} implies that, just as in the single-particle case, the field in a multi-particle system is governed by the potential possibilities of finding particles in the vicinities of different spatial points. These possibilities are indeed continuously distributed.  
This means that, for a quantum electric field interacting with a multi-electron nonrelativistic system, the outcomes of each run of a field-strength measurement in an individual member of the ensemble depend on potential possibilities that manifest across all members of the ensemble.  
Consequently, the paradox arises in this case as well.

Let us note that, when discussing the electric field, we refer only to the contribution generated by charged quantum particles.  
Our reasoning, based on the constraint equation \eqref{Maxwell}, is clearly unrelated to the contribution to the electric field strength that arises from the time variation of the magnetic field.

According to \eqref{Dla_valsnogo_stanu_ro}, a general electric-field configuration that can be realized with nonzero probability has the form
\begin{equation}\label{Plus_magmetic}
	\vec{E}(\vec{r}) 
	= \frac{e}{4\pi} \int d\vec{r}'\, 
	\frac{\rho(\vec{r}-\vec{r}\,')}{|\vec{r}\,'|^{2}}\,\vec{e}_{\vec{r}\,'}
	\;+\;
	\overrightarrow{\mathbf{rot}}\!\left(\vec{K}(\vec{r})\right).
\end{equation}
Here, $\vec{K}(\vec{r})$ is an arbitrary vector function for which the curl exists at every point $\vec{r}$.  
The second term is not a potential field and therefore describes the possibility of observing an electric field generated by time-varying magnetic fields.

The contribution $\overrightarrow{\mathbf{rot}}\bigl(\vec{K}(\vec{r})\bigr)$, unlike the first term in \eqref{Plus_magmetic}, is independent of the potential possibilities of observing charged quantum particles.  
However, like the first term, it is independent of any particular realization of these potential possibilities.  
Moreover, the first term in \eqref{Plus_magmetic} is deterministic because the \enquote{charge density} $e\rho(\vec{r})$ is deterministic, whereas the second term is random.  
Consequently, different realizations of the electric field differ in the value of the second term.

In situations where photon radiation is negligible and the field may be treated as electrostatic (for example, inside an atom), only the first contribution in \eqref{Plus_magmetic} remains.  
Let us now focus on this particular case.

\begin{figure}[!htbp]
	\centering
	\includegraphics[width=1.0\linewidth]{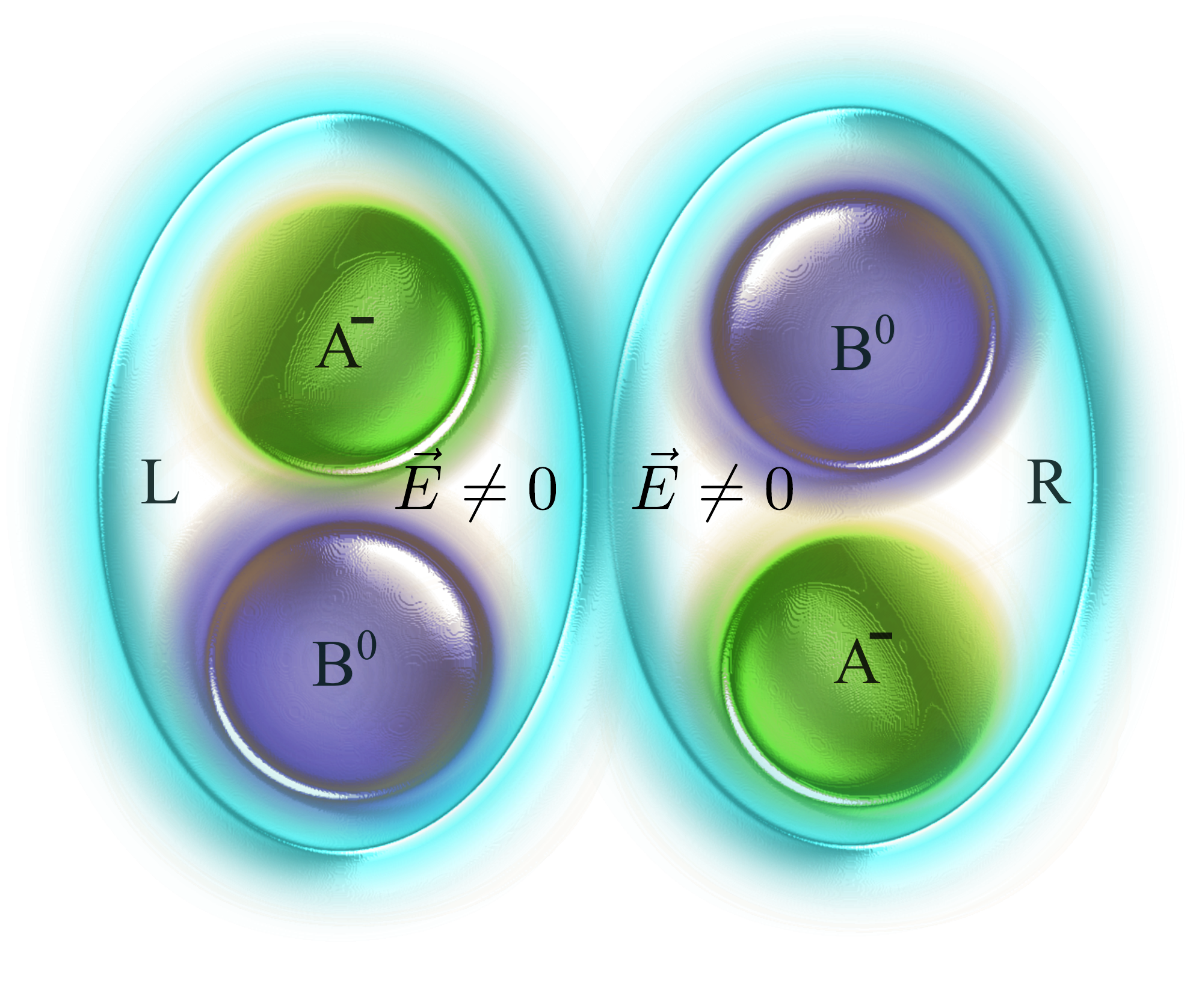}
	\caption{\enquote{Exchange} configurations in the scattering of a charged particle with a neutral one.}
	\label{fig:aminusbnol}
\end{figure}

Due to the fact that the electromagnetic field of a quantum system is determined by potentialities, we encounter another paradoxical situation. Consider the scattering of an electrically charged particle $A^{-}$ with a neutral particle $B^{0}$. In the asymptotic state after scattering, the expansion of the wave function contains \enquote{exchange} terms schematically illustrated in Fig.~\ref{fig:aminusbnol}.

Due to the presence of these \enquote{exchange} terms, the potential possibilities of observing the charged particle arise in both regions $L$ and $R$ shown in Fig.~\ref{fig:aminusbnol}, even though these regions are spatially well separated. The charged particle will ultimately be detected in only one of them. However, since the electric field is determined by potential possibilities, one may expect the field to be nonzero in both regions. Although the spatial region in which the charged particle can potentially be observed is large, the corresponding \enquote{charge density} associated with this potentiality becomes exceedingly small. Consequently, the resulting field in each region is extremely weak. As a result, an experimental verification (or refutation) of this effect may be difficult. Nevertheless, such an experiment could help clarify the situation regarding the paradox under consideration.

 \section{Conclusion}

The main conclusion of this work is that the role of potential possibilities in quantum dynamics deserves significantly greater attention. This is essential for developing an approach that consistently unites the principles of quantum theory with those of special relativity. Quantum field theory is not a complete answer to this problem, taking into account its well-known and long-standing difficulties \cite{Haag:1955ev}.

One possible origin of these difficulties is that relativistic principles are formulated for actual events rather than for potential possibilities. For instance, a typical situation in quantum mechanics is that an arbitrary state contains potential possibilities of observing the same particle at the same moment of time but at different spatial points. Even for arbitrarily large separations between these points, the existence of such possibilities does not contradict the relativistic speed limit, since no transport of any sort takes place between them.
This example illustrates that relativistic principles must be applied differently to potential possibilities than to actual events. This raises the question of how relativistic constraints should be implemented at the level of potential possibilities.As seen from the considerations presented in this work, these potential possibilities play a significant role in quantum dynamics. Given this role, addressing this question may clarify the form that relativistic quantum dynamics could take.

In this work, we did not consider the well-known thought experiments involving boxes \cite{debroglie:jpa-00236174,Hardy,Norsen2005EinsteinsB}, because the presence of external interventions in those scenarios substantially changes the situation compared to the one analyzed in \cite{EPRPhysRev.47.777,schredinger_1935,schredinger_1936,BohrNPhysRev.48.696,FurryPhysRev.49.393,AharonovBohmPhysRev.108.1070,PhysicsPhysiqueFizika.1.195,CANTRELL1978499}. 
Moreover, the boundary conditions reflecting the presence of boxes and partitions significantly alter the analytic properties of the probability amplitudes. 
This topic is discussed in more detail in Appendix~3.

\section*{Appendix~1: Hamiltonian symmetry and exchange states}

\subsection*{Symmetry under the transformation $\hat I_J$}

Our goal is to show that the Hamiltonian ${{\hat{H}}^{\left(1\otimes 2\right)}}$ of any isolated nonrelativistic multiparticle system \(1 \otimes 2\) remains invariant under the transformation \eqref{Peretvorenna_IJ}. We represent the Hamiltonian using Jacobi coordinates \cite{Cornilledoi:10.1142/5272,faddeev1993quantum}. The complete set of these coordinates, denoted by \(J\) \eqref{J}, consists of the center-of-mass position vector \(\vec{R}\) together with the set of relative position vectors \(y\) \eqref{y}.

To justify the symmetry of the Hamiltonian with respect to the transformation \eqref{Peretvorenna_IJ}, we rely on two properties of Jacobi coordinates:

\begin{enumerate}
	\item Just as in Cartesian coordinates, the kinetic-energy operator ${{\hat{H}}_{0}}$ remains a linear combination of Laplacians with respect to the components of each vector in the set \(J\). In particular, it is a linear combination of second derivatives with respect to the individual Jacobi coordinates.
	
	\item Under a translation of the entire system, only the center-of-mass coordinate \(\vec{R}\) shifts by the translation vector, while all relative coordinates in the set \(y\) remain unchanged.
	
\end{enumerate}

The first property implies that the kinetic-energy operator of the system \(1 \otimes 2\) is invariant under the transformation \eqref{Peretvorenna_IJ}. Indeed, each relative coordinate changes sign under this transformation, but the corresponding second derivatives remain unchanged, while the Laplacian with respect to the center-of-mass coordinate is unaffected, since the center-of-mass coordinate itself does not change.

The interaction operator \(\hat{H}^{\text{int}}_{1 \otimes 2}\) describes all interactions between particles within each subsystem, as well as all interactions between particles belonging to different subsystems. To establish its invariance under the transformation \eqref{Peretvorenna_IJ}, we can consider other symmetry properties of this operator. These properties follow from the symmetries of the total Hamiltonian \(\hat{H}^{(1\otimes 2)}\), and also from those of its kinetic part \(\hat{H}_{0}\).

Since the system \(1 \otimes 2\) is isolated, the symmetries of the total Hamiltonian 
\({\hat{H}}^{(1 \otimes 2)}\) follow from the symmetries of space and time. This is a consequence of 
the absence of external interactions. Given that the kinetic-energy operator 
\({\hat H}_{0}\) does not account for interactions, it is invariant under the same spatial symmetries 
as the Hamiltonian of the entire isolated system. Consequently, the interaction operator 
contained in \({\hat{H}}^{(1 \otimes 2)}\) must be invariant under these symmetries as well.

In particular, the uniformity of space implies symmetry under parallel translations, while the isotropy of space implies symmetry under spatial rotations. Both symmetries restrict the set of quantities that may appear as arguments of the interaction operator. The translational symmetry forbids any dependence of 
\(\hat{H}^{\mathrm{int}}_{1\otimes 2}\) on the center-of-mass position vector \(\vec{R}\), 
which changes under translations, whereas the interaction operator must remain invariant. As a consequence of rotational symmetry, \(\hat{H}^{\mathrm{int}}_{1\otimes 2}\) may depend only on mutual scalar products of the vectors from the set \(y\), as well as on their contractions with other vectors and tensors with respect to the rotation group. For instance, \(\hat{H}^{\mathrm{int}}_{1\otimes 2}\) may include multipole moments or polarization tensors.

To analyze dependencies on these contractions, we need to take into account another spatial symmetry of 
\(\hat{H}^{(1\otimes 2)}\), \(\hat{H}_0\), and, consequently, \(\hat{H}^{\mathrm{int}}_{1 \otimes 2}\) — 
the symmetry with respect to spatial inversion. Under this symmetry, vectors and tensors can be either 
true vectors (tensors), such as the dipole moment, or pseudovectors (pseudotensors), such as spin. 
We consider the case in which we can neglect the weak interaction due to the smallness and rarity of its effects.     
In this case, the interaction operator must be invariant under spatial inversion. Therefore, only contractions that 
remain unchanged under this transformation are allowed as arguments of \(\hat{H}^{\mathrm{int}}_{1 \otimes 2}\).

These contractions cannot depend 
on the center-of-mass position vector \(\vec{R}\), as a consequence of the translational symmetry 
of the interaction operator; therefore, their transformation properties are independent of 
how \(\vec{R}\) transforms. Since all Jacobi coordinates except \(\vec{R}\) transform under 
\eqref{Peretvorenna_IJ} in the same way as under inversion, the allowed 
contractions constructed from these coordinates remain unchanged under 
\eqref{Peretvorenna_IJ}.

In addition, mutual scalar products of the relative position vectors from the set \(y\) are 
quadratic in these vectors and therefore remain unchanged when each vector change sign. 
Hence, all possible arguments of the interaction operator map into themselves as a result of the 
transformation \eqref{Peretvorenna_IJ}. Consequently, this operator is symmetric with respect to 
this transformation.

Since both operators \({\hat H}_{0}\) and \(\hat{H}^{\mathrm{int}}_{1 \otimes 2}\) remain unchanged 
under the transformation \eqref{Peretvorenna_IJ}, we conclude that this transformation is a 
symmetry of the full Hamiltonian \({\hat{H}}^{(1 \otimes 2)}\).

\subsection*{Comparison with Spatial Inversion}

Since the symmetry transformation $\hat{I}_J$ in \eqref{Peretvorenna_IJ} can be used to analyze the relative significance of interactions between the subsystems of the system $1 \otimes 2$ and the measuring apparatus, we examine here several of its properties. It is convenient to do so by comparing them with the properties of spatial inversion.

The essential difference between the transformations $\hat{I}_J$ and the spatial inversion $\hat{I}$ is that inversion is defined for each point of three-dimensional Euclidean space, whereas $\hat{I}_J$ is defined for each configuration of the multiparticle system $1 \otimes 2$. Such a configuration is the set of position vectors of the particles that may potentially be observed.

As in \eqref{Peretvorenna_IJ}, we denote by $N_{1}$ and $N_{2}$ the numbers of particles in subsystems~1 and~2, respectively. A possible configuration therefore contains $N_{1}+N_{2}$ position vectors,
\begin{equation}\label{Configuration}
	\{\, \vec{r}_{i} \mid i = 1,2,\dots, N_{1}+N_{2} \,\}.
\end{equation}

Thus, while $\hat{I}$ is a transformation defined on three-dimensional space, 
$\hat{I}_J$ is a transformation defined on a $3(N_{1}+N_{2})$-dimensional 
configuration space.

Despite this significant difference, a certain connection between these transformations can nevertheless be established. Indeed, one may state that under the transformation $\hat{I}_J$, all points of any configuration are mapped onto the same points as under spatial inversion with respect to the center of mass of that configuration.

To support this statement, let us note that each position vector $\vec{r}_{i}$ of any configuration \eqref{Configuration} can be written as a linear combination of the Jacobi coordinates introduced in \eqref{J}:
\begin{equation}\label{Radius_vectori_cherez_Riy}
	\vec{r}_{i}
	= \mu_{i,\vec{R}}\,\vec{R}
	+ \sum_{j=1}^{N_{1}+N_{2}-2} \mu_{ij}\,\vec{y}_{j}
	+ \mu_{i,\vec{Y}}\,\vec{Y}.
\end{equation}
Here $\mu_{i,\vec{R}}, \mu_{ij}$, and $\mu_{i,\vec{Y}}$ are certain coefficients. For the present purpose, the only relevant coefficients are $\mu_{i,\vec{R}}$, while the remaining coefficients play no role in the argument below.

To determine the values of the coefficients $\mu_{i,\vec{R}}$, we use the
transformation properties of all vectors in \eqref{Radius_vectori_cherez_Riy}
under spatial translations. The vectors $\vec{r}_{i}$ and $\vec{R}$ transform
in the same way: under any translation by a vector $\vec{a}$ they change as
$\vec{r}_{i} \to \vec{r}_{i} + \vec{a}$ and $\vec{R} \to \vec{R} + \vec{a}$.
By contrast, the relative position vectors $y$ in \eqref{y} remain invariant
under translations. Consequently, the transformation laws of $\vec{r}_{i}$ and
$\vec{R}$ can be identical only if $\mu_{i,\vec{R}} = 1$ for all $i$.

As a consequence,
\begin{equation}\label{CM_inversion}
	\vec{r}_{i} - \vec{R} =
	\sum_{j=1}^{N_{1}+N_{2}-2} \mu_{ij}\,\vec{y}_{j}
	+ \mu_{i,\vec{Y}}\,\vec{Y}.
\end{equation}
Using the definition of the transformation \eqref{Peretvorenna_IJ}, we see that 
under $\hat{I}_J$ each difference $\vec{r}_{i} - \vec{R}$ for a given configuration
\eqref{Configuration} changes sign, while the vector $\vec{R}$ remains unchanged. 
Exactly the same transformation behavior would occur under the spatial inversion 
$\hat{I}$ taken with respect to the point $\vec{R}$.
 
Therefore, while spatial inversion has a single inversion center common to all configurations, the transformation \(\hat{I}_J\) assigns to each configuration its own inversion center determined by its center of mass.  

\subsection*{Spatial structure of exchange states}

To determine whether both subsystems $1$ and $2$ can interact with the apparatus, we must describe the spatial arrangement of the regions $D_{1}(t)$ and $D_{2}(t)$ relative to the region $D_A$ (Fig.~\ref{fig:exchangeterms}), where particles of subsystems~1 and~2 may interact with particles of the apparatus~$A$.

For this purpose, it is sufficient to consider the expectation values of the center-of-mass position vectors of subsystems~1 and~2 in a pair of \enquote{exchange} states, $\lvert b \rangle$ and $\hat{I}_J \lvert b \rangle$. Our goal is to express the expectation values in one of these states in terms of those in the other.

As before, for a given configuration~\eqref{Configuration}, we denote by $\vec{R}_1$ and $\vec{R}_2$ the center-of-mass position vectors of subsystems~1 and~2, respectively. Their expectation values in the two \enquote{exchange} states are
\begin{equation}\label{Average_R1R2}
	\begin{aligned}
		\langle \vec{R}_{k} \rangle_{b} &= \langle b \lvert \vec{R}_{k} \rvert b \rangle ,\\
		\langle \vec{R}_{k} \rangle_{\hat{I}_J b} &= \langle \hat{I}_J b \lvert \vec{R}_{k} \rvert \hat{I}_J b \rangle,
		\qquad k = 1, 2.
	\end{aligned}
\end{equation}

Regarding these expectation values, we must take into account that 
$\vec{R}_1$ and $\vec{R}_2$ are not members of the set $J$~\eqref{J} 
of chosen Jacobi coordinates. Therefore, they are not among the 
integration variables in the integrals~\eqref{Average_R1R2}.
However, for any configuration, the vectors $\vec{R}_1$ and $\vec{R}_2$ 
can be expressed in terms of the elements of the set $J$. 
In particular, as follows from~\eqref{R_i_y}, $\vec{R}_1$ and $\vec{R}_2$ 
are functions of $\vec{R}$ and $\vec{Y}$.

Given this, we may consider the images $\hat{I}_J \vec{R}_k$, $k = 1, 2$, 
of the vectors $\vec{R}_1$ and $\vec{R}_2$ under the transformation $\hat{I}_J$. 
The expectation values \eqref{Average_R1R2} can be expressed in terms of 
the expectation values of $\hat{I}_J \vec{R}_k$. To do so, we treat the 
transformation \eqref{Peretvorenna_IJ} as a change of variables in the 
integrals \eqref{Average_R1R2}. Performing the change of variables 
$J \to \hat{I}_J J$ in these integrals, we obtain
\begin{equation}\label{IJB}
	\big\langle \vec{R}_{k} \big\rangle_{\hat{I}_J b}
	= \langle b \lvert \hat{I}_J \vec{R}_{k} \rvert b \rangle .
\end{equation}

Using Eqs.~\eqref{R_i_y}, \eqref{Peretvorenna_IJ}, and \eqref{IJB}, we find
\begin{equation}\label{exhange_charachter}
	\begin{aligned}
		& {{\left\langle {{{\vec{R}}}_{1}} \right\rangle }_{{{{\hat{I}}}_{J}}b}}={{\left\langle {{{\vec{R}}}_{2}} \right\rangle }_{b}}+{{\left\langle {\vec{q}} \right\rangle }_{b}}, \\ 
		& {{\left\langle {{{\vec{R}}}_{2}} \right\rangle }_{{{{\hat{I}}}_{J}}b}}={{\left\langle {{{\vec{R}}}_{1}} \right\rangle }_{b}}+{{\left\langle {\vec{q}} \right\rangle }_{b}}, \\ 
		& {{\left\langle {\vec{q}} \right\rangle }_{b}}=\frac{{{M}_{2}}-{{M}_{1}}}{{{M}_{1}}+{{M}_{2}}}\left( {{\left\langle {{{\vec{R}}}_{2}} \right\rangle }_{b}}-{{\left\langle {{{\vec{R}}}_{1}} \right\rangle }_{b}} \right). \\ 
	\end{aligned}
\end{equation}

Equation~\eqref{exhange_charachter} demonstrates the exchange character 
of the transformation \eqref{Peretvorenna_IJ}. As seen from 
Eq.~\eqref{exhange_charachter}, the exchange is always accompanied 
by an additional shift directed toward the initial location of the 
center-of-mass expectation value of the subsystem with the greater mass.

\begin{figure}[!htbp]
	\centering
	\includegraphics[width=0.7\linewidth]{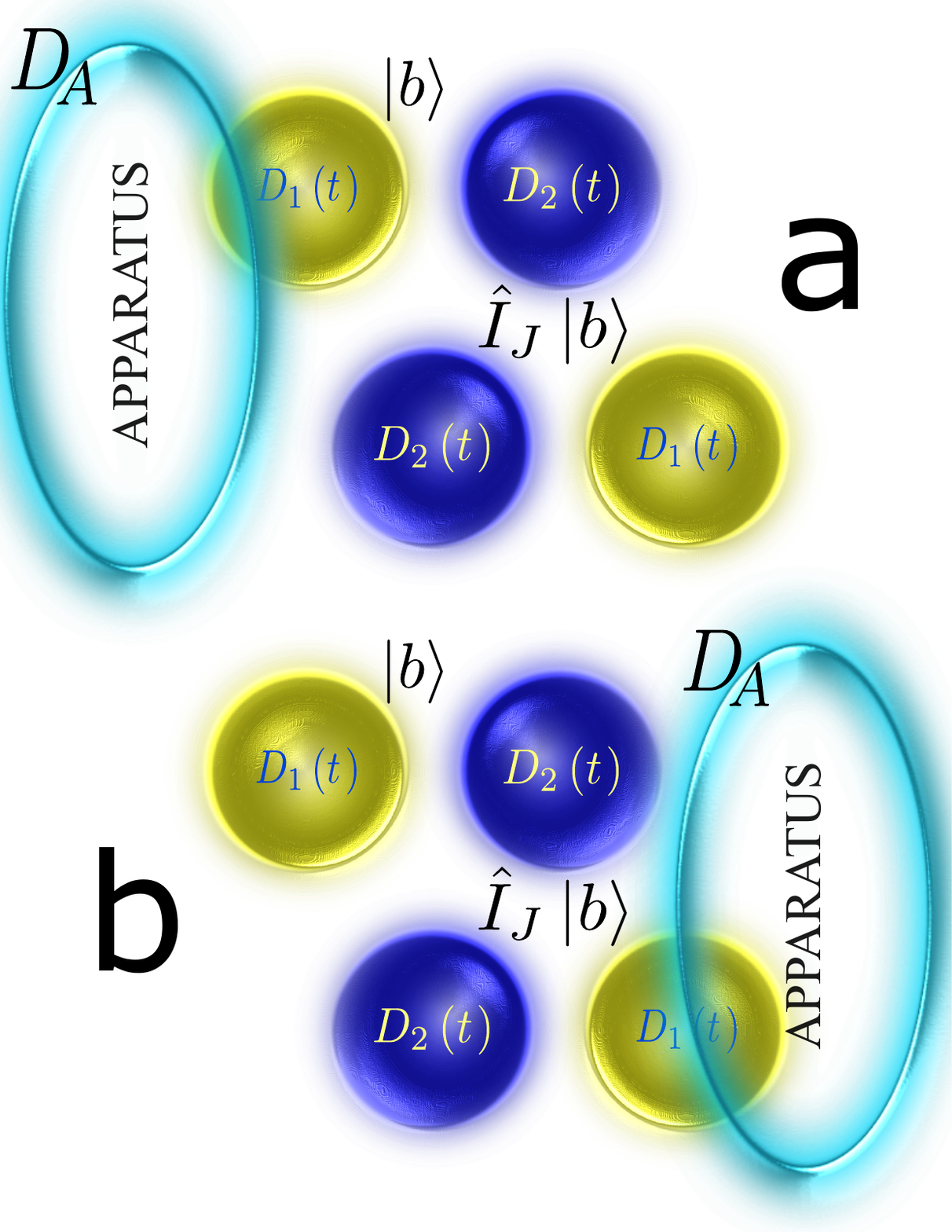}
	\caption{Configuration of the regions $D_{1}(t)\cup D_{2}(t)$ 
		for the states $\lvert b\rangle$ and $\hat{I}_{J}\lvert b\rangle$, 
		together with two possible placements of the apparatus.}
	\label{fig:arrangementapparatusexchange}
\end{figure}

For example, if $M_{2} > M_{1}$, we obtain the configuration of regions 
in the asymptotic state schematically shown in 
Fig.~\ref{fig:arrangementapparatusexchange}. As seen from this figure, 
regardless of where the apparatus is placed, the subsystem with the 
smaller mass can be observed in the region $D_{A}$ earlier than the 
other subsystem.

In the asymptotic state, 
$\bigl\lvert \langle \vec{R}_{2} \rangle_{b} - \langle \vec{R}_{1} \rangle_{b} \bigr\rvert$ 
is large. If $M_{2} \neq M_{1}$, then the magnitude of the shift vector 
$\bigl\lvert \langle \vec{q} \rangle_{b} \bigr\rvert$ in 
Eq.~\eqref{exhange_charachter} is also large. This raises the question 
of whether, for example, the region $D_{2}(t)$ in the state 
$\hat{I}_{J}\lvert b\rangle$ (see Fig.~\ref{fig:arrangementapparatusexchange}(a)) 
can intersect with the interaction region $D_{A}$ when the apparatus is 
located on the left, as shown in Fig.~\ref{fig:arrangementapparatusexchange}(a).

To address this question, let us consider an estimate based on real 
experimental data. For instance, in the experiment of 
Ref.~\cite{PhysRevX.13.021031}, a Bose--Einstein condensate of rubidium 
atoms was split into two parts separated by 
$\lvert \langle \vec{R}_{2} \rangle_{b} - \langle \vec{R}_{1} \rangle_{b} \rvert 
\sim 100\,\mu\text{m}$. Each condensate contained approximately $700$ 
rubidium atoms, so the dimensionless parameter in 
Eq.~\eqref{exhange_charachter} was 
$\bigl\lvert (M_{2}-M_{1})/(M_{2}+M_{1}) \bigr\rvert \sim 10^{-3}$. 
Hence, $\lvert \langle \vec{q} \rangle_{b} \rvert \sim 10^{-7}\,\text{m}$.

The measurement utilized laser radiation in the optical range, so the 
interaction region \(D_A\) of the apparatus is at least on the order of the 
optical wavelength—comparable to 
$\lvert \langle \vec{q} \rangle_{b} \rvert$. With these parameters, both 
regions $D_{1}(t)$ and $D_{2}(t)$ in the \enquote{exchange} states intersect 
the region $D_{A}$.

The parameters from Ref.~\cite{PhysRevX.13.021031} are used here solely 
as an illustration of experimentally achievable values, not as evidence 
that the apparatus interacted with both subsystems in that particular 
experiment. In that setup, interaction with both subsystems 
was ensured by the measurement method itself. At the same time, our 
estimate shows that, in addition to the interaction responsible for the 
observed effect and for determining the arrangement of the subsystems, 
the formation of the post-measurement state could also have been 
influenced by interaction with the unobserved \enquote{exchange}  configuration.

\subsection*{\enquote{Exchange} states and state formation time under measurement}

As discussed above, in a typical EPR-type scenario
\cite{AharonovBohmPhysRev.108.1070,CANTRELL1978499}
the probability of observing the particles of the composite system
$1 \otimes 2$ is initially significantly nonzero only within a small
spatial region $D$ (Fig.~\ref{fig:aparatusfar}), which is located far
from the measuring apparatus. As a consequence, in the initial state
the probability of detecting any particle of the system $1 \otimes 2$
in the vicinity of the apparatus is negligible.

However, the time evolution of the initial entangled state leads to an
increase in the probability of observing the particles of the system
near the apparatus. This increase is a consequence of the existence of
probability flows during the time evolution of the entangled state
(Fig.~\ref{fig:twoapparatusmeasurement}). The presence of the
\enquote{exchange} terms $\lvert b\rangle$ and
$\hat{I}_{J}\lvert b\rangle$ in the decomposition of the asymptotic
state gives rise to probability flows directed toward the apparatus
for particles of both subsystems. For particles of one subsystem,
this flow is provided by the state $\lvert b\rangle$, whereas for
particles of the other subsystem it is provided by the
\enquote{exchange} term $\hat{I}_{J}\lvert b\rangle$
(Fig.~\ref{fig:twoapparatusmeasurement}).

\begin{figure}[!htbp]
	\centering
	\includegraphics[width=1.0\linewidth]{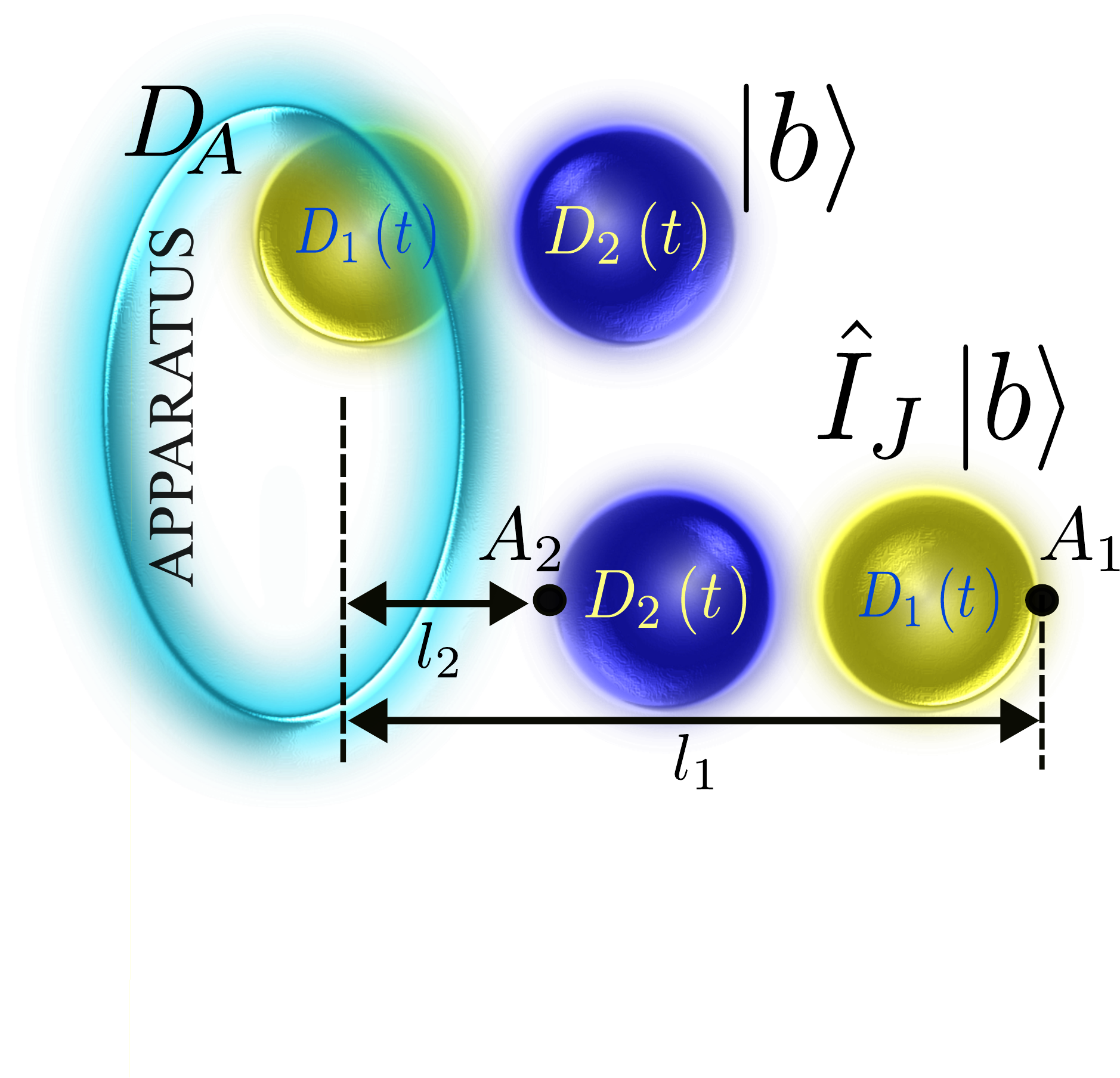}
	\caption{The relationship between the distances $l_{1}$ and $l_{2}$ that the probabilities of observing particles of subsystems $1$ and $2$ must overcome during the formation of the state after measurement.}
	\label{fig:configurationbijba1a2}
\end{figure}

In the general case, however, the magnitudes of these probability
flows are different. Indeed, as seen from
Fig.~\ref{fig:arrangementapparatusexchange} and as discussed above,
the region of the most likely observation of the subsystem with the
smaller mass is located closer to the apparatus than the analogous
region for the subsystem with the larger mass. As before, we continue
to consider the case $M_{2} > M_{1}$. Therefore, the region $D_{1}(t)$
is closer to the apparatus than $D_{2}(t)$
(Fig.~\ref{fig:configurationbijba1a2}). This implies that, in the
vicinity of the apparatus $D_{A}$, the probability flow associated
with particles of subsystem~1 is larger than the corresponding flow
for particles of subsystem~2.

As a consequence of this relation between the magnitudes of the
probability flows, the region $D_{1}(t)$ intersects the apparatus
vicinity $D_{A}$ earlier than $D_{2}(t)$. This means that the particles
of subsystem~1 begin to interact significantly with the apparatus
earlier than the particles of subsystem~2. Note, however, that this
does not imply that local interactions between particles of
subsystem~2 and the apparatus are strictly excluded.

This follows from the fact that the definition
\eqref{Integral_normuvanna_eps} does not imply that the probability of
observing particles outside the region
$D_{1}(t)\cup D_{2}(t)$ is strictly zero. According to this definition,
these probabilities are bounded from above by the probability
measurement error $\epsilon$ in
\eqref{Integral_normuvanna_eps}. Moreover, the analytic properties of
the probability amplitude, discussed above, require it to be nonzero
everywhere, except possibly at a discrete set of isolated points.
Consequently, the interaction terms between the particles of
subsystem~2 and the apparatus in
Eq.~\eqref{Classical_apparatus_interaction} are nonzero. However, their
magnitudes are much smaller than those associated with particles of
subsystem~1.

Thus, the observable measurement outcome is most likely determined by
the interaction between the particles of subsystem~1 and the
apparatus. Consequently, one of the eigenvalues of the measured
dynamical variable associated with subsystem~1 is most likely to be
observed.

The formation of this observable result takes a finite time interval.
For example, let us assume that the detection of subsystem~1 occurs via
the observation of its track. Track formation in a detector requires a
certain transfer of energy from the measured system to the particles
of the detector. For instance, a chemical reaction in a photographic
layer requires energy absorption, and track formation involves the
realization of such reactions in a microscopically large number of
molecules.

Since the apparatus is a classical object, its energy must change
continuously. Therefore, the absorption of the energy required for
track formation cannot be instantaneous and necessarily takes a
finite time.

The existence of the \enquote{exchange} terms leads to an increase in
this time interval. Indeed, let us adopt the assumption that the energy
required for track formation is absorbed by the apparatus together
with the absorption of subsystem~1 itself. In this case, track
formation is completed only when the absorption of subsystem~1 is
essentially complete.

Here, \enquote{complete absorption} means that the probability of
observing particles of subsystem~1 outside the detector region has
become negligible. Only at this stage can the apparatus absorb the
maximal amount of energy associated with subsystem~1, which is
necessary for the formation of a macroscopic track.

As discussed in the Introduction, these absorption processes proceed
via flows of probability and energy directed toward the apparatus
(Fig.~\ref{fig:configurationbijba1a2}). In particular, the probability
flow provides the transport of a significant portion of the
observation probability for particles of subsystem~1 over a large
distance $l_{1}$ (Fig.~\ref{fig:configurationbijba1a2}) from the point
$A_{1}$ to the apparatus. Therefore, the necessity to \enquote{gather} all the energy required for track formation within the apparatus implies that the corresponding observation probability must also be \enquote{gathered} there. This, in turn, through the existence of the \enquote{exchange}
state $\hat{I}_{J}\lvert b\rangle$
(see Fig.~\ref{fig:configurationbijba1a2}), requires probability
transport over a large distance, which inevitably leads to an increase
in the time required for the formation of the measurement outcome.

Let us now take into account that, during this sufficiently long time
interval, together with the processes discussed above, there also
exists a probability flow directed toward the apparatus and associated
with subsystem~2, as mentioned previously. This flow transports the
probability of observing particles of subsystem~2 over a distance
$l_2$ (Fig.~\ref{fig:configurationbijba1a2}), which is much smaller than
the distance $l_1$. This relation, $l_2 \ll l_1$, leads to the
possibility that the region $D_2$ may intersect with the apparatus
region $D_A$ (Fig.~\ref{fig:configurationbijba1a2}). It is sufficient to
transport the probability of observing particles of subsystem~2 from
the vicinity of the point $A_2$
(Fig.~\ref{fig:configurationbijba1a2}) in order to ensure a local
interaction of these particles with the apparatus during the
considered time interval of measurement outcome formation.

The realization of this possibility depends on the magnitude of the
probability flow directed from the point $A_2$ toward the apparatus.
As is well known \cite{Flugge2012practical}, the probability current
density contains the mass of the quantum system in the denominator.
Therefore, if the mass of subsystem~2 is large, the associated
probability flow can be small. A natural example of such a situation
is provided by the ionization of an atom or a molecule. In this case,
subsystem~1 is an electron, whereas subsystem~2 is a heavy ion.

In a linear combination of continuum-spectrum states of the ionized
system, probability flows associated with both the electron and the
ion are present. Among other directions, these flows include components
directed toward the distant detector as well as away from it. For the
electron, the flow directed toward the detector leads to the
possibility of its detection within the considered time interval.
By contrast, the probability flow associated with the heavy ion,
although nonzero, is negligible.

Hence, in this case there is significant interaction only between the electron and the apparatus. However, as discussed above, the time evolution of the system composed of the electron and the ion as an isolated system does not lead to spatial separation between them. The electric attraction between these particles reinforces this effect. This time evolution of the isolated system occurs before the probability of observing the electron near the apparatus increases to values sufficient to produce significant interaction, which would break the system's isolation. Thus, both possibilities exist simultaneously: to observe the electron near the apparatus and to observe it near the ion.

If the apparatus absorbs the electron as a consequence of their
interaction, as considered above, then the probability of observing the
electron decreases, among other locations, in the vicinity of the ion.
This decrease affects the ion through its local electric interaction
with the electron. Hence, although a significant direct local
interaction between the apparatus and the ion is impossible, an
indirect interaction mediated by the electron is possible instead.
There is some analogy here with field theory, where an interaction
between two distant particles is realized through two local
interactions between the field and each of the particles.

In our case, there are two local interactions: between the electron and
the apparatus, and between the electron and the ion. The local time
evolution of the state ensures the transport of dynamical variables
between the distant apparatus and the ion, in a manner similar to that
provided by a physical field.   

In the present case, we can again consider the distances $l_1$ and $l_2$
(Fig.~\ref{fig:configurationbijba1a2}). The first of these is the distance
over which the probability of observing the electron must be transported
toward the apparatus by the corresponding probability flow. The second distance is now understood as the distance over which an
exchange of physical quantities between the ion and the apparatus is
realized via probability flows arising during the time evolution of the
hybrid system composed of the electron, the ion, and the apparatus.
In the case under consideration, $l_1 > l_2$, but not $l_1 \gg l_2$, in contrast to the situation discussed above.
 
The relation $l_1 > l_2$ again provides the possibility of an indirect
interaction between the apparatus and the ion, mediated by local flows
of various dynamical variables between them. To determine whether this
possibility is realized, we must again compare the magnitudes of these
flows. To this end, we take into account that the probability amplitude
of the hybrid state of the system under consideration is defined on a
linear space $L$, which is a direct orthogonal sum of three subspaces,
\[
L = L_e \oplus L_i \oplus L_A .
\]
Here, $L_e$ is the subspace of the electron coordinates, $L_i$ is the
subspace of the coordinates of the particles composing the ion, and
$L_A$ is the subspace of the coordinates of the particles composing the
apparatus.

Let us note that all these flows transfer physical quantities through
changes in the observation probability of the electron. This implies
that the most significant components of the flows are their projections
onto the subspace $L_e$. Consequently, all these projections contain in
the denominator the same mass, namely the electron mass, in contrast to
the previous situation. As a result, the magnitudes of these projected
flows are of the same order.

Taking this into account, together with the relation $l_1 > l_2$, we
conclude that the flows between the apparatus and the ion have
sufficient time to affect the system state after the measurement, while
the observable outcome is being formed.
   
Therefore, even if the masses of subsystems~1 and~2 are substantially
different, the processes that occur in the system composed of both
subsystems and the apparatus during the measurement affect both
subsystems rather than only one of them. Hence, the paradox does not
arise in this case, just as in the cases considered previously.

The finite duration required for the formation of the observable outcome also suggests that local interactions between the apparatus and a distant system can influence the system in a specifically quantum way. This effect can be analyzed by considering the time evolution of states using the well-known method of Path Integrals \cite{feynman2010quantum}.

\begin{figure}[!htbp]
	\centering
	\includegraphics[width=1.0\linewidth]{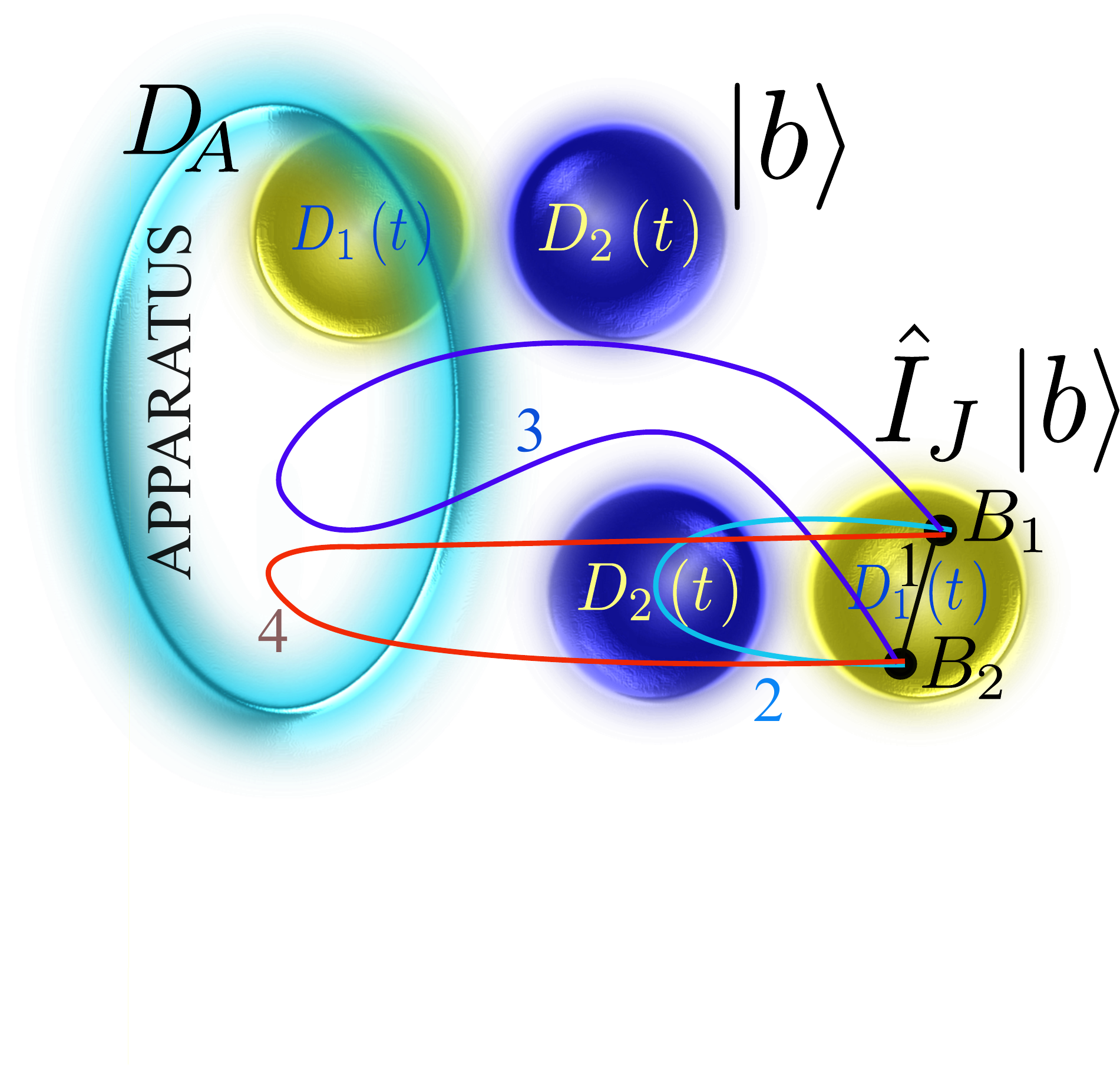}
	\caption{The presence of trajectories $2, 3, 4$ determines the impact of interactions with the apparatus and with subsystem $2$ on the dynamics of the state within the distant region ${{D}_{1}}\left( t \right)$ in the state ${{\hat{I}}_{J}}\left| b \right\rangle$. }
	\label{fig:feynmanpathbijbb1b2}
\end{figure}

Let us consider the transition amplitude of a certain particle of
subsystem~1 from point $B_{1}$ to point $B_{2}$ within the region
$D_{1}(t)$, as shown in Fig.~\ref{fig:feynmanpathbijbb1b2}. According to
Ref.~\cite{feynman2010quantum}, this amplitude can be obtained by summing
over all possible paths of the particle's coordinates as functions of
time. Thus, we can apply reasoning similar to that used in
Ref.~\cite{Wilson:1974sk} to our analysis.

Among all such paths, there are paths similar to those labeled
$1$, $2$, $3$, and $4$ shown schematically in
Fig.~\ref{fig:feynmanpathbijbb1b2}. Since paths~3 and~4 intersect the
region $D_{A}$, the action functional evaluated along these paths
necessarily accounts for the local interaction between the particle
and the apparatus through the corresponding terms in the Lagrangian of
the quantum system interacting with the apparatus. By analogous
reasoning, the presence of paths such as~2 and~4 accounts for the local
interaction between subsystems~1 and~2. 

Given that we are dealing with a quantum situation, the fact that these
paths may be far from the classical trajectory determined by the
principle of stationary action does not imply a cancellation of their
contributions to the transition amplitude.

As a consequence, the transition amplitude from the pre-measurement
state to the post-measurement state necessarily incorporates both the
local interactions of each subsystem with the apparatus and the local
interaction between the subsystems themselves. These interactions
therefore influence the formation of the post-measurement state.

The significance of this influence increases with increasing duration
of the measurement. Indeed, let us consider the contribution of a fixed
long path, such as path~3 or~4
(Fig.~\ref{fig:feynmanpathbijbb1b2}), to the transition amplitude
evaluated over different time intervals.

The endpoint conditions of a given path determine characteristic
values of the time derivatives of the coordinates along that path.
When the time interval is short, a long path necessarily corresponds to
large magnitudes of these derivatives. As a consequence, the kinetic
contribution to the action evaluated along such a path is much larger
than the corresponding contribution of the potential energy.
Consequently, the effects of interaction are suppressed by the large
kinetic term in the action integral.
         
Correspondingly, an increase in the time interval leads to a decrease in
the magnitudes of the characteristic values of the time derivatives.
As a result, the kinetic contribution to the action evaluated along the
path under consideration decreases, and the potential term in the
action is no longer suppressed.

Given that the existence of the \enquote{exchange} states
(Fig.~\ref{fig:configurationbijba1a2}) leads to the necessity of
transporting the observation probability over large distances
($l_1$ in Fig.~\ref{fig:configurationbijba1a2}), the measurement time
must be sufficiently large. This property provides the possibility for
local interactions between the particles of the apparatus and the
particles of the quantum system $1 \otimes 2$ to affect the
post-measurement state even without an intersection of the regions
$D_1(t)$ and $D_2(t)$ with the apparatus region $D_A$.

By analogous reasoning, a similar conclusion applies to the influence
of local interactions between the particles of subsystem~1 and those
of subsystem~2 when paths such as~2 and~4 in
Fig.~\ref{fig:feynmanpathbijbb1b2} are taken into account.

\subsection*{State formation time and lifetime of temporary spatial separation}

In Sec.~1, we considered a hypothetical situation in which, during the time evolution of a composite quantum system \(1 \otimes 2\), an approximate spatial separation between subsystems~1 and~2 occurs for a finite time interval. This means that, during this interval, the probability of observing particles belonging to different subsystems at distances that allow for significant interaction becomes small, although it does not become identically zero. In such a hypothetical case, the interaction between particles of different subsystems could be neglected to a good approximation.

However, as shown in Sec.~1, this situation cannot be realized in the asymptotic state at infinite time and is therefore necessarily temporary. Let us assume that, during the finite time interval in which this hypothetical separation is approximately realized, we attempt to perform a measurement on one of the subsystems, either~1 or~2. The question then arises whether the interaction between the subsystems can be neglected during the measurement process. The answer depends on the relation between the duration of the temporary approximate spatial separation and the duration of the measurement process. The purpose of this subsection is to compare these two durations.

First, let us note that the hypothetical existence of even an approximate spatial separation between the subsystems implies that the state under consideration is not an eigenstate of the total momentum operator of subsystem~1 nor of subsystem~2.
Indeed, in a total-momentum eigenstate the probability density must be spatially uniform. In the present case, however, the probability of observing the centers of mass of the subsystems at small separations is smaller than the probability corresponding to larger separations. Consequently, the probability density is not uniform, and the state cannot be a momentum eigenstate.

Given this, and taking into account that the interaction between the subsystems is assumed to be negligible, we conclude that the regions \(D_1(t)\) and \(D_2(t)\) must expand. This expansion occurs via probability flows in all directions. In particular, there are flows directed from \(D_1(t)\) toward \(D_2(t)\) and from \(D_2(t)\) toward \(D_1(t)\). As a consequence of these flows, either the spatial separation will disappear in time, or, at some stage of the expansion, the interaction between particles of different subsystems will become significant and prevent further expansion. In both cases, the neglect of the interaction will eventually cease to be valid.

As can be seen in Fig.~\ref{fig:configurationbijba1a2}, the distance between the regions \(D_1(t)\) and \(D_2(t)\) in the state \(\hat{I}_{J}\left| b \right\rangle\) is shorter than the distance between \(D_1(t)\) and the apparatus. Therefore, the probability flows that describe the expansion of the regions \(D_1(t)\) and \(D_2(t)\) have to cover a shorter distance to eliminate the approximate spatial separation than the probability flow from the region \(D_1(t)\) to the apparatus.

Thus, even if, before the measurement, the interaction between particles of the different subsystems could be approximately neglected, this approximation cannot remain valid during the measurement. In other words, if the particles of subsystems~1 and~2, due to their characteristics (for example, charges or spins), are able to interact, then it is impossible to perform a measurement in the system \(1 \otimes 2\) without interaction between particles belonging to different subsystems.

\section*{Appendix~2: Emergence of \enquote{exchange} terms in relativistic scattering}

In relativistic scattering processes, the statement concerning the existence of \enquote{exchange} terms in the decomposition of the asymptotic state can be justified for several reasons.

One of these reasons is the property of crossing symmetry of the scattering amplitude~\cite{PhysRev.96.1433,PhysRev.130.436,Scattering_Theory,book:667497,PhysRev.112.1344,Caron-Huot:2023ikn}. Although this is a purely theoretical result, it has been extensively used in the description of experimental data, as demonstrated, for example, in Refs.~\cite{book:667497,Rekalo:1999mt,Kohara:2019qoq,PhysRevD.102.054029,PhysRev.95.1612}. Beyond scattering processes, crossing symmetry has also been applied to the description of double beta decay~\cite{Primakoff:1959chj}.

We restrict ourselves here to the case of scalar particle scattering and label the particles participating in the scattering process as particle~1 and particle~2.
For the scattering amplitude \(A\) to describe elastic scattering in all three crossing channels~\cite{Scattering_Theory}, it must be a function of the three Lorentz invariants \(s\), \(t\), and \(u\),
\begin{equation}\label{Mandelstam}
	s = (P_1 + P_2)^2, \quad
	t = (P_1 - P_1')^2, \quad
	u = (P_1 - P_2')^2.
\end{equation}
Here \(P_1\) and \(P_2\) are the four-momenta of particles~1 and~2, respectively, in the initial state of the \(s\)-channel, \(P_1'\) and \(P_2'\) are the four-momenta of the particles in the final state of the scattering process within the same channel, and the squares denote Lorentz-invariant scalar products of the corresponding four-vectors with themselves in Minkowski space.

The requirement of crossing symmetry for an expression of the form \(A(s,t,u)\) is that it remain invariant under all permutations of its arguments \(s\), \(t\), and \(u\).
As a consequence, if we assign the variable \(s\) a fixed value \(s_1\) corresponding to an experiment performed in the \(s\)-channel, we obtain a function
\begin{equation}\label{s_channel}
	A_s(t,u) = A(s=s_1,t,u),
\end{equation}
which is symmetric under permutations of \(t\) and \(u\).
From the definition~\eqref{Mandelstam}, it follows that a permutation of \(t\) and \(u\) corresponds to a permutation of \(P_1'\) and \(P_2'\).
That is, if the scattering amplitude is nonzero for a final state in which particle~1 has four-momentum \(P_1'\) and particle~2 has four-momentum \(P_2'\), then it is also nonzero for the final state in which particle~1 has four-momentum \(P_2'\) and particle~2 has four-momentum \(P_1'\).
Therefore, in the decomposition of the final asymptotic state in terms of four-momentum eigenstates, together with any given basis state there necessarily appears the corresponding \enquote{exchange} state in which the momentum eigenvalues of the two particles are interchanged.

\begin{figure}[!htbp]
	\centering
	\includegraphics[width=1.0\linewidth]{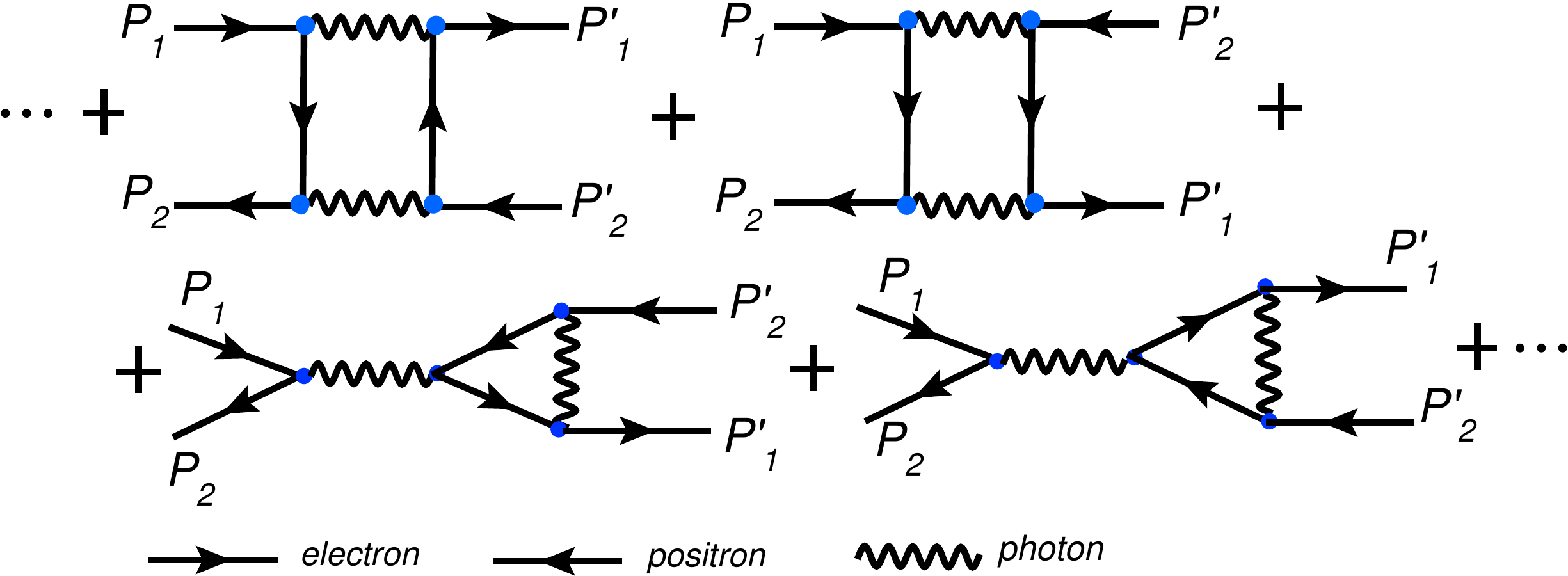}
	\caption{Examples of electron--positron scattering diagrams with exchanged external lines}
	\label{fig:exchangediagram}
\end{figure}

A second reason arises within dynamical scattering models that admit a Feynman-diagram representation, for example, in quantum electrodynamics.
In such models, the elastic scattering amplitude is given by a sum over an infinite set of Feynman diagrams \(\mathcal{F}\).
If we take an arbitrary elastic scattering diagram from the set \(\mathcal{F}\) and interchange the outgoing lines \(P_1'\) and \(P_2'\), we generally do not obtain another valid diagram from \(\mathcal{F}\), since such an interchange may violate charge conservation at certain vertices.
However, there exist diagrams for which this interchange is allowed, yielding another diagram belonging to \(\mathcal{F}\).
Some examples of such diagrams are shown in Fig.~\ref{fig:exchangediagram}.

Since the scattering amplitude must be represented as a sum over all allowed diagrams, this requirement implies, in particular, that one must include diagrams corresponding to all admissible connections of the external lines.
For elastic scattering diagrams containing two incoming and two outgoing external lines, this leads to the inclusion of diagrams that are related to each other by an exchange of the outgoing lines.
As a consequence, the resulting scattering amplitude describes a process whose final asymptotic state decomposition necessarily contains \enquote{exchange} basis states.

\begin{figure}[!htbp]
	\centering
	\includegraphics[width=1.0\linewidth]{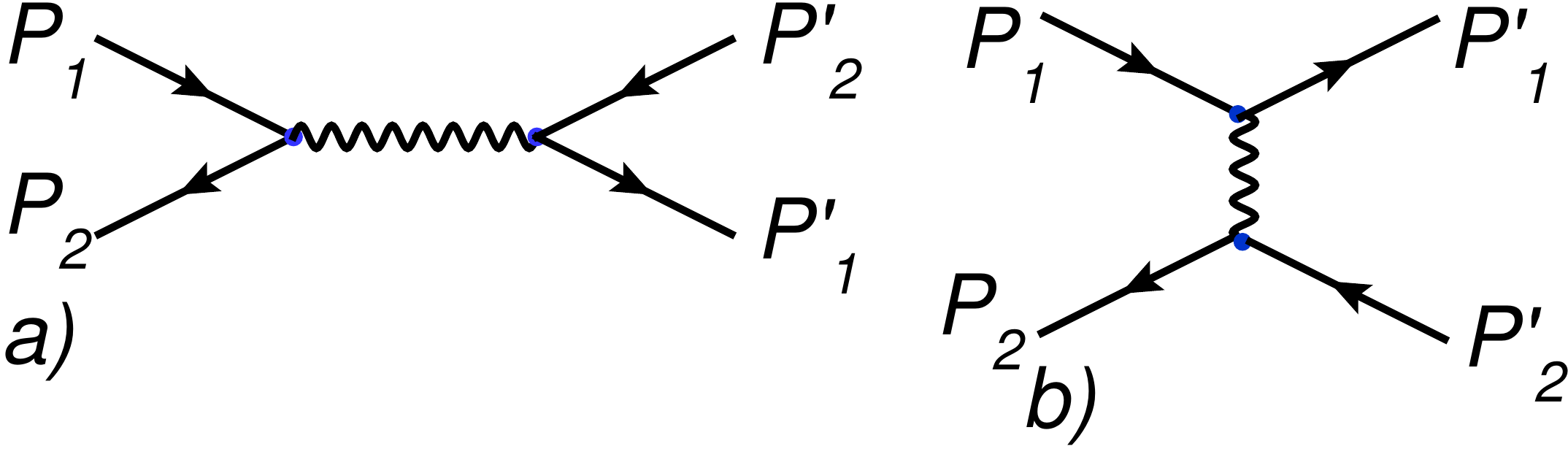}
	\caption{Example of a QED diagram symmetric under the exchange of lines \(P_1'\) and \(P_2'\) (a) and a diagram for which such an exchange is not allowed (b)}
	\label{fig:annigilationdiagram}
\end{figure}

Another mechanism leading to \enquote{exchange} contributions is related to the domain of definition of the scattering amplitude.
Even diagrams such as Fig.~\ref{fig:annigilationdiagram}(b), for which the exchange of \(P_1'\) and \(P_2'\) does not generate another valid diagram, can still contribute to \enquote{exchange} terms in the asymptotic state.
Indeed, exchanging the external lines in a diagram corresponds to exchanging all characteristics of the particles, including not only their four-momenta but also internal quantum numbers such as charge, spin, or polarization.
Changes in these internal quantum numbers may render a given diagram forbidden, whereas an exchange of four-momenta alone does not.
If \(Q_1\) and \(Q_2\) are four-vectors such that the assignment \(P_1' = Q_1\), \(P_2' = Q_2\) satisfies energy--momentum conservation, then the assignment \(P_1' = Q_2\), \(P_2' = Q_1\) satisfies it as well.
Therefore, if the configuration \(P_1' = Q_1\), \(P_2' = Q_2\) belongs to the domain of definition of the scattering amplitude, then the \enquote{exchange} configuration \(P_1' = Q_2\), \(P_2' = Q_1\) belongs to this domain as well.
As a consequence, the analytic expression corresponding to the diagram in Fig.~\ref{fig:annigilationdiagram}(b) yields nonzero contributions to the scattering amplitude for both assignments of the final-state momenta, although their magnitudes may differ substantially.

The relative magnitudes of the \enquote{exchange} terms in the asymptotic state depend on the dynamics of the scattering process.
This can be observed, for example, by comparing the results of Ref.~\cite{PhysRevLett.34.233} for the processes \(e^{-}e^{+} \to e^{-}e^{+}\) and \(e^{-}e^{+} \to \mu^{-}\mu^{+}\).
The ratio of the corresponding contributions can change significantly with collision energy, as demonstrated by comparing the low-energy measurements reported in Ref.~\cite{PhysRevLett.34.233} with the high-energy data from LEP~\cite{LEP-2}.
Nevertheless, the cross sections for both \enquote{exchange} configurations, corresponding to scattering angles \(\theta\) and \(180^\circ - \theta\), remain nonzero.
As discussed in Sec.~1, this provides experimental support for the considerations presented there concerning the presence of \enquote{exchange} contributions in the asymptotic entangled state of the composite system.

\section*{Appendix 3. Analysis of thought experiments with boxes}

Our aim here is to examine the differences between the standard EPR situation
\cite{EPRPhysRev.47.777,schredinger_1935,schredinger_1936,BohrNPhysRev.48.696,FurryPhysRev.49.393,AharonovBohmPhysRev.108.1070,PhysicsPhysiqueFizika.1.195}
and the well-known thought experiments with boxes
\cite{debroglie:jpa-00236174,Hardy,Norsen2005EinsteinsB}.

In these thought experiments, a single quantum particle is confined within a box
whose walls are impermeable to the particle.
An impermeable partition is then inserted into the box, dividing it into two
separate boxes.
Subsequently, the boxes are separated by a large distance, and a measurement is
performed in one of them.
This measurement either reveals the presence of the particle or does not.
Since the outcome of a measurement performed in one box determines the outcome
of a possible measurement performed in the distant box, a paradoxical situation
arises.

Two main differences can be emphasized that distinguish this situation from the
standard EPR scenario.

First, whereas the standard EPR considerations address isolated quantum systems,
the box thought experiments involve nonisolated systems.
An external intervention occurs during the insertion of an impermeable partition
into the box containing a quantum particle. Another external influence takes place when the boxes are transported over large distances relative to each other.
While in the standard EPR situation the spatial separation is assumed to result
from the time evolution of an isolated system, the boxes cannot be separated
without external intervention.
Therefore, in the analysis of box experiments, these external interactions must
be incorporated into the system Hamiltonian and into the boundary conditions.
As a consequence, the resulting Hamiltonian differs from that of the isolated
system appearing in Eq.~\eqref{Shredinger12}.

Second, whereas all terms in Eq.~\eqref{Shredinger12} are assumed to be analytic
functions, the impermeable partitions and the walls of the boxes correspond to
infinite potential barriers.
Moreover, when the boxes are transported over large distances, not only does the
height of the potential barrier become infinite, but its width becomes infinite
as well.
Thus, in experiments with boxes there are not only additional terms in the
Hamiltonian describing external interactions, but these terms also render the
Hamiltonian singular.

The extent to which the properties of a quantum system with a singular
Hamiltonian differ from those of a \enquote{regular} quantum system can be seen,
for example, from the analysis presented in Ref.~\cite{2002AmJPh..70..307G}.
As shown in Ref.~\cite{2002AmJPh..70..307G}, after the partition is inserted,
the probability amplitude does not split between the two parts of the box, in
contrast to what is assumed in other thought experiments
\cite{Norsen2005EinsteinsB,debroglie:jpa-00236174,Broglie1964TheCI,Hardy}.
Instead, due to the singular nature of the potential, the probability amplitude
becomes localized in one part of the box, while in the other part the probability
of observing the particle vanishes.
Consequently, there are no potential possibilities for the particle to be
present in the two parts of the box, but only in one of them, contrary to what is
claimed, for example, in Ref.~\cite{debroglie:jpa-00236174}.
That is, after the insertion of the partition, exactly the situation described
in Ref.~\cite{debroglie:jpa-00236174} as the only \enquote{reasonable}
interpretation is realized.
However, in contrast to the interpretation presented in
Ref.~\cite{debroglie:jpa-00236174}, this situation arises as a direct
consequence of the quantum dynamical description with a singular infinite
potential, rather than from assumptions that go beyond standard quantum mechanics.

\begin{figure}[!htbp]
	\centering
	\includegraphics[width=0.75\linewidth]{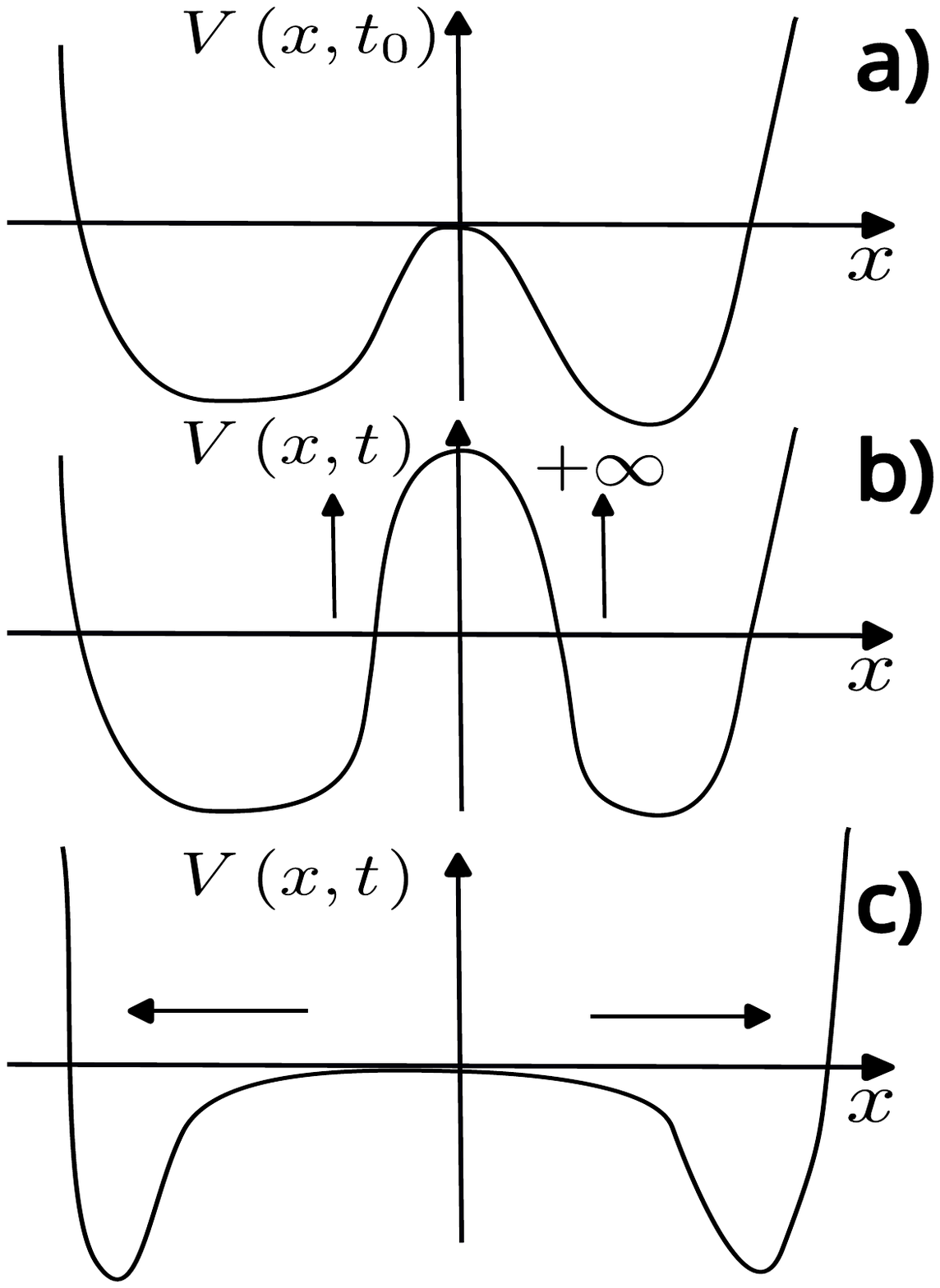}
	\caption{Potential with two minima (a) and its possible transformations over time into a potential with an infinitely high barrier (b) or a potential with an infinitely wide barrier (c).}
	\label{fig:potentialvwithtwominimums}
\end{figure}

This situation can be generalized to a more general one-dimensional problem.
Let us consider the state of a particle that, at the initial time, is
subject to a potential with two minima, as shown in
Fig.~\ref{fig:potentialvwithtwominimums}(a).
Since the barrier separating the two minima has a finite height and width, the
particle can tunnel between the two regions.
Suppose that, due to an external influence, the height of the barrier increases
with time, as shown in Fig.~\ref{fig:potentialvwithtwominimums}(b), or that its
width increases, as shown in Fig.~\ref{fig:potentialvwithtwominimums}(c).
Let us assume that in both cases the barrier height or width tends to infinity,
so that tunneling vanishes in the limit, similarly to the box experiment.

The question is whether, in this limit, the probability of observing the
particle splits between the two minima, or whether it becomes localized in one
of them, as in the particular case analyzed in
Ref.~\cite{2002AmJPh..70..307G}.
One might expect that the result obtained in the general situations illustrated
in Fig.~\ref{fig:potentialvwithtwominimums}(b) and (c) is similar to that found
in the specific case discussed in Ref.~\cite{2002AmJPh..70..307G}.

Suppose that, in the absence of tunneling, the probability of observing the particle
is distributed between the regions of the two minima.
Let us further assume that, during a measurement, the apparatus detects the particle
in the region of one of the minima.
Since tunneling is impossible, the probability amplitude cannot become localized
in the region where the apparatus is situated.
This would imply that a measurement performed with another apparatus in the region
of the other minimum could yield a non-zero probability of detecting the same particle
a second time.
Consequently, one would erroneously observe two particles instead of one.

As we can see, these external influences make the situation in the thought
experiments with boxes significantly different from that considered in the standard
EPR scenario \cite{EPRPhysRev.47.777,schredinger_1935,schredinger_1936,BohrNPhysRev.48.696,FurryPhysRev.49.393,AharonovBohmPhysRev.108.1070,PhysicsPhysiqueFizika.1.195}.

\clearpage 
\bibliographystyle{unsrtnat}
\bibliography{./references-windowsUTF8}

\end{document}